\documentclass[fleqn,usenatbib,useAMS]{mnras}

\usepackage{graphicx}
\usepackage{amsmath}
\usepackage{amssymb}
\usepackage{multicol}
\usepackage{bm}
\usepackage{pdflscape}

\newcommand{\Ka}{\,K$\alpha$ }

\usepackage[T1]{fontenc}
\usepackage{ae,aecompl}

\usepackage{newtxtext,newtxmath}
\title[Coronal Parameters of BAT AGN]{Fundamental X-ray Corona Parameters of \textit{Swift}/BAT AGN}

\author[Hinkle \& Mushotzky]{Jason T. Hinkle$^{1,2}$\thanks{E-mail: jhinkle6@hawaii.edu} and Richard Mushotzky$^{2,3}$
\\
$^{1}$Institute for Astronomy, University of Hawai`i, 2680 Woodlawn Dr., Honolulu, HI 96822, USA
\\
$^{2}$Department of Astronomy, University of Maryland, College Park, MD 20742, USA
\\
$^{3}$Joint Space-Science Institute, University of Maryland, College Park, MD 20742, USA}

\begin{document}
\label{firstpage}
\pagerange{\pageref{firstpage}--\pageref{lastpage}}
\maketitle

\begin{abstract}
While X-ray emission from active galactic nuclei (AGN) is common, the detailed physics behind this emission is not well understood. This is in part because high quality broadband spectra are required to precisely derive fundamental parameters of X-ray emission such as the photon index, folding energy, and reflection coefficient. Here we present values of such parameters for 33 AGN observed as part of the 105 month Swift/BAT campaign and with coordinated archival XMM-Newton and NuSTAR observations. We look for correlations between the various coronal parameters in addition to correlations between coronal parameters and physical properties such as black hole mass and Eddington ratio. Using our empirical model, we find good fits to almost all of our objects. The folding energy was constrained for 30 of our 33 objects. When comparing Seyfert 1 - 1.9 to Seyfert 2 galaxies, a K-S test indicates that Seyfert 2 AGN have lower Eddington ratios and photon indices than Seyfert 1 - 1.9 objects with p-values of $5.6 \times 10^{-5}$ and $7.5 \times 10^{-3}$ respectively. We recover a known correlation between photon index and reflection coefficient as well as the X-ray Baldwin effect. Finally, we find that the inclusion of the high energy Swift BAT data significantly reduces the uncertainties of spectral parameters as compared to fits without the BAT data.
\end{abstract}

\begin{keywords}
black hole physics -- galaxies: active -- galaxies: Seyfert -- quasars: general -- X-rays: galaxies
\end{keywords}

\section{Introduction}
Active galactic nuclei (AGN) are the luminous centers of galaxies where matter is actively accreting onto the central supermassive black hole (SMBH) \citep[e.g.,][]{rees84, antonucci93, urry95}. AGN are fundamentally multi-wavelength objects, with strong emission in the radio \citep[e.g.,][]{urry95, best05}, infrared \citep[e.g.,][]{sanders88, sanders96, stern05}, optical \citep[e.g.,][]{boroson92, kaspi00, kewley06}, ultraviolet \citep[e.g.,][]{murray95, haardt96}, X-ray \citep[e.g.,][]{reynolds97b, fabian12}, and gamma-ray \citep[e.g.,][]{hartman99, fermi15} wavebands. While the strength of emission in each energy band varies depending on the class of AGN, luminous X-ray emission is nearly ubiquitous across the entire family of AGN \citep[e.g.,][]{elvis78, mushotzky93}.

X-ray emission from AGN is commonly explained by a comptonizing region near the accretion disk surrounding the SMBH. Physical models for this region include a hot corona around the accretion disk \citep[e.g.,][]{haardt91}, a hot inner flow of electrons \citep[e.g.,][]{zdziarski04, done07b}, and the base of a jet along the rotation axis of the black hole \citep[e.g.,][]{matt91, martocchia96, henri97, markoff05}. Given the diversity of AGN X-ray spectra, it may be true that each of these processes are important in different objects.

The putative X-ray spectrum of an AGN is comprised of several main components. The underlying continuum radiation is roughly approximated by a power law, caused by inverse Compton (IC) scattering of lower energy UV photons from the accretion disk by hot, energetic electrons in the so-called corona \citep[e.g.,][]{antonucci93, nandra94}. At energies below $\sim$ 1 keV, there is a soft excess, the origin of which is largely unclear \citep[e.g.,][]{gierlinski04, sobolewska07, done07a, boissay16}. However, some recent observational \citep[e.g.,][]{jin12, petrucci18} and theoretical \citep[e.g.,][]{rozanska15, ballantyne20, petrucci20} studies have suggested the existence of a warm (kT $\sim 1$ keV), optically thick ($\tau \sim 10-20$) corona as the source of the soft excess. Furthermore, some studies have argued that the soft excess may be well-modeled by relativistically blurred ionized reflection \citep[e.g.,][]{crummy06, garcia19, xu21}. Between a few keV and up to hundreds of keV there is a reflection hump created by X-rays being IC scattered (or reflected) off the colder accretion disk or more distant material \citep[e.g.,][]{pounds90, george91, haardt93, nandra94, magdziarz95}. 

For most AGN, the strongest emission line seen in the X-ray is the fluorescent Fe \Ka line \citep[e.g,][]{fabian00, reynolds03}, which is often broad and exhibits relativistic effects due to its creation close to the SBMH \citep[e.g.,][]{pounds90, laor91, nandra97, reynolds97a, elvis00}. However, there is also a relatively narrower Fe \Ka component \citep{shu10}, which presumably originates from more distant material. In some cases, the line profile of broad Fe \Ka emission can be used to constrain the spin of the black hole \citep[e.g.,][]{reynolds08, bambi17, jones20, abarr21}.

Despite how pervasive X-ray emission from AGN is, the number of plausible physical scenarios suggest that the fundamental physics behind this X-ray emission is not well understood. This includes information on the shape, size, and location of the X-ray corona. Two of the most fundamental parameters that can be used to constrain the above characteristics are the photon index of the underlying power law and the cutoff (or folding) energy of the continuum. The folding energy, which is generally between 50 and 300 keV, is thought to be related to the temperature of the hot comptonizing electrons in the corona \citep{fabian15, tortosa18}. Additionally there seems to be a relationship between the folding energy and the photon index \citep[][]{dadina07, perola02, panessa11, deRosa12, molina13, ricci17}.

The lack of strong constraints on the physics of the X-ray emitting region is in part due to the lack of high signal to noise ratio (S/N) data on AGN in the requisite energy ranges at E $\gtrsim$ 75 keV where the effects of a folding energy are best defined. This requires the analysis of hard X-ray data in order to constrain the folding energy. Several previous studies have made use of the Nuclear Spectroscopic Telescope Array \citep[NuSTAR;][]{harrison13} satellite to probe the hard X-ray band up to roughly 60 keV. Studies such as \citet{brenneman14}, \citet{balokovic15}, \citet{fabian15}, \citet{matt15}, \citet{marinucci14a}, \citet{marinucci14b}, \citet{marinucci16} and \citet{tortosa17}, have provided constraints on coronal properties using \textit{NuSTAR} and other X-ray satellites such as Suzaku \citep{mitsuda07}, XMM-Newton \citep{jansen01}, and Swift \citep[][]{gehrels04}. To reduce uncertainties on the fundamental X-ray corona parameters and provide stronger constraints on correlations between them, it is important to include higher energy data (i.e. above ~75 keV). For this reason, we incorporate data from the Swift Burst Alert Telescope \citep[BAT; ][]{barthelmy05}, in addition to NuSTAR and XMM-Newton data in this paper.

As we will show, the addition of the BAT data results in smaller uncertainties on the values of the reflection coefficient, folding energy and photon index in the simplest model, \textsc{pexrav}, which adequately describes the data. While more sophisticated models exist (e.g. \textsc{borus} \citep{balokovic18} and \textsc{mytorus} \citep{murphy09}) the additional numbers of degrees of freedom often result in poorly constrained parameters for many sources, severely limiting the sample size. Since we find that \textsc{pexrav} provides an acceptable fit to the high energy continuum data for virtually all of our sources with reasonable parameters, we have decided to systematically use this relatively simple model rather attempt to fit more sophisticated models to this sample.

The Swift satellite \citep[][]{gehrels04} was launched in November 2004. One of its instruments, the Swift BAT \citep{barthelmy05} surveys the entire sky between 14 and 195 keV \citep[][]{oh18}, with a primary intent of quickly detecting gamma-ray bursts to initiate rapid follow-up. Because of the large field-of-view (FOV) of the BAT detector, there is a wealth of hard X-ray data on X-ray sources across the sky, including over a thousand AGN of various types. 

There are many X-ray telescopes that operate at less than 10 keV, an important energy range for studying the soft excess and Fe \Ka line properties of AGN. For our purposes, the XMM-Newton satellite provides the best combination of sensitivity, bandpass, and spectral resolution. The XMM-Newton satellite, launched in December 1999, has two X-ray instruments, a grating spectrometer and the European Photon Imaging Camera (EPIC). The EPIC instrument has three detectors, two MOS CCDs \citep{turner01} and one PN CCD \citep{struder01}.

The NuSTAR satellite \citep{harrison13} was launched in June 2012. It is comprised of two Wolter 1 telescopes which focus photons onto two separate focal plane modules (FPMs). Its unique multi-layer design gives NuSTAR a nominal energy range from 3 to 79 keV, a bandpass that was largely understudied in the past. NuSTAR is vital to our analysis in this paper because it links the lower energy data of XMM-Newton to the high energy data of the Swift BAT.

In this study we make use of archival data from the Swift BAT, NuSTAR, and XMM-Newton telescopes. We use stacked spectra from the BAT, two spectra (one per FPM) from NuSTAR, and whenever possible, three EPIC spectra from XMM-Newton to provide high-fidelity, high S/N spectra of the AGN in our sample. To understand the complex inner regions of AGN, it is vital to look for relationships between fundamental spectral parameters and physical quantities such as the Eddington ratio and SMBH mass. 

In this paper, we present a study of the derived spectral parameters of 33 AGN and expected correlations between them. This paper is organized as follows. Section \ref{sec:sample} will describe our sample of AGN. Section \ref{sec:methods} will detail our analysis methods and fitting procedures. Section \ref{sec:results} will display our results and Section \ref{sec:discussion} will provide discussion on these results. Section \ref{sec:conclusion} will summarize the paper and its main conclusions. Throughout the paper we assume a cosmology of $H_0$ = 70.0 km s$^{-1}$ Mpc$^{-1}$, $\Omega_{M} = 0.27$, and $\Omega_{\Lambda} = 0.73$.

\section{Sample} \label{sec:sample}

We selected our current sample of AGN from the 105-Month Swift-BAT All-sky Hard X-Ray Survey\footnote{\url{https://swift.gsfc.nasa.gov/results/bs105mon/}} \citep{oh18}. Compiled from almost nine years of observations, this survey has a sensitivity of 8.40 x 10$^{-12}$ erg s$^{-1}$ cm$^{-2}$ in the 14 -- 195 keV band over 90\% of the sky with eight-channel spectra averaged over the 105 month duration of the survey. However, AGN are known to be variable at all wavelengths \citep[e.g.,][]{huchra83, ulrich97, peterson04, drake09}. Therefore, in order to avoid issues caused by changing spectral shapes over the duration of the BAT 105-month survey, we restrict our analysis to sources with simultaneous XMM-Newton and NuSTAR observations. While the BAT data is stacked over several years, the earlier analyses of \citet{derosa08} and \citet{ricci11} using stacked INTEGRAL data provide a pathway for a similar analysis with more sensitive and higher S/N hard X-ray data from the Swift BAT. From the larger set of 1099 objects classified in the BAT 105-month survey as Seyferts, `Unknown AGN', and `Beamed AGN', we found 42 objects that had simultaneous XMM-Newton and NuSTAR observations with good quality data. 

Within this larger group of objects, we exclude several from our analysis, mainly due to known high amplitude spectral variability that caused offsets between the simultaneous XMM-Newton and NuSTAR data from the stacked BAT data. For NGC 3783, the line-of-sight column density is known to have changed significantly over time and therefore does not fit the assumptions of our simple models \citep{mao19}. NGC 1566 is a known changing-look AGN \citep[][]{parker19}, and it is likely that our simple model cannot capture the more complex physics occurring in this object. Mrk 335 has been shown to have long-term spectral trends in its X-ray light curves \citep{grupe12}. NGC 1365 is known to have variable line-of-sight absorption, with changes on sub-day timescales \citep{rivers15}. The spectral shape of Mrk 1044 is variable over the period of the BAT observations \citep{mallick18}. Additionally, for Mrk 1044 we find that when fitting the 2013 XMM-Newton and 2018 XMM-Newton spectra, we find a significant change in photon index, from $2.36 \pm 0.04$ in 2013 to $3.27 \pm 0.01$ in 2018, and thus does not fit our assumptions. 1H 0323+342 is variable in both X-ray flux and hardness ratio over time, suggesting a changing spectral shape \citep{mundo20}. 1H0419-577 is variable in both X-ray flux and spectral shape over time, in addition to a strong soft excess \citep{fabian05, jiang19}. Fairall 49 shows large spectral variability even on short timescales \citep{iwasawa16}. Finally, NGC 5548 is variable on roughly day timescales with significant changes in spectral parameters \citep[e.g.,][]{chiang00}.

Additionally, there were several BAT AGN with simultaneous XMM-Newton and NuSTAR data that we excluded from our sample due to low S/N data. As NuSTAR data links the low and high energy data, it fundamentally drives our ability to neatly constrain parameters. Therefore we impose a cutoff of 4000 counts in the NuSTAR band, below which we do not include an object in our sample. With fewer than 4000 NuSTAR counts, the constraints on fundamental parameters become poor, either with uncertainties several times larger than the rest of the sample or with a large number of upper/lower limits. The AGN excluded due to low NuSTAR counts were 3C 234.0, IC 588, IC 751, NGC 3718, AM 0224-283, SDSS J103315.71+525217.8, ESO 244-IG 030, and ESO 317- G 041.

While we only include AGN with simultaneous XMM-Newton and NuSTAR observations, there may be other objects in the BAT 105-Month Survey that have reasonably constant spectral shapes, and therefore the analysis presented in this paper could be applied to them even with XMM-Newton and NuSTAR observations at different times. Thus by requiring simultaneous observations, our selection criteria likely biases our sample towards well-studied and/or bright AGN.

For each object we used the NASA/IPAC Extragalactic Database\footnote{\url{https://ned.ipac.caltech.edu/}} \citep[NED;][]{helou91} to determine the classification and redshift of each AGN. Throughout the paper, we group our sources into Seyfert 1 - 1.9 objects and Seyfert 2 objects. We also used the High Energy Astrophysics Science Archive Research Center (HEASARC) nH calculator \citep[][]{hip4i16} to compute the Galactic column density along the line of sight. This allowed us to fit for the column density of the AGN host galaxy separately from the column density contribution of the Milky Way.

\section{Methods} \label{sec:methods}

\subsection{Data Acquisition and Reduction}
The data for this paper were obtained from the various archives corresponding to the different telescopes used.  The XMM-Newton Science Archive\footnote{\url{http://nxsa.esac.esa.int/nxsa-web/\#search}} has reduced spectra for each of the point sources in the observed FOV. For our XMM-Newton MOS and PN spectra, we used the spectra for the central AGN point source. We used the corresponding background and ancillary files from the XMM-Newton Science Archive in addition to the appropriate canned response functions. 

We obtained NuSTAR data from the HEASARC Browse server\footnote{\url{https://heasarc.gsfc.nasa.gov/db-perl/W3Browse/w3browse.pl}}. For the NuSTAR observations, we ran the \textsc{nuproducts} function on the cleaned event files to create appropriate spectral, background, ancillary, and response files. The source and background extraction regions varied slightly for each object but in general the source regions had radii of $\sim 1$ arcminute and the background region was an annulus with outer radius of 4 - 5 arcminutes with the corresponding source region subtracted. The source regions were centered on the AGN. Whenever possible, the background region was an annulus centered on the AGN. Otherwise, the background region was a source-free circle of 4 - 5 arcminutes in radius.

For our Swift BAT spectra, we used the stacked spectra for the 105 month duration of Swift observations from the 105-Month Swift-BAT All-sky Hard X-Ray Survey. In addition, a response matrix appropriate for all the BAT spectra was used. For reproducibility, we list the observation ID numbers for the AGN in our sample in Table \ref{tab:obsid}.

\begin{table} 
\caption{XMM-Newton, NuSTAR, and Swift BAT observation identification numbers for the sources in our sample.}
\label{tab:obsid}
\begin{tabular}{lccc}
\hline
Object & BAT ID & XMM-Newton ID & NuSTAR ID \\ 
\hline
Mrk1501 & 8 & 0795620101 & 60301014002  \\ 
Mrk1148 & 36 & 0801890301 & 60160028002 \\ 
Fairall9 & 73 &0741330101 & 60001130003 \\ 
Mrk359 & 77 & 0830551001 & 60402021006 \\ 
NGC931 & 129 & 0760530201 & 60101002002 \\ 
NGC1052 & 140 & 0790980101 & 60201056002 \\ 
NGC1068 & 144 & 0740060401 & 60002033002 \\ 
3C109 & 212 & 0795600101 & 60301011002 \\ 
3C120 & 226 & 0693781601 & 60001042003 \\ 
Ark120 & 266 & 0721600401 & 60001044004 \\ 
ESO362-18 & 269 & 0790810101 & 60201046002 \\ 
Mrk3 & 325 & 0741050101 & 60002049002 \\ 
IRAS09149-6206 & 447 & 0830490101 & 60401020002 \\ 
NGC3227 & 497 & 0782520201 & 60202002002 \\
NGC3998 & 579 & 0790840101 & 60201050002 \\ 
NGC4151 & 595 & 0679780301 & 60001111005 \\ 
Mrk766 & 608 & 0763790401 & 60101022002 \\ 
3C273 & 619 & 0414191101 & 10002020003 \\
NGC4579 & 1409 & 0790840201 & 60201051002 \\ 
NGC4593 & 631 & 0740920201 & 60001149002 \\ 
NGC4785 & 1411 & 0743010101 & 60001143002 \\ 
Mrk273 & 1430 & 0722610201 & 60002028002 \\ 
Mrk841 & 753 & 0763790501 & 60101023002 \\ 
Mrk1392 & 754 & 0795670101 & 60160605002 \\ 
3C382 & 984 & 0790600101 & 60202015002 \\
SwiftJ2127.4+5654 & 1111 & 0693781801 & 60001110005 \\ 
IIZw171 & 1143 & 0795620201 & 60301015002 \\ 
NGC7314 & 1157 & 0790650101 & 60201031002 \\ 
Mrk915 & 1161 & 0744490401 & 60002060002 \\ 
MR2251-178 & 1172 & 0763920601 & 60102025004 \\ 
NGC7469 & 1182 & 0760350801 & 60101001014 \\ 
Mrk926 & 1183 & 0790640101 & 60201029002 \\
NGC7582 & 1188 & 0782720301 & 60201003002 \\ 
\hline
\end{tabular}
\end{table} 

\subsection{Spectral Fitting} \label{sec:specfit}
We fit our spectra using version 12.10.1f of XSPEC \citep[][]{arnaud96}. We used two methods to fit the AGN in our sample. The choice of which model to use depended on which better fit the soft X-ray emission. The first approach, which generally worked for Type 1 AGN is a continuum comprised of a blackbody and an exponentially cut-off power law with reflection, photoelectric absorption, and a nominal Fe \Ka line: \textsc{constant$\times$TBabs$\times$TBabs$\times$(pexrav+zbbody+zgauss)}. The second, which generally worked for Type 2 AGN, is a continuum comprised of an exponentially cut-off power law with reflection, photoelectric absorption with a variable covering fraction, and a nominal Fe \Ka line: \textsc{constant$\times$TBabs$\times$TBpcf$\times$(pexrav+zgauss)}. Generally we fit the spectra between 0.5 keV and the upper end of the BAT bandpass. For some sources, we ignored below certain energies mainly to avoid the effects of a strong soft X-ray component or significant line emission over the soft X-ray bandpass. The Appendix details sources for which the fitted energy range begins above 0.5 keV. Using the estimated pile-up fractions from the XMM-Newton spectral reductions, we find no sources with a pile-up fraction above 2\%, with the mean and median fractions below 1\%. None of our sources are bright enough for significant NuSTAR pile-up.

For each approach, the constant allowed for an offset between the different telescopes and instruments used in our analysis. To constrain absorption, we used \textsc{TBabs} assuming the abundances of \citet{wilms2000}. By using two separate absorption components, with one frozen to the Galactic value, we can constrain the column density in the AGN host galaxy. The \textsc{pexrav} component \citep[][]{magdziarz95} is an exponentially cut off power law reflected off of neutral material. While this ``simple'' model may not perfectly encapsulate the detailed physics behind this emission, it works well for the varying S/N of our data and provides us with constraints on the fundamental parameters we wish to measure. The \textsc{zgauss} component was used to fit the Fe \Ka line.  

As mentioned previously, the salient difference between the two approaches is the parameter corresponding to the soft X-ray emission. The \textsc{zbbody} component is a redshifted blackbody, which, while it may not be an accurate physical description of the soft excess, fits the emission well in most cases \citep[e.g.,][]{jiang18}. The \textsc{TBpcf} parameter is a partial covering model, where clouds of gas in our line of sight absorb some fraction of the light from the AGN. Some Seyfert 1 - 1.9 objects were fit using the partial covering model rather than the blackbody model based on the existence of line emission in the soft X-ray suggesting an ionized absorber. As the goal of our paper is to characterize these sources with simple phenomenological models, we prioritize a uniform approach that works rather than finding the best fit for each object with more complex models. 

As opposed to \textsc{pexmon}, where the Fe \Ka line is included as part of the model, we fit the Fe \Ka line as a separate Gaussian component. We consider the partial covering model to be phenomenological in that it well describes objects not only with partial covering but independent soft components due to photoionized gas emission due to star formation and X-ray binaries. It is only in the very highest S/N observations that X-ray CCD data can constrain such physical components. Thus, for the sake of uniformity we have not added the other possibilities.

To estimate uncertainties on our best-fit values, we used the XSPEC routine \textsc{steppar} to find the 90\% confidence interval ($\Delta \chi^2 = 2.706$). For several of our objects we ran a Markov Chain Monte Carlo program in XSPEC for comparison. We found that in some cases, the two approaches yielded similar uncertainties, but for others the MCMC uncertainties were significantly smaller than those obtained using \textsc{steppar}. Thus, to avoid underestimating our uncertainties, we used \textsc{steppar} for the remainder of our analysis.

\begin{figure*}
 \includegraphics[width=0.325\textwidth]{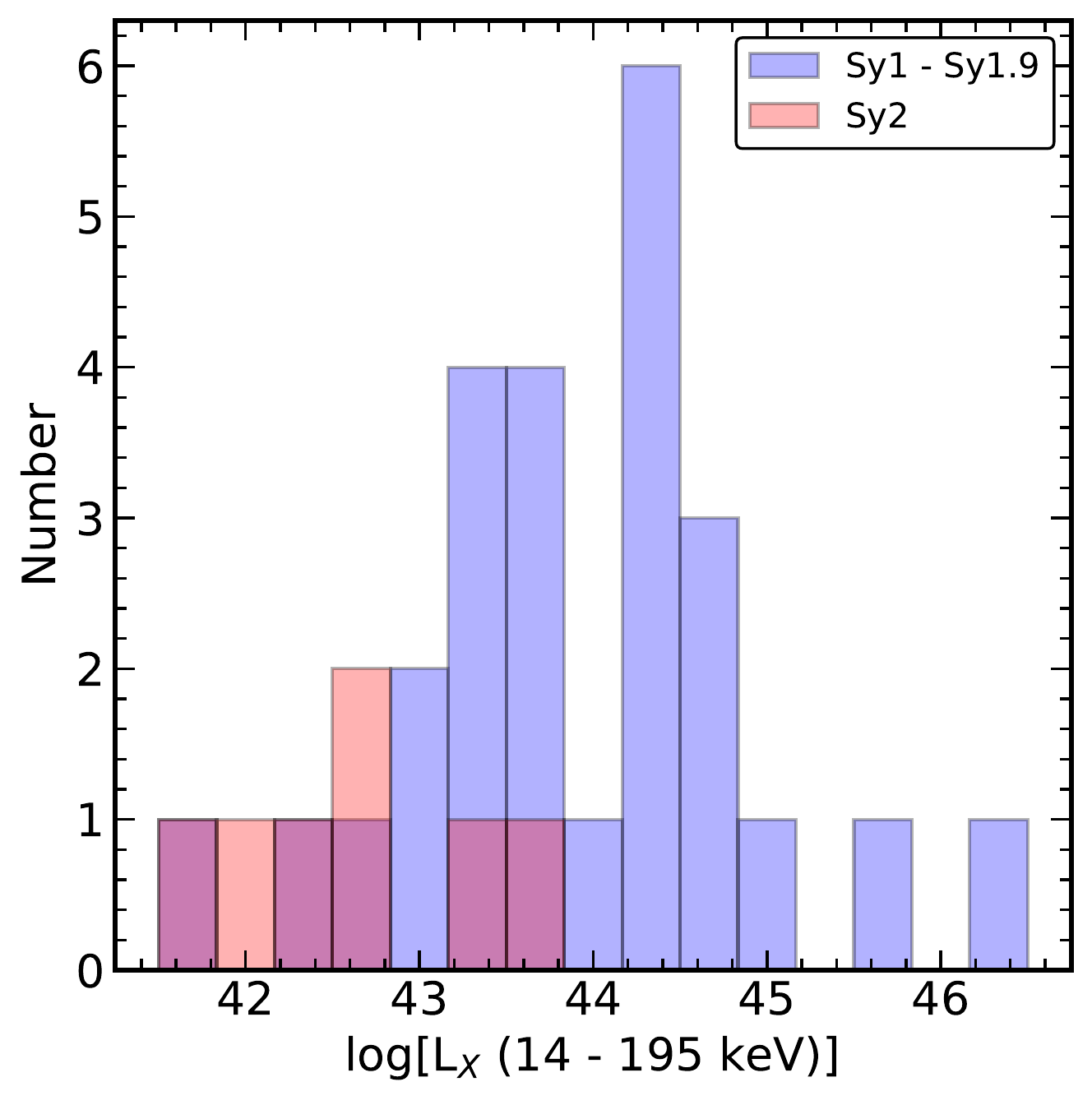}\hfill
 \includegraphics[width=0.335\textwidth]{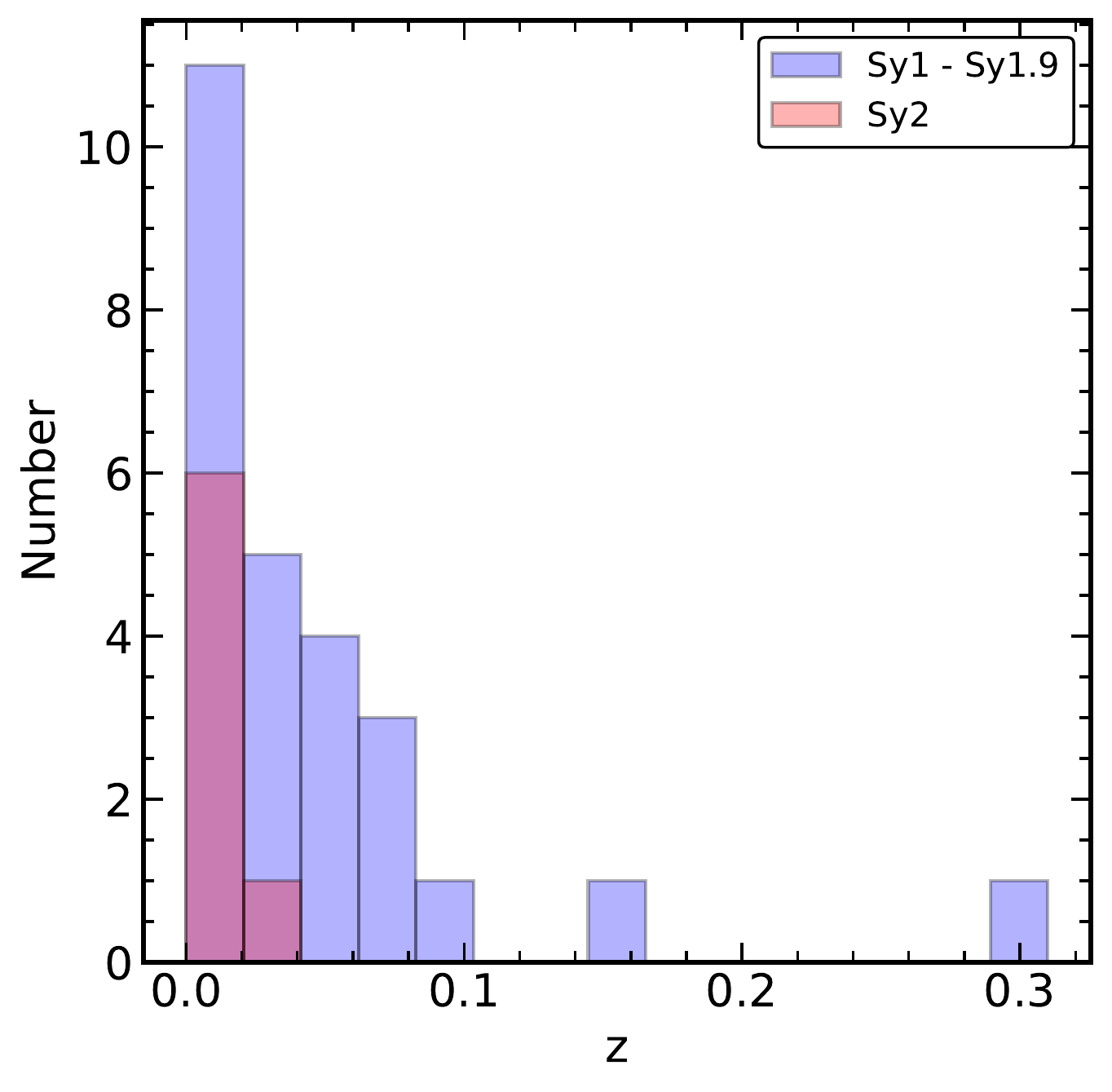}\hfill
 \includegraphics[width=0.34\textwidth]{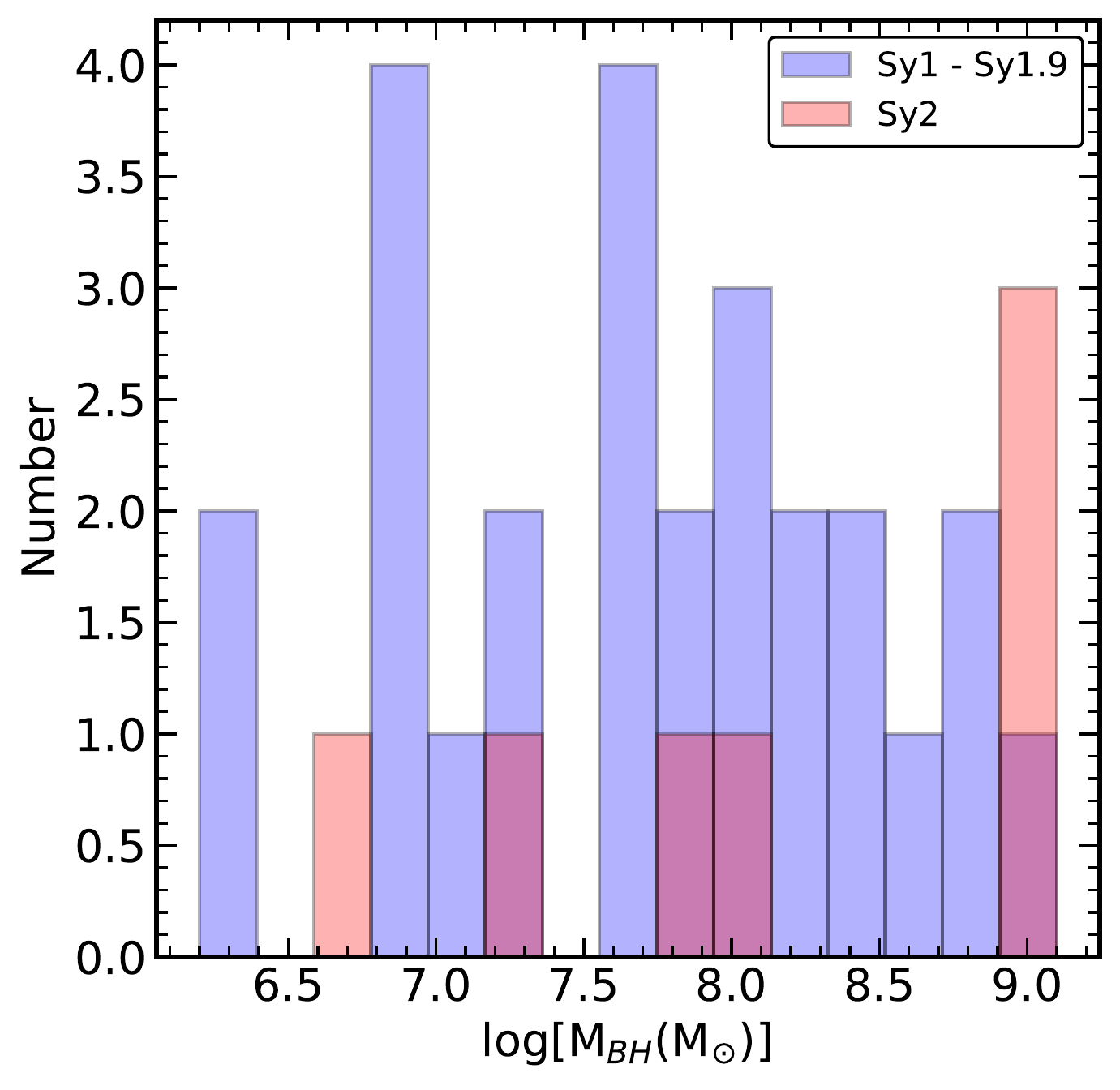}
 \caption{\textit{Left panel}: Histogram of hard X-ray luminosity in the 14 -195 keV band as measured by the Swift BAT. In each panel Seyfert 1 - 1.9 objects are shown in blue and Seyfert 2 AGN are shown in red. Seyfert 2 AGN are less luminous than the Seyfert 1 - 1.9 AGN. \textit{Middle panel}: Histogram of host galaxy redshift. The large majority of AGN are local with z < 0.1.  \textit{Right panel}: Histogram of central supermassive black hole mass, which is relatively flat.}
 \label{fig:hists}
\end{figure*}

\subsection{Physical Parameters of Sample}
Whenever possible, we obtained black hole masses for the objects in our sample from BASS \citep[BAT AGN Spectroscopic Survey;][]{koss17, ricci17}. Unfortunately BASS only included parameters for the Swift BAT 70-Month Hard X-ray Survey\footnote{\url{https://swift.gsfc.nasa.gov/results/bs70mon/}}, which does not include some of the objects in this paper. When not in BASS, we found black hole masses in the literature if possible. In this pursuit, we made use of the AGN Black Hole Mass Database \citep{bentz15}, to search for well-studied AGN with virial masses. For several of the AGN, there were no published black hole masses. We next searched the literature for velocity dispersion measurements and applied the M-$\sigma$ relation of \citet{gultekin09}. Finally, if none of these were possible, we applied the scaling relation between bulge near-infrared (NIR) luminosity and SMBH mass of \citet{marconi03}. We assumed that the AGN contributes 33\% of the total flux in the 2MASS passbands (the median value in \citealp{burtscher15}).

For each of the sources in our sample, we compute the bolometric luminosity from the 14 - 195 keV luminosity following \citet{winter12}. The Eddington ratios were calculated using these bolometric luminosities and the appropriate black hole mass. Three of the objects in our sample, 3C 109, 3C 273 and MR2251-178, were found to be super-Eddington. For the case of 3C 109 however, we note that recent X-ray analysis has called the virial SMBH mass into question \citep{chalise20}. Additionally, the black hole mass for MR2251-178 comes from NIR scaling relations, so the super-Eddington ratio found here is tenuous. Several of the AGN in our sample have Eddington ratios below $1 \times 10^{-3}$, where there is a predicted change in the nature of accretion and where transitions between high and low states in galactic black holes are often seen \citep[e.g,][]{merloni03, done07b}. We note that these objects do not appear to have anomalous properties compared to the entire sample.

Distributions of the key physical parameters of our sample are shown in Figure \ref{fig:hists}. These are the hard X-ray luminosity (in the 14 - 195 keV band), redshift, and black hole mass. The luminosities of the AGN in our sample range over 5 orders of magnitude, with Seyfert 2 objects tending to have lower luminosities, consistent with the properties of the entire BAT sample \citep{ricci17}. The overwhelming majority of the AGN in our sample are nearby, with a redshift of less than 0.1. Only two AGN in our sample are at higher redshift: 3C 109 (z = 0.3056) and 3C 273 (z = 0.1583). The black hole masses in our sample are roughly evenly distributed between $10^{6.3}$ and $10^{9.0}$ M$\odot$, with no significant difference between Seyfert 1 - 1.9 and Seyfert 2 AGN.

It is important to note that the available black hole masses in the literature are often derived using different methods. While we include references and calculated uncertainties for the black hole masses when possible, this likely introduces noise into any of our attempts to look at correlations with black hole mass. For instance, \citet{guo20} suggest that AGN variability can introduce a $\sim 0.3$ dex scatter in single epoch SMBH masses. We mitigate these issues as much as possible by calculating bolometric luminosity and Eddington ratio using a consistent approach.

\section{Results} \label{sec:results}

\subsection{Coronal Parameters}

With our phenomenological model we are able to derive a number of fundamental parameters for the X-ray emitting corona. These include the photon index, folding energy, and reflection coefficient. Additionally, we obtain constraints on line of sight column density, equivalent width of the Fe \Ka line, and soft excess parameters. In this section we will summarize our results, look for correlations between parameters, and note the differences between the Seyfert 1 - 1.9 and Seyfert 2 classes of objects in our sample. Table \ref{tab:Kendall} lists the results of using Kendall $\tau$ test to search for correlations between various parameters. Table \ref{tab:K-S} shows the results of a K-S test \citep{massey51} comparing the Seyfert 1 - 1.9 and Seyfert 2 classes of AGN. Table \ref{tab:param} lists the properties of the AGN in our sample as well as physical and fundamental parameters of the X-ray corona.  Unless otherwise stated, errors and limits in this paper are reported to the 90\% confidence level.

\begin{figure*}
\centering
 \includegraphics[width=.49\textwidth]{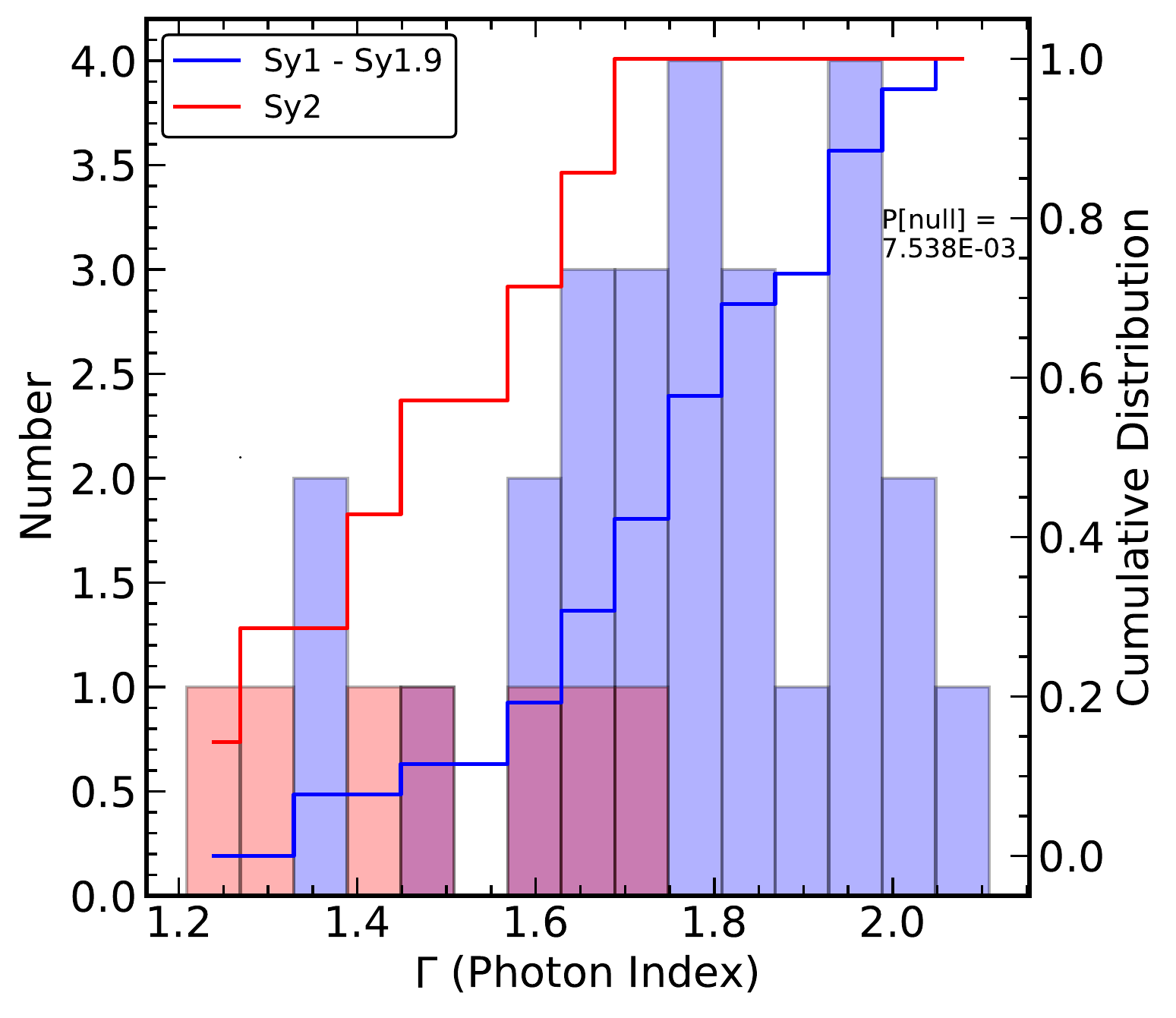}\hfill
 \includegraphics[width=.49\textwidth]{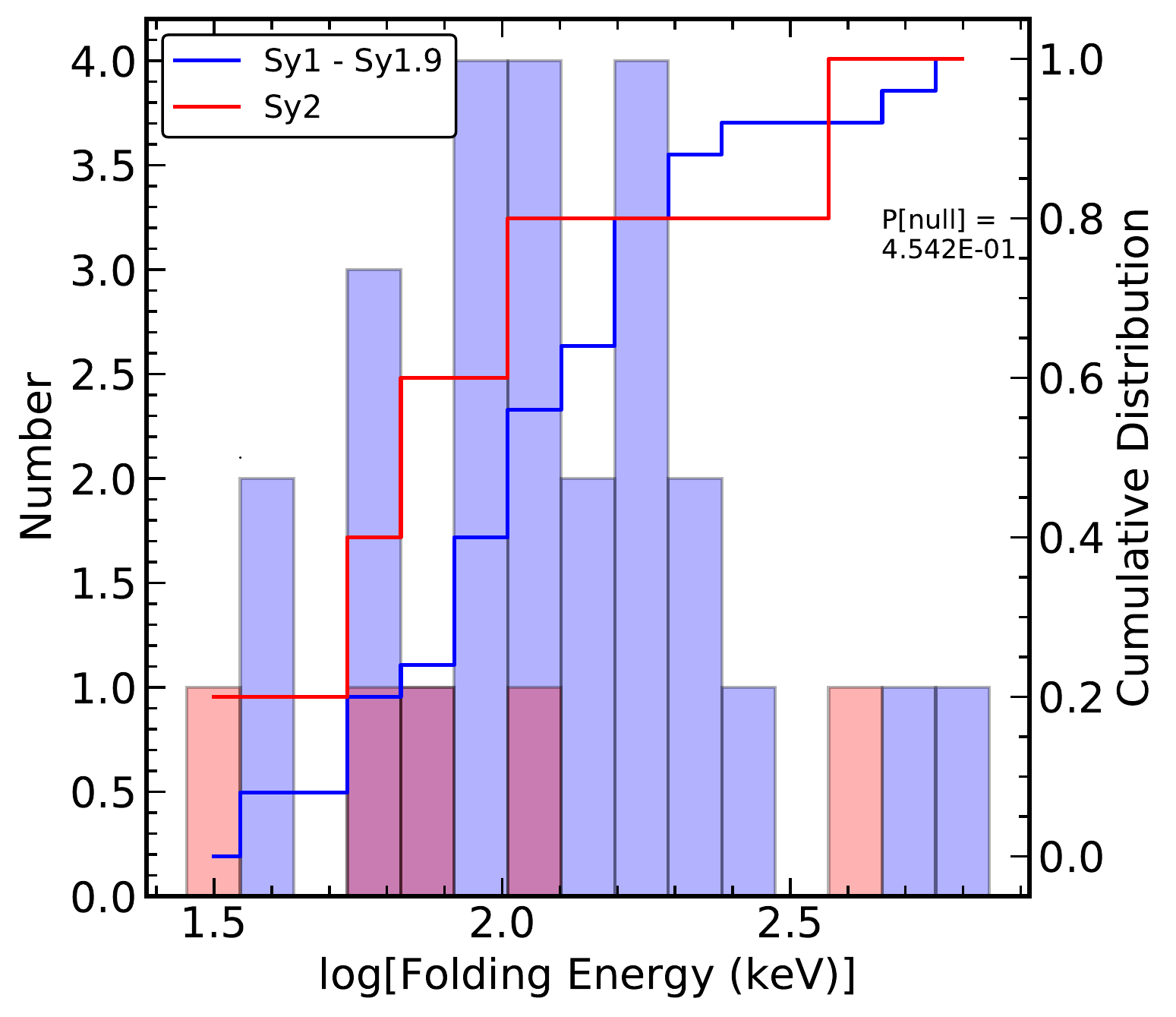} \\
 \includegraphics[width=.49\textwidth]{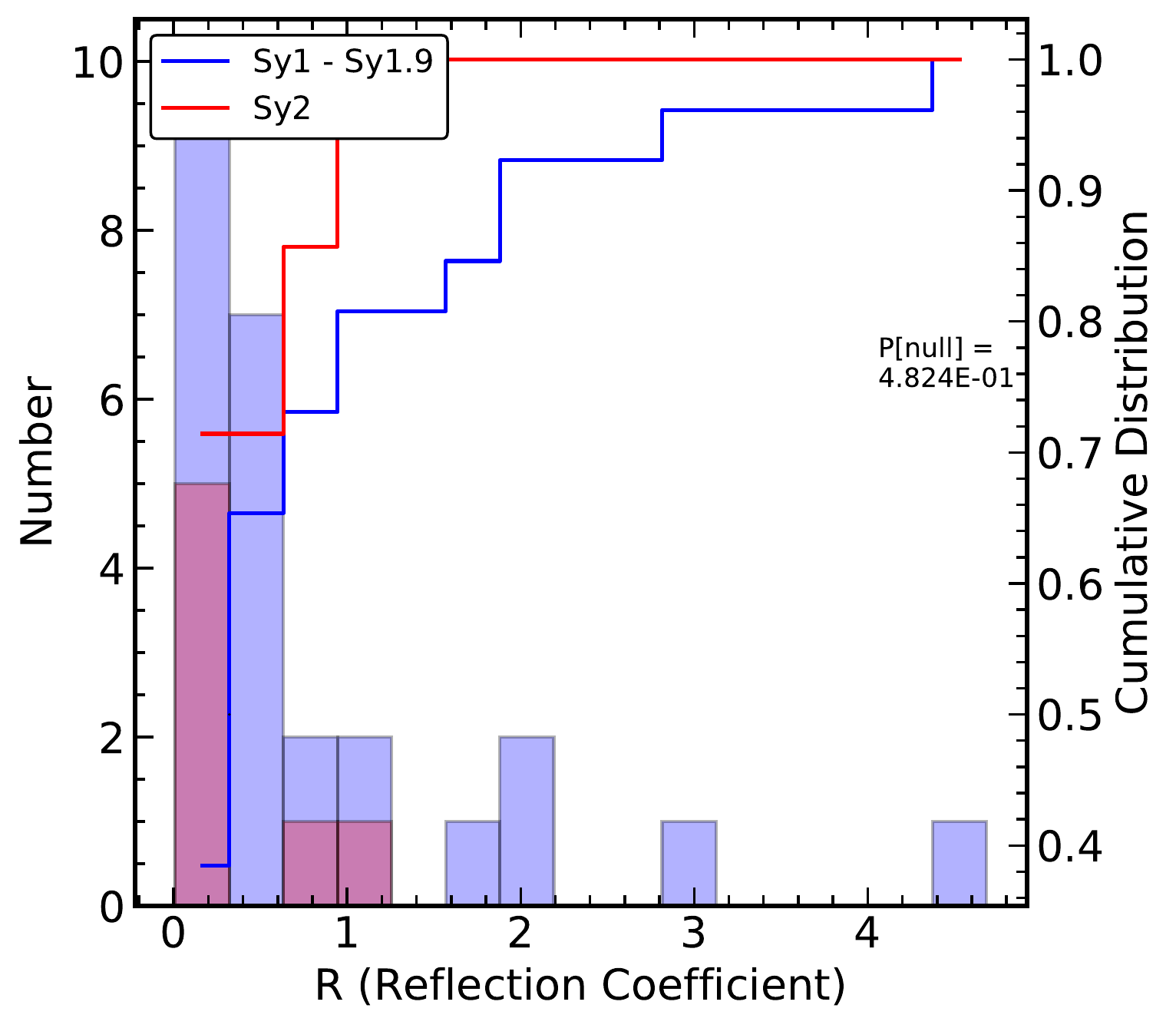}\hfill
 \includegraphics[width=.48\textwidth]{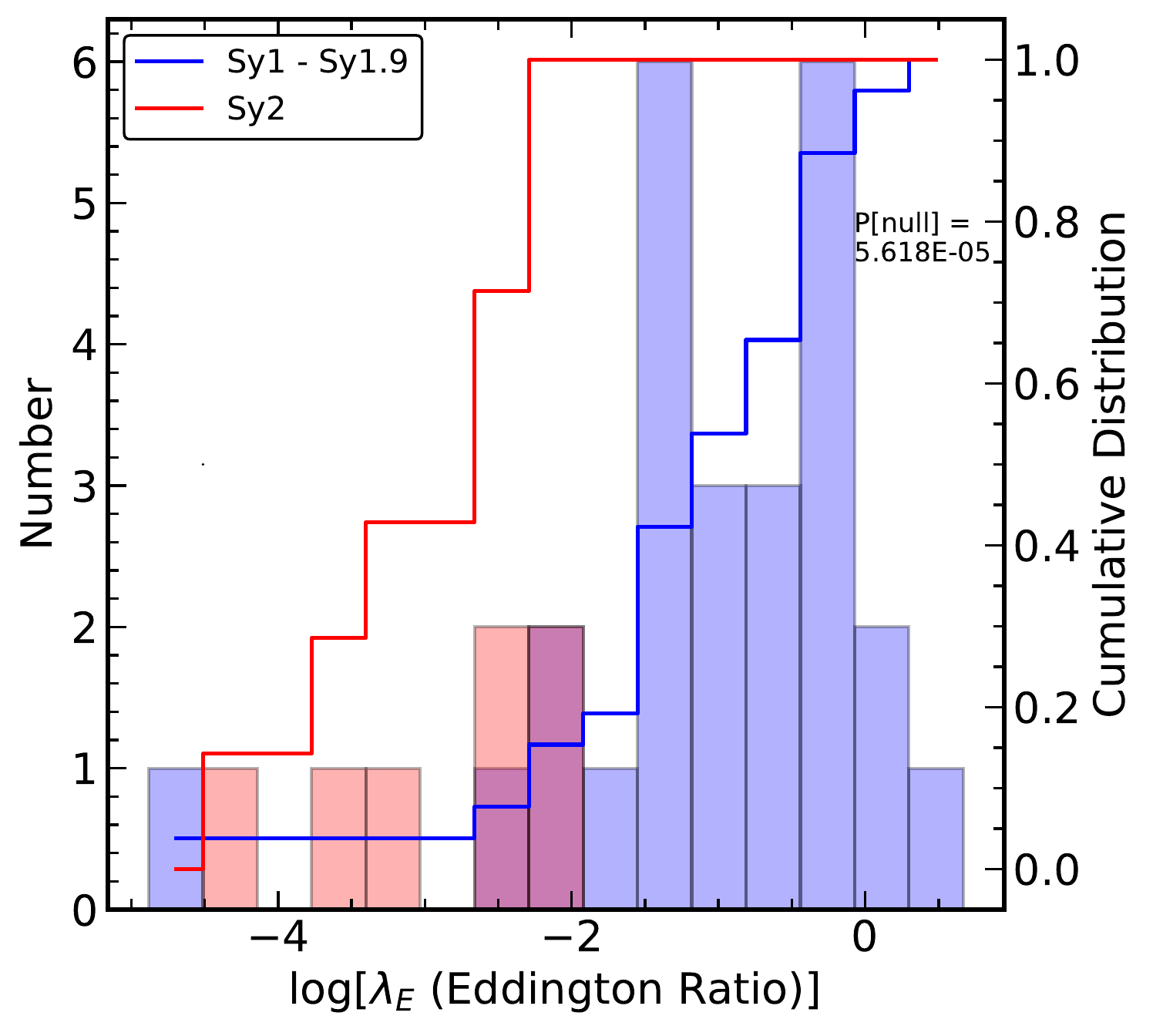}
 \caption{\textit{Upper left panel}: Histogram of photon index and corresponding cumulative distribution function. In each panel Seyfert 1 - 1.9 objects are shown in blue and Seyfert 2 AGN are shown in red. The value of P[null] from the K-S test indicates that the distributions of photon index for the different Seyfert classes are likely to be  drawn from different distributions. \textit{Upper right panel}: Histogram of the logarithm of the folding energy and corresponding cumulative distribution function for constrained values only. \textit{Lower left panel}: Histogram of the reflection coefficient and corresponding cumulative distribution function. \textit{Lower right panel}: Histogram of the logarithm of the Eddington ratio and corresponding cumulative distribution function. The Seyfert 2 AGN have significantly lower Eddington ratios than the Seyfert 1 - 1.9 objects as indicated by the K-S test.}
 \label{fig:KS_tests}
\end{figure*}

Figure \ref{fig:KS_tests} shows the distributions of photon index, folding energy, reflection coefficient, and Eddington ratio separated into Seyfert 1 - 1.9 and Seyfert 2 classes. For each of these parameters we use the K-S test to compute the probability of the null hypothesis that the samples are drawn from the same distribution. There are two parameters for which the Seyfert 1 - 1.9 and Seyfert 2 objects show significantly different distributions. These are the photon index and the Eddington ratio. The Seyfert 2 galaxies tend to have harder spectral slopes than the Seyfert 1 - 1.9 galaxies, consistent with the entire BAT sample \citep{ricci17}. The Eddington ratios of Seyfert 2 AGN is lower than for Seyfert 1 - 1.9 galaxies, consistent with previous results \citep[e.g.,][]{marinucci12}. The Eddington ratio difference is also supported by the recent discoveries of optical changing-look AGN, where as the source brightens emission lines become broader, which has been suggested to be a result of an increased accretion rate \citep[e.g.,][]{shappee14, denney14, yang18}.

The median column density for our sample was $3.6 \times 10^{20}$ cm$^{-2}$, including physically reasonable upper limits. For some of our objects, the best-fit column density was extremely low, so we froze it to $1 \times 10^{19}$ cm$^{-2}$ (a physically plausible limit) when fitting for the other parameters. The median column density for Seyfert 1 - 1.9 objects was $1.1 \times 10^{20}$ cm$^{-2}$. This is less than the median column density for Seyfert 2 galaxies, which was $37.4 \times 10^{22}$ cm$^{-2}$. This is as expected, as the unified model of AGN \citep[e.g.,][]{antonucci93} suggests that our line of sight to the central SMBH in Seyfert 2 galaxies is obscured by a dusty torus. The K-S test between the two Seyfert classes yields a p-value of $5.5 \times 10^{-3}$, supporting this conclusion. Similarly, as expected, all of the objects for which the measured column density was consistent with zero were Seyfert 1 - 1.9 galaxies. Only one of the AGN in our sample, NGC 1068, is Compton thick, with a column density of $(7.40^{+0.54}_{-0.62})\times 10^{24}$ cm$^{-2}$.

The median photon index for our sample was $1.72$. Figure \ref{fig:KS_tests} shows that the distribution of photon indices is consistent with those expected from Comptonization \citep{zdziarski85}. As seen in previous work \citep[e.g.,][]{ricci17}, the median photon index is slightly harder for the Seyfert 2s in our sample, $1.46$, as compared to $1.77$ for the Seyfert 1 - 1.9 AGN. This can additionally be seen in Figure \ref{fig:KS_tests}, where the K-S test indicates that the photon indices for the different Seyfert classes may be drawn from different distributions, with a p-value of $7.5 \times 10^{-3}$.

\begin{figure}
 \includegraphics[width=\columnwidth]{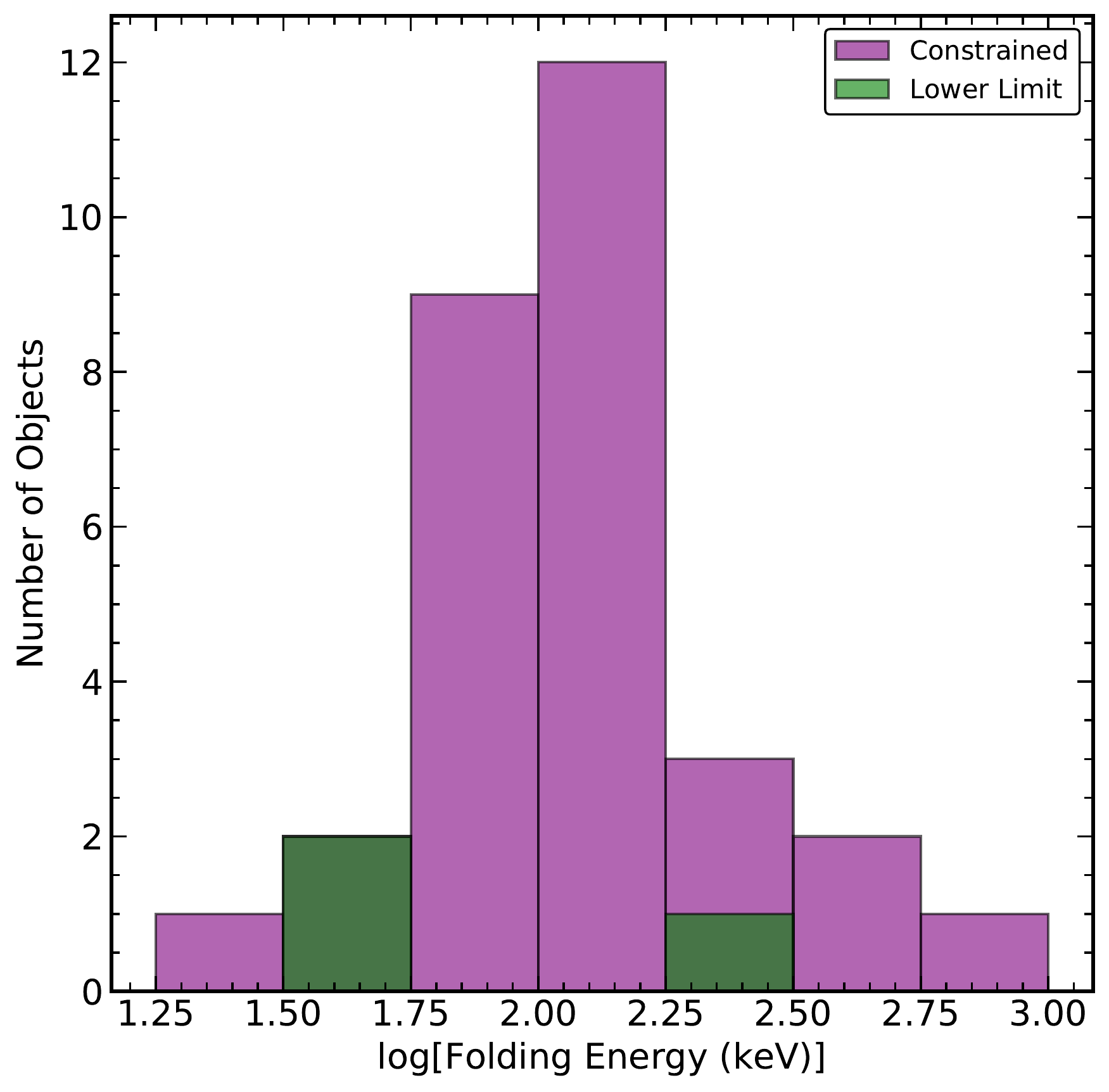}
 \caption{Histogram of the logarithm of the folding energies. The purple bars represent values of the folding energies that were constrained in the fit. The green bars are lower limits for those objects for which the folding energy could not be constrained. The lower limits on folding energy lie at slightly lower values than the constrained energies. The majority of objects have folding energies less than 300 keV and there are no objects with limits above 1022 keV (twice the electron rest mass).}
 \label{fig:folde_hist}
\end{figure}

For a large majority of our sample (30 out of 33 objects), we are able to constrain the folding energy rather than simply obtaining a lower limit. The median constrained folding energy for our sample was $112.6$ keV, somewhat higher than the median lower limit of $53.0$ keV. The minimum and maximum constrained folding energies were $28.4^{+7.7}_{-4.0}$ and $700.2_{-122.7}^{+187.8}$ keV respectively, indicating the very wide range of this parameter and its fractional uncertainties. Figure \ref{fig:folde_hist} shows the distribution of folding energies, both for constrained values and lower limits. As expected from observations of the cosmic X-ray background \citep{gilli07}, they tend to cluster below approximately 300 keV. This is consistent with what has been seen in previous studies \citep{ghisellini93, stern95, fabian15}. Figure \ref{fig:KS_tests} indicates that there is no clear separation in the distributions of Seyfert 1 - 1.9 and Seyfert 2 AGN with respect to folding energy. Interestingly, the lower limits on folding energy are disproportionately found for Seyfert 2s. Two of the seven Seyfert 2 AGN in our sample have lower limits on folding energy, as compared to one of twenty-seven for the Seyfert 1 - 1.9 objects. We suspect this is due to the smaller effective bandpass over which Seyfert 2 AGN are fit due to the high columns, thus allowing more uncertainty in the other derived parameters (photon index and reflection coefficient).

The reflection coefficient from our fit represents the solid angle of material that the X-rays in the reflection hump see, in units of pi. The median reflection coefficient for our sample was $0.41$, including physically reasonable upper limits. For some of our objects, the best-fit reflection coefficient was extremely low, so we froze it to 0.001 when fitting for the other parameters. While some of these objects still have a detected Fe \Ka line, this is likely either due to a high Fe abundance or material with a low Compton optical depth but a moderate column density. A K-S test of the reflection coefficients gives a p-value of $0.48$ indicating that the distributions of Seyfert 1 - 1.9 and Seyfert 2 reflection coefficients are statistically consistent. This is surprising, as the column density distributions are different and some models link these variables \citep[e.g.,][]{murphy09}. Similar to the folding energy, upper limits on reflection coefficient are found more often for Seyfert 2 AGN (4/7) than Seyfert 1 - 1.9 (7/26) objects. The constrained reflection coefficients for the AGN in our sample are qualitatively similar to previous results in the literature \citep[e.g.,][]{zdziarski99, vasylenko15, lubinski16, lanz19}.

\begin{figure}
 \includegraphics[width=\columnwidth]{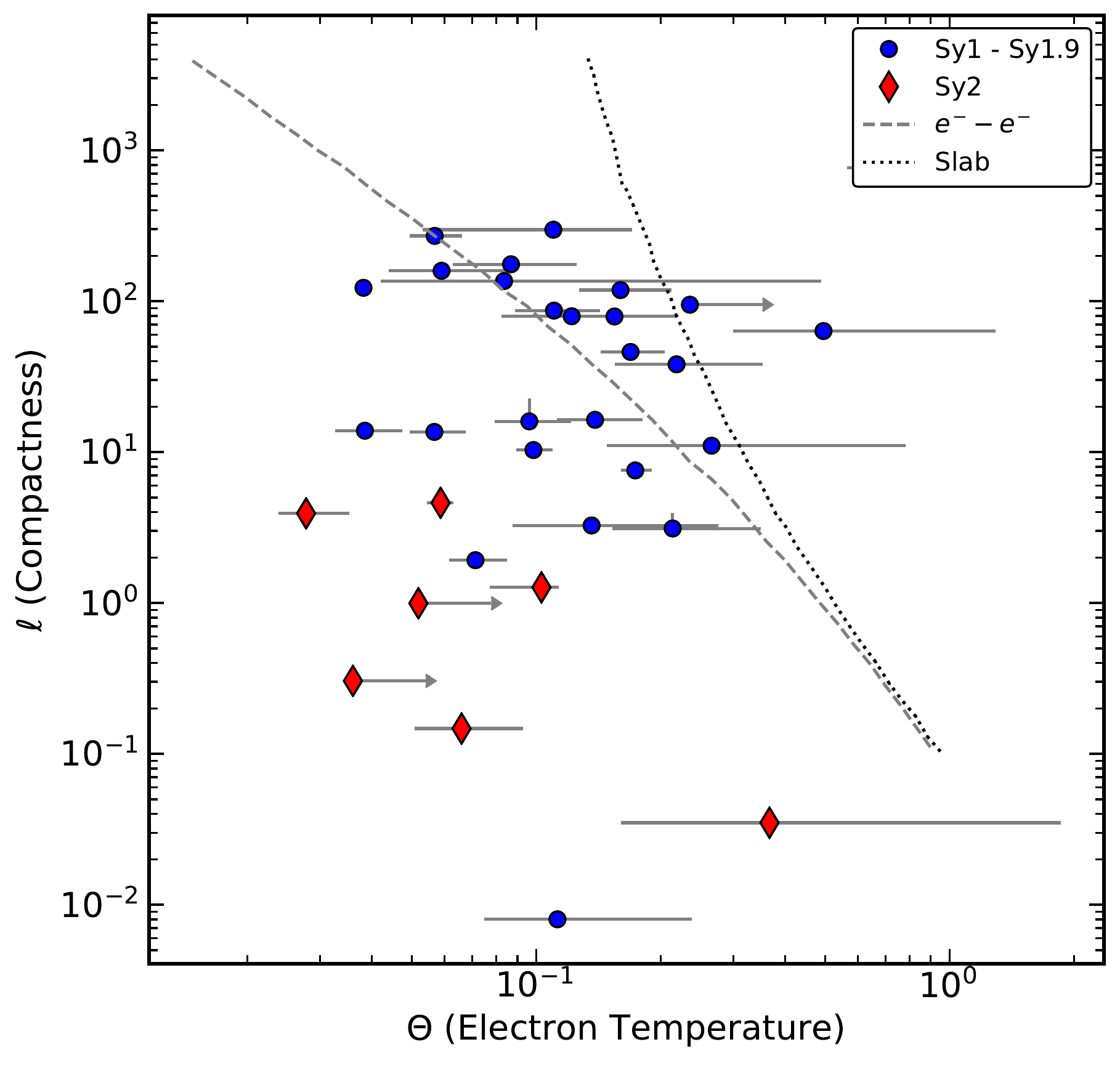}
 \caption{Compactness as compared to electron temperature. The blue circles represent Seyfert 1 through Seyfert 1.9 classifications and the red diamonds are Seyfert 2 galaxies. The gray dashed line is the e-e coupling line from \citet{ghisellini93} and the black dotted line is the expectation for a slab geometry \citep{stern95}. The majority of our sample lies below the slab line \citep{stern95}, with only one AGN with a constrained folding energy lying above the slab line.}
 \label{fig:comp_theta}
\end{figure}

To study the relationship between physical size and luminosity, we calculate the compactness parameter \citep[][]{guilbert83, fabian15} If present, we use the gravitational radii presented in \cite{fabian15}, and otherwise followed their assumption that the radius of the X-ray emitting region is ten gravitational radii. The median compactness parameter was 16, with a minimum value of $8.0 \times 10^{-3}$ and a maximum value of 767. We compute the electron temperature using $kT_{e} = E_{fold} / 2$ and the corresponding $\Theta = kT_{e} / m_{e}c^{2}$ \citep{petrucci01, fabian15, middei19}. While this scaling between folding energy and electron temperature is dependent on optical depth (with the above appropriate for optical depths less than one), the S/N of many of our objects is not high enough to directly fit for electron temperature using a Comptonization model. In Figure \ref{fig:comp_theta} we compare the compactness to $\Theta$. As expected, the majority of sources lie along the $e^{-}-e^{-}$ coupling line \citep{ghisellini93, fabian15} and below the pair balance line for a slab geometry \citep{stern95, fabian15}. This indicates the importance of pair production in regulating the X-ray corona of AGN and the resulting spectral shape \citep{fabian15}. Only two of the AGN in our sample lie above the slab line, with one having a lower limit on the electron temperature, potentially indicating poor constraints on the underlying Comptonized spectrum.

\subsection{Soft Excess Parameters}
While not the primary focus of our analysis, for most of our AGN it is necessary to include a model component to fit for the soft excess. This allowed us to properly constrain the higher energy portion of the spectrum from which the information on photon index, folding energy, and reflection coefficient are derived. Similarly, we include a redshifted Gaussian in our model to fit the strong Fe \Ka line. We have recorded each of the best fit soft excess and  Fe \Ka line parameters and present them in Table \ref{tab:soft_fe}.

As detailed in Section \ref{sec:specfit}, we fit the soft X-ray emission of the AGN in our sample with two different approaches. The first was a redshifted blackbody. This approach was mainly used for Seyfert 1 - 1.9 AGN. For this parameter we recorded the temperature and the luminosity of the blackbody component. The median temperature is $0.14$ keV, which is consistent with previous work using this model \citep[e.g.,][]{gierlinski04}. The median luminosity of the blackbody is log[L(erg s$^{-1}$)] = $43.4$. As expected, this is significantly less than the higher energy emission caused by IC scattering of softer photons.

The other approach we used to model the soft X-ray contribution was a partial covering model. The median covering fraction for our entire sample was $0.93$. For the Seyfert 1 - 1.9 objects, the median was $0.89$, lower that then median covering fraction for the Seyfert 2 AGN at $0.94$, again expected from the AGN unification model \citep{antonucci93}. The covering fractions for the AGN in our sample are largely consistent with or slightly higher than previous results \citep[e.g.,][]{mor09, ramos11, lanz19, zhao20b}. Furthermore, the lower covering fractions for Seyfert 1 - 1.9 objects as compared to Seyfert 2s is in agreement with earlier studies, both in the IR \citep{mor09, ramos11} and the X-ray \citep{lubinski16}. However, a K-S test for the Seyfert 1 - 1.9 and Seyfert 2 AGN shows that the two classes are consistent with being drawn from the same distribution with a p-value of $0.43$, possible due to the low number of Seyfert 1 - 1.9 objects fit with this model.

\subsection{Correlations Between Parameters}

There exist many theoretically expected correlations between the X-ray corona parameters and the physical properties of the AGN \citep[e.g.,][]{zdziarski99, ricci17, tortosa18}. Some of these correlations are based on the predicted geometries of the X-ray corona and the location of X-ray emission. In addition, there are expected correlations between X-ray properties and the driving physical parameters of the AGN such as mass of the central SMBH and the Eddington ratio. In this section, we search for such correlations and compare to previous analyses.

\begin{table} 
\caption{Kendall $\tau$ correlation test values for different combinations of parameters. Throughout the paper, we consider correlations with p-values less than 0.05 as significant. Considering the $\sim10$ correlations we search for in our study, if we make our threshold for significance 0.005, only the X-ray Baldwin effect and the correlation between photon index and reflection coefficient are recovered. The parameters are $\Gamma$ (photon index), E$_{fold}$ (e-folding energy), R (reflection coefficient), M$_{BH}$ (SMBH mass), $\lambda_{edd}$ (Eddington ratio), EW (Fe \Ka equivalent width), L$_{bol}$ (bolometric luminosity), $\ell$ (compactness parameter), and N$_{H}$ (hydrogen column density).}
\label{tab:Kendall}
\centering
\begin{tabular}{cccc}
\hline
Parameter 1 & Parameter 2 & $\tau$ & p-value\\ 
\hline
$\Gamma$ & E$_{fold}$ & 1.97E-01 & 1.11E-01 \\ 
$\Gamma$ & R & 3.64E-01 & 2.61E-03 \\ 
$\Gamma$ & M$_{BH}$ & $-$6.82E-02 & 5.90E-01 \\
$\Gamma$ & $\lambda_{edd}$ & 1.29E-01 & 3.02E-01 \\
$\Gamma$ & EW & $-$2.61E-01 & 3.30E-02 \\
$\Gamma$ & L$_{bol}$ & 8.15E-02 & 5.05E-01 \\
E$_{fold}$ & R & $-$1.67E-01 & 1.79E-01 \\ 
E$_{fold}$ & M$_{BH}$ & 1.52E-01 & 2.23E-01 \\
E$_{fold}$ & $\lambda_{edd}$ & 7.58E-02 & 5.49E-01 \\
E$_{fold}$ & $\ell$ & 4.92E-02 & 7.01E-01 \\
R & EW & 1.02E-01 & 4.15E-01 \\ 
EW & N$_{H}$ & 3.42E-01 & 5.28E-03 \\ 
EW & L$_{bol}$ & $-$3.81E-01 & 1.84E-03 \\ 
\hline
\end{tabular}
\end{table}

\begin{figure}
 \includegraphics[width=\columnwidth]{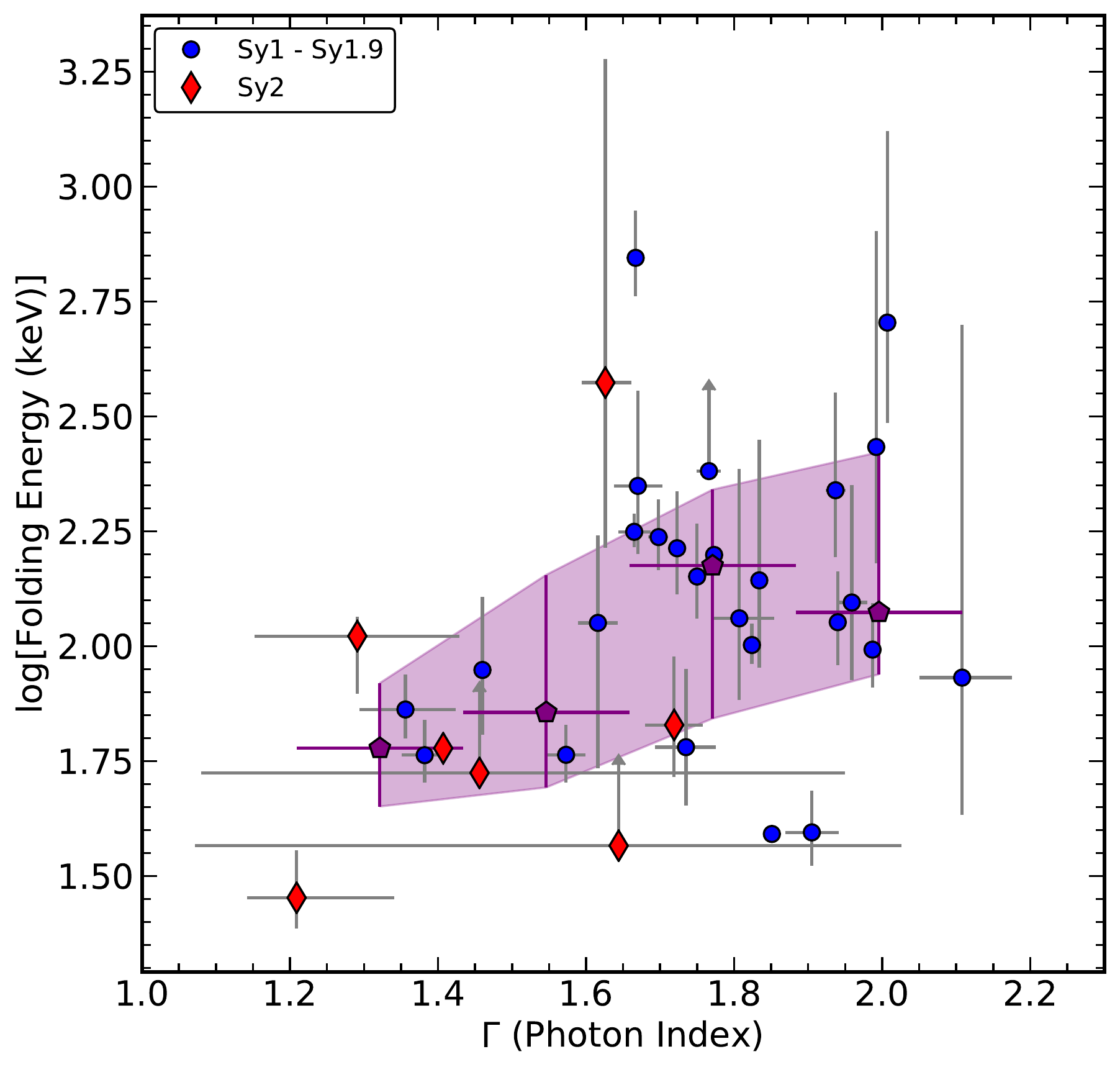}
 \caption{Folding energy as compared to the photon index. The blue circles represent Seyfert 1 through Seyfert 1.9 classifications and the red diamonds are Seyfert 2 galaxies. The purple pentagons are binned values of the folding energy and the shading corresponds to 1$\sigma$ scatter on this binned point.}
 \label{fig:folde_gamma}
\end{figure}

\begin{figure}
 \includegraphics[width=\columnwidth]{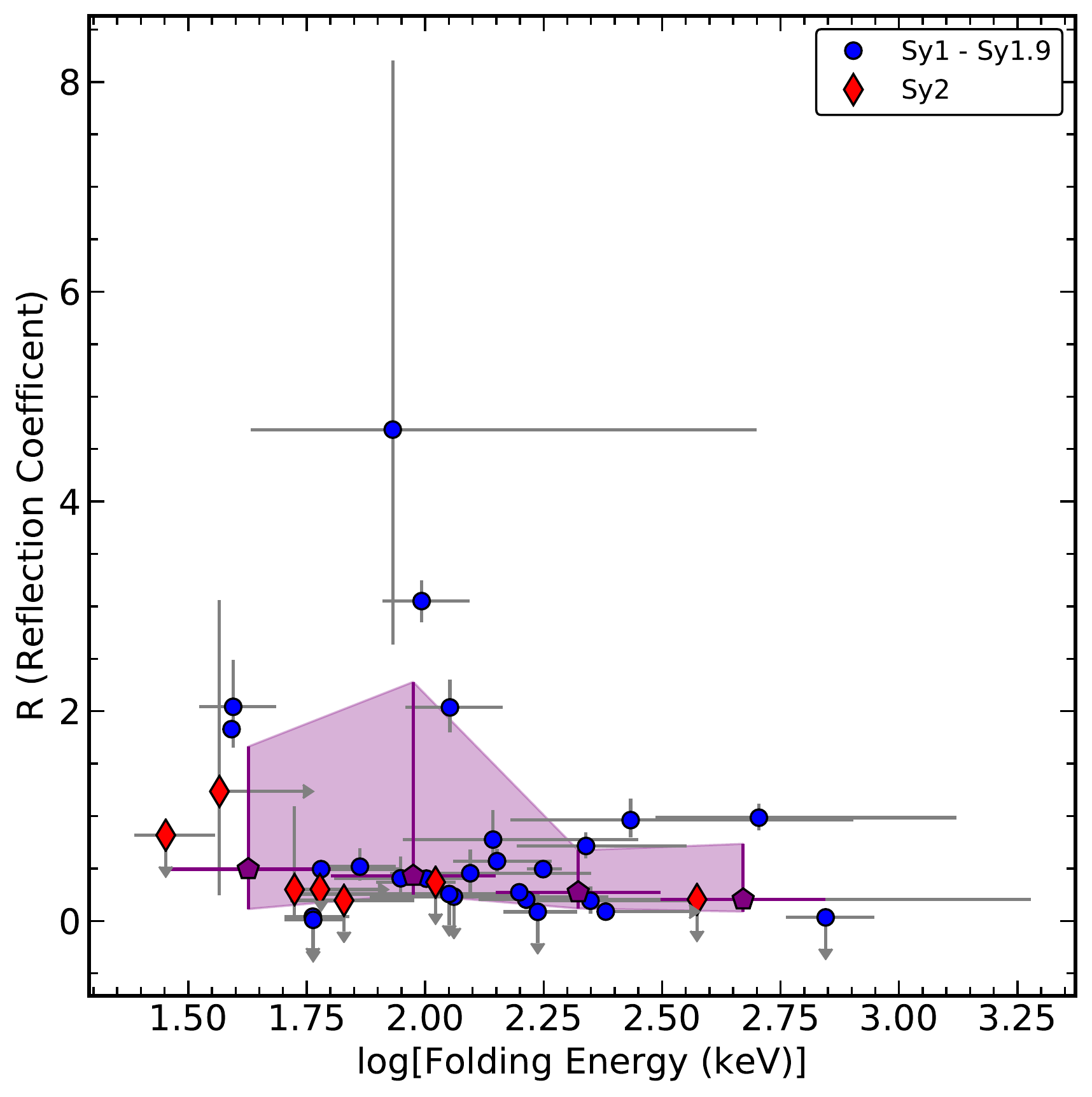}
 \caption{Reflection coefficient as compared to the folding energy. The blue circles represent Seyfert 1 through Seyfert 1.9 classifications and the red diamonds are Seyfert 2 galaxies. The purple pentagons are binned values of the folding energy and the shading corresponds to 1$\sigma$ scatter on this binned point.}
 \label{fig:refl_foldE}
\end{figure}

For each combination of parameters, we used the Kendall $\tau$ correlation test to look for evidence of correlations between the parameters, with the results shown in Table \ref{tab:Kendall}. We chose to use this correlation test over the Spearman correlation test, due to the robustness of the Kendall $\tau$ test to smaller samples and outliers. We first did this with all our AGN in the sample. If the corresponding p-value was larger than 0.05, we determined that a correlation did not exist. However, considering the $\sim10$ correlations we search for in this study, if we make our threshold for significance 0.005, of all the correlations found at a p-value of 0.05, only the X-ray Baldwin effect and the relationship between photon index and reflection coefficient are recovered. If this was the case, we repeated the Kendall $\tau$ correlation test on the Seyfert 1 - 1.9 and Seyfert 2 objects individually. If there was a correlation between any sets of parameters, we used a Theil-Sen estimator to find the slope of the correlation and 10,000 bootstrap iterations with replacement to obtain uncertainties on the slope. By finding the median slope between all pairs of values and computing our uncertainties from a bootstrap approach, we are robust to outliers and limits in our data. Additionally, we binned each of our combinations of parameters in four equal-sized horizontal bins by taking the median of the vertical parameter and plus/minus one sigma error bars. In the case of iron equivalent width (see Section \ref{sec:iron}) we excluded the outlier NGC 1068 from our binning procedure to ensure even sampling. These stacked points are shown in purple in our figures to guide the eye, with shading corresponding to the uncertainties on the stacked values. 

Figure \ref{fig:folde_gamma} shows a weak correlation between the photon index and folding energy, albeit with large scatter. We note that the existence of a correlation between these parameters has been found in other studies, but there is some dispute in the literature as to whether this is a real correlation \citep[e.g.,][]{ricci17, tortosa17, derosa08, kamraj18}. Additionally, the behavior of these two variables even for single AGN is not fully understood \citep[e.g.,][]{kang21}. Answering these questions will require even larger samples with broadband X-ray data similar to our sample, since this correlation requires accurate estimates of the folding energy. 

In Figure \ref{fig:refl_foldE}, we show the relationship between the reflection coefficient and the folding energy. The objects with the highest reflection coefficients tend to also have lower folding energies. At the highest values of folding energy, above $\sim 100$ keV, the reflection coefficients are neatly collapsed onto a sequence with values $\lesssim 1$.

\begin{figure*}
 \includegraphics[width=\textwidth]{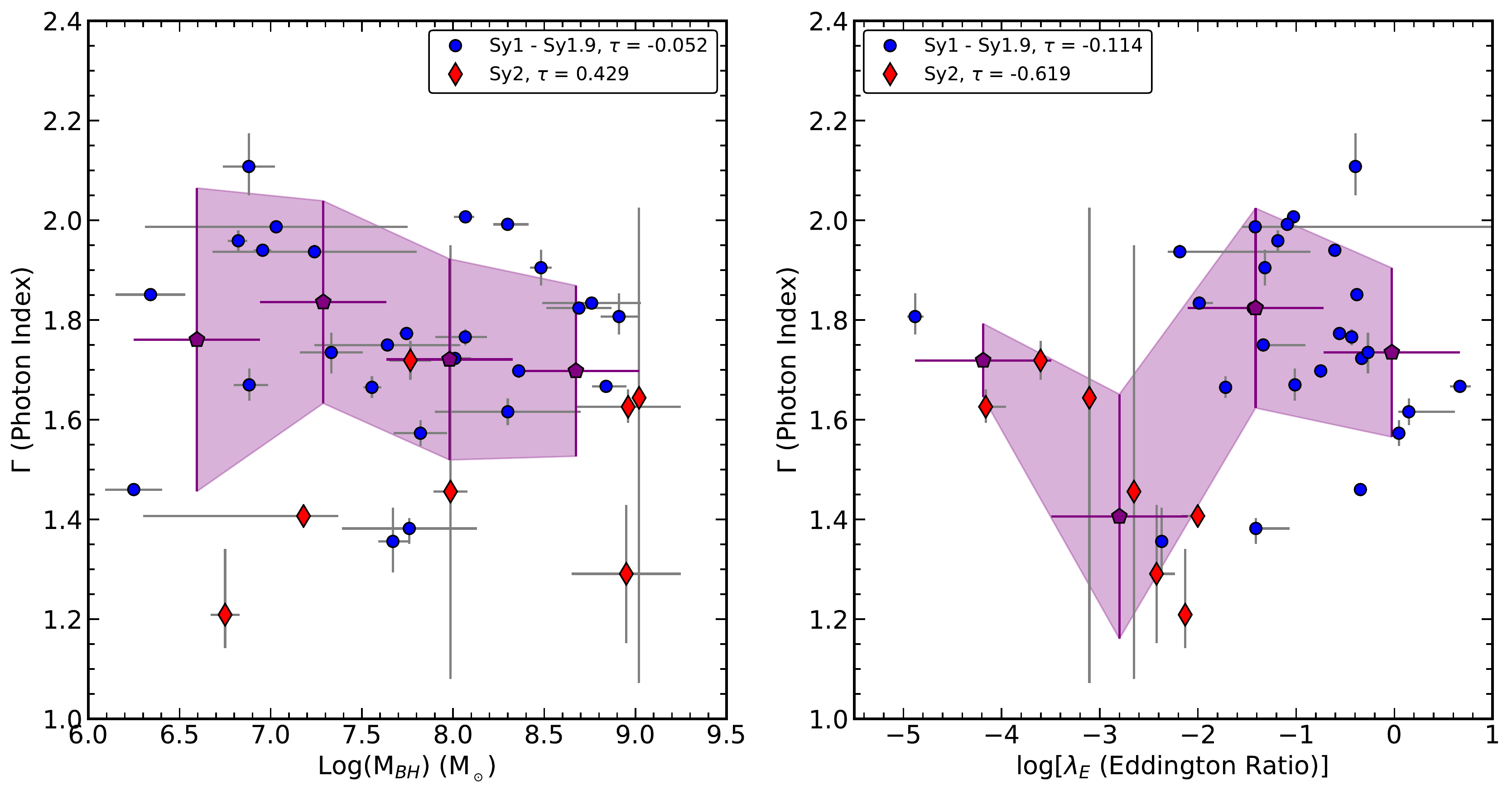}
 \caption{\textit{Left panel:} Photon index as compared to the SMBH mass, with no strong correlation. The blue circles represent Seyfert 1 through Seyfert 1.9 classifications and the red diamonds are Seyfert 2 galaxies. \textit{Right panel:} Photon index as compared to the Eddington ratio, again showing no correlation. The purple pentagons are binned values of the photon index and the shading corresponds to 1$\sigma$ scatter on this binned point. In both cases, the correlations seen in the Seyfert 2 AGN are stronger than the the Seyfert 1 - 1.9 AGN, but still not significant.}
 \label{fig:gamma_mbh_edd}
\end{figure*}

It is expected that AGN coronal parameters should be correlated with driving physical parameters such as the SMBH mass and Eddington ratio. Figure \ref{fig:gamma_mbh_edd} shows the derived photon indices versus the SMBH masses and Eddington ratios. We do not find any correlations with photon index and either black hole mass or Eddington ratio. In both cases, the Seyfert 2 AGN have a higher Kendall $\tau$ value, although neither yields a significant p-value. 

\begin{figure*}
 \includegraphics[width=\textwidth]{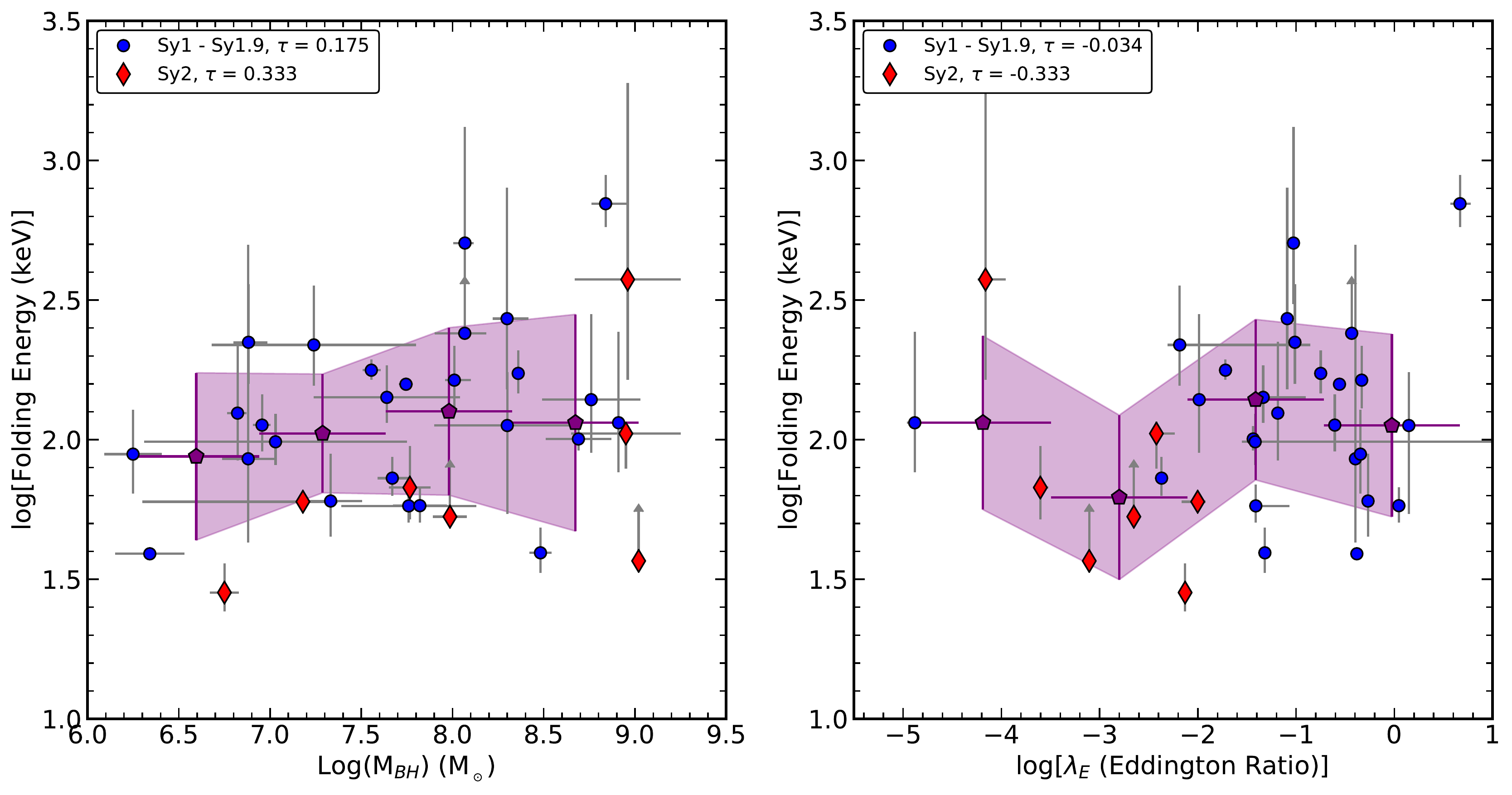}
 \caption{\textit{Left panel:} Folding energy as compared SMBH mass. \textit{Right panel:} Folding energy as compared to the Eddington ratio. In both panels, the blue circles represent Seyfert 1 through Seyfert 1.9 classifications and the red diamonds are Seyfert 2 galaxies. The purple pentagons are binned values of the folding energy and the shading corresponds to 1$\sigma$ scatter on this binned point. In both cases, the correlations seen in the Seyfert 2 AGN are stronger than the the Seyfert 1 - 1.9 AGN, but still not significant.}
 \label{fig:folde_mbh_edd}
\end{figure*}

Figure \ref{fig:folde_mbh_edd} shows the derived folding energies versus the SMBH masses and Eddington ratios. When considering all the objects in our sample, there are no strong correlations between the folding energy and these physical parameters. However, if we only look at Seyfert 2 objects, there is a weak positive correlation between folding energy and black hole mass and a moderate negative correlation between folding energy and Eddington ratio, in agreement with \citet{ricci18}. Several of the objects with the highest values of folding energy have near- or super-Eddington luminosities. Nonetheless, there is still significant scatter in the folding energies even close to an Eddington ratio of one. There appears to be a similar weak correlation in terms of black hole mass, with the highest folding energies occurring in AGN with the highest central SMBH masses. 

\begin{figure}
 \includegraphics[width=\columnwidth]{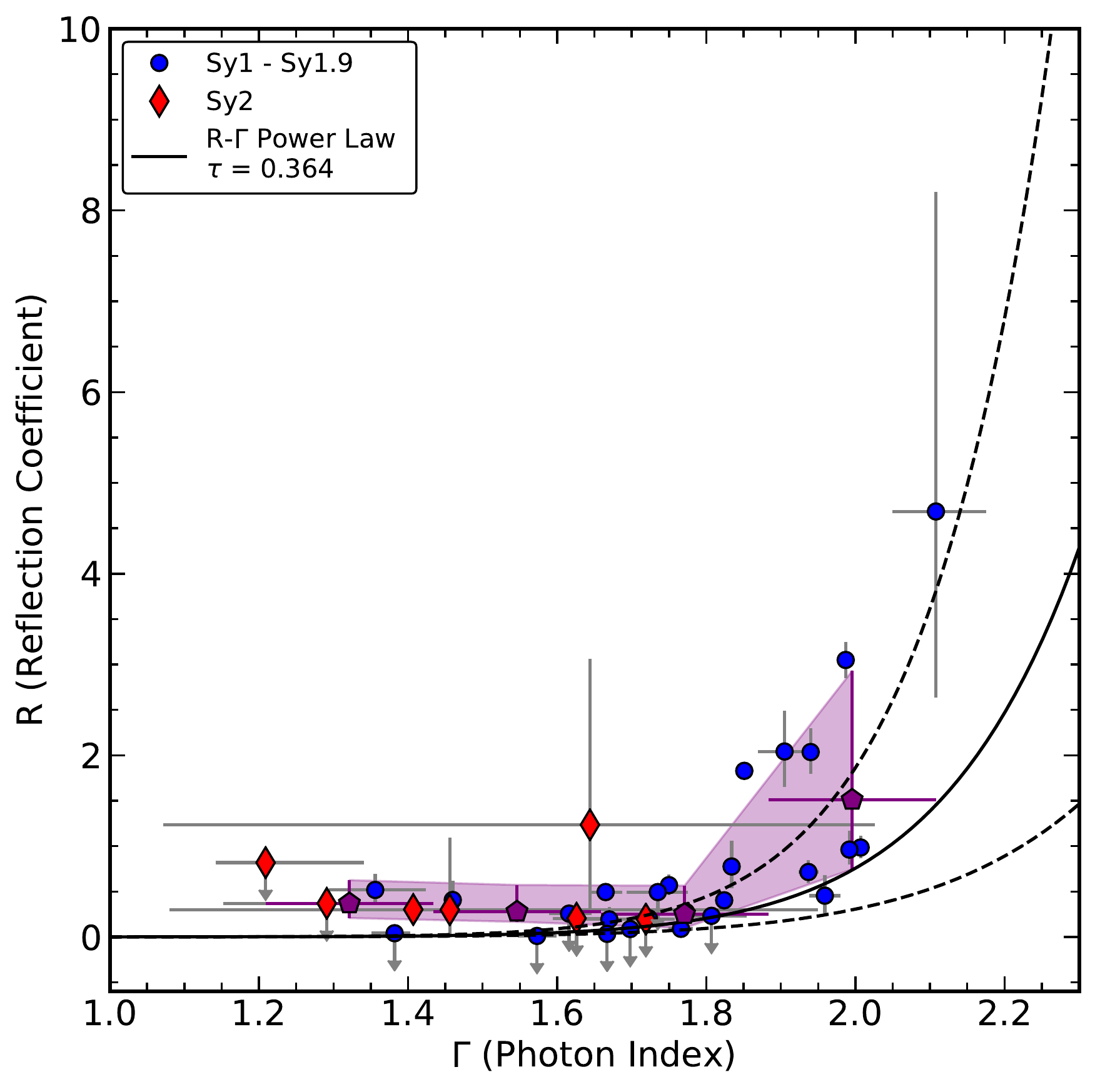}
 \caption{Reflection coefficient as compared to photon index. The blue circles represent Seyfert 1 through Seyfert 1.9 classifications and the red diamonds are Seyfert 2 galaxies. The black solid line and dashed lines are the best-fit power-law and corresponding uncertainties respectively from \citet{zdziarski99}, which qualitatively agree with our results. The purple pentagons are binned values of the reflection coefficient and the shading corresponds to 1$\sigma$ scatter on this binned point.}
 \label{fig:refl_gamma}
\end{figure}

Previous studies of AGN samples have found a power-law relationship between reflection coefficient and the photon index, which has been interpreted as the reflecting material being a source of soft photons that are IC scattered to higher energies \citep[e.g.,][]{zdziarski99, beloborodov99, vasylenko15, lubinski16}. We show this relationship in Figure \ref{fig:refl_gamma} with the power-law relationship of \citet{zdziarski99}. We find a p-value of $2.6 \times 10^{-3}$ for this correlation. Our results qualitatively agree with these previous studies, although the increase in reflection coefficients for our sample begins at a slightly flatter photon index. All of the AGN for which the reflection coefficients are higher than $\sim 2$ are sources with soft spectra. This is consistent with recent studies \citep{ezhikode20}.

\begin{table} 
\caption{Kolmogorov-Smirnov test results comparing the Seyfert 1 - 1.9 and Seyfert 2 AGN for the various parameters measured in our sample.}
\label{tab:K-S}
\centering
\begin{tabular}{cccc}
\hline
Parameter & D & P[null]\\ 
\hline
$\Gamma$ & 6.65E-01 & 7.54E-03 \\ 
E$_{fold}$ & 3.57E-01 & 3.81E-01 \\ 
R & 3.30E-01 & 4.82E-01 \\ 
N$_{H}$ & 6.76E-01 & 5.46E-03 \\
f$_{cvr}$ & 5.50E-01 & 4.29E-01 \\
M$_{BH}$ & 4.29E-01 & 1.90E-01 \\ 
$\lambda_{edd}$ & 8.85E-01 & 5.62E-05 \\ 
L$_{bol}$ & 5.99E-01 & 2.13E-02 \\
$\ell$ & 8.46E-01 & 1.65E-04 \\
E$_{K\alpha}$ & 2.75E-01 & 6.99E-01 \\
$\sigma_{K\alpha}$ & 2.91E-01 & 6.26E-01 \\
EW & 7.31E-01 & 2.35E-03 \\ 
\hline
\end{tabular}
\end{table}

\subsection{Iron \Ka Line Characteristics} \label{sec:iron}

The fluorescent Fe \Ka line is produced by the reprocessing of radiation by material near the central SMBH \citep[e.g.,][]{matt91, sulentic98}. The location of Fe \Ka line production very close to the SMBH is supported by its broad profile and evidence of general relativistic effects \citep[e.g.][]{fabian00, reynolds03, reynolds08}. For many of the AGN in our sample, it was possible to constrain the energy and width of the Fe \Ka line. For Mrk 1448 and 3C 273, the objects that did not have a clear Fe \Ka line and for which we could not constrain a line width, we froze the energy at a rest-frame energy $6.403$ keV and width at $0.1$ keV and fit for the normalization. It should be noted that if there is an intrinsically broad line that is undetected in these objects, our limits obtained with a narrow line will be underestimated \citep{reynolds97c}. All of the objects for which we had to freeze a line energy were Seyfert 1 - 1.9 AGN.

The median energy of the Fe \Ka line was $6.41$ keV, slightly blue-shifted from the rest energy of $6.403$ keV, but consistent with the calibration of the XMM-Newton detectors. Using a K-S test, the Fe \Ka line energies are consistent between the two classes with a p-value of $0.70$. The median line width in our sample was $0.11$ keV. Again, the line widths of the Seyfert 1 - 1.9 and Seyfert 2 classes are consistent based on a K-S test. Each class of AGN has objects whose line widths were consistent with zero, suggesting either an extremely narrow or weak Fe \Ka line. 

\begin{figure*}
 \includegraphics[width=\textwidth]{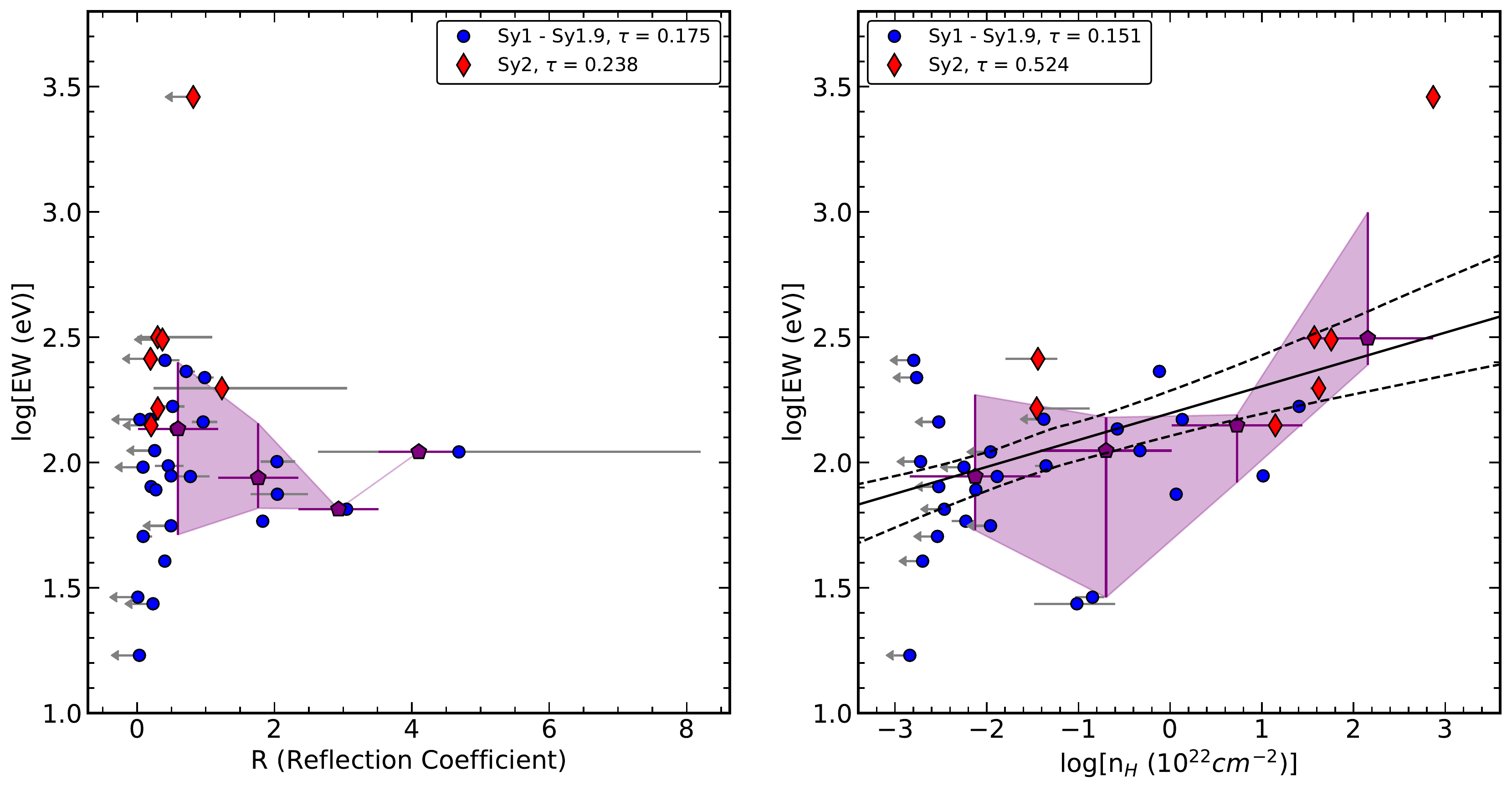}
 \caption{\textit{Left panel:} EW as compared to the reflection coefficient. The blue circles represent Seyfert 1 through Seyfert 1.9 classifications and the red diamonds are Seyfert 2 galaxies. \textit{Right panel:} EW as compared to the column density. The black solid line is the Theil-Sen estimator fit to the Seyfert 2 AGN and the dashed lines are plus/minus one sigma from our bootstrap runs.}
 \label{fig:ew_refl_nH}
\end{figure*}

To compare the strength of the Fe \Ka lines in our AGN, we measured the equivalent width (EW). This allowed us to place constraints on the presence of this emission even for the AGN where the fit did not converge on an energy for the Fe \Ka line. The median EW for our entire sample was $110.2$ eV. The Seyfert 1 - 1.9 AGN had generally weaker Fe \Ka emission, with a median EW of $92.1$ eV. The Seyfert 2 AGN had stronger emission, at $259.2$ eV. The EW distributions for these classes are distinct as show by the p-value of $2.4 \times 10^{-3}$ from a K-S test. In Figure \ref{fig:ew_refl_nH}, we compare the EW to the reflection coefficient and column density. There is no correlation between EW and reflection coefficient, but there is a positive correlation between EW and the column density. We also find that the Seyfert 2 AGN tend to occupy different regions of parameter space than the Seyfert 1 - 1.9 objects. There is one outlier in our sample in terms of EW, NGC 1068 with an equivalent width of 2875 eV. NGC 1068 also has highest column density of any of the objects in our sample, over an order of magnitude higher than the next highest AGN. As is shown in Figure \ref{fig:ew_refl_nH}, many of the objects with high EWs also have large obscuration along the line of sight.

\begin{figure}
 \includegraphics[width=\columnwidth]{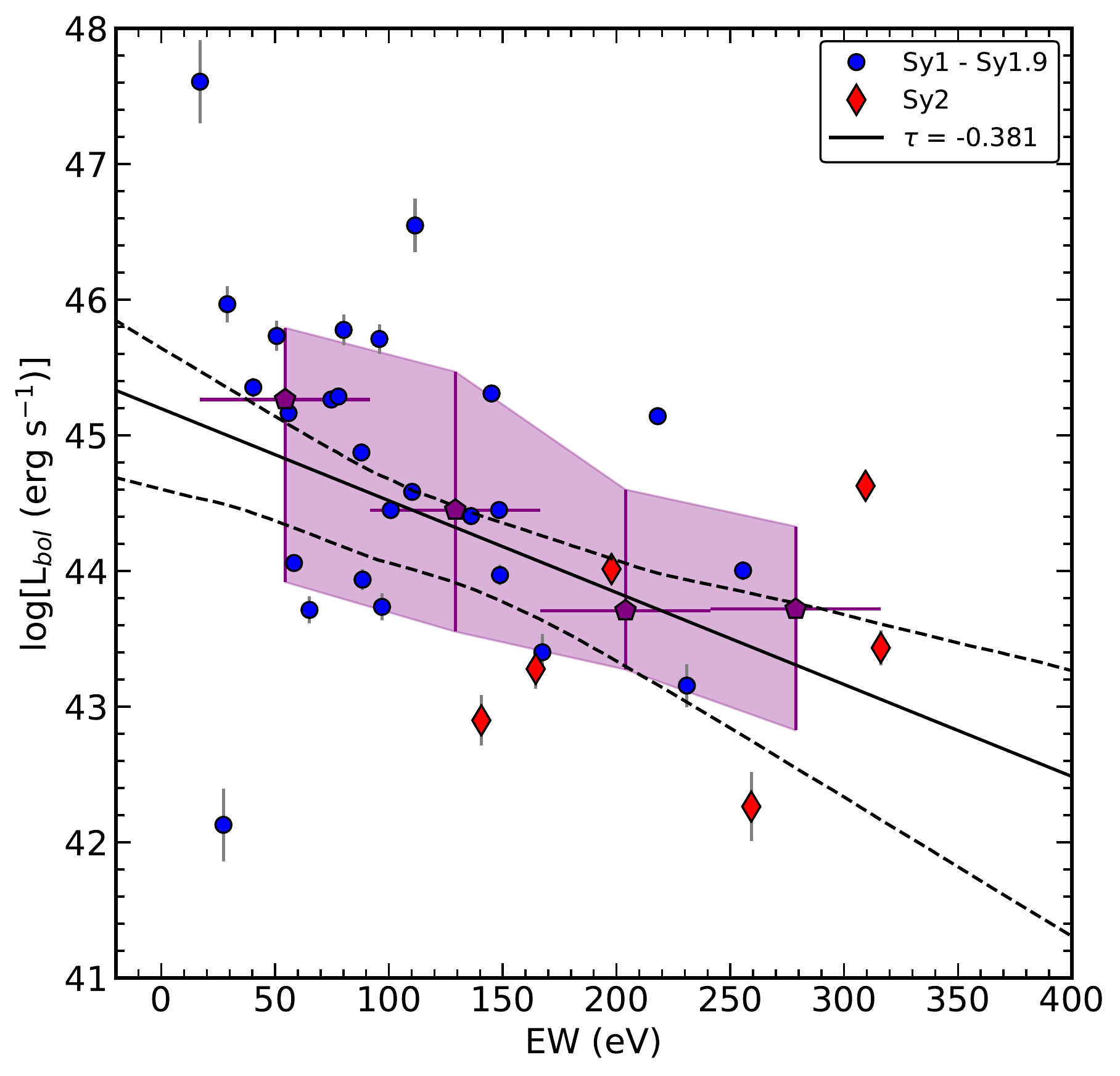}
 \caption{EW of the Fe \Ka line as compared to the bolometric luminosity. The blue circles represent Seyfert 1 through Seyfert 1.9 classifications and the red diamonds are Seyfert 2 galaxies. The black solid line is the Theil-Sen estimator fit to the Seyfert 2 AGN and the dashed lines are plus/minus one sigma from our bootstrap runs. The purple pentagons are binned values of the bolometric luminosity and the shading corresponds to 1$\sigma$ scatter on this binned point. Not shown in this plot is the outlier NGC 1068, with an extremely large EW of 2875 eV.}
 \label{fig:ew_Lbol}
\end{figure}

In Figure \ref{fig:ew_Lbol} we compare the EW of the Fe \Ka line to the bolometric luminosity, calculated from the \citet{winter12} scaling of the the Swift BAT 14 - 195 keV luminosity. Similar to \citet{iwasawa93} and \citet{nandra97}, we find a significant anti-correlation between the Fe \Ka line EW and X-ray luminosity. This relationship, otherwise known as the X-ray Baldwin Effect, is the strongest correlation we find between any two parameters in our sample, with a Kendall $\tau = -0.38$ and a corresponding p-value of $1.8 \times 10^{-3}$.

\begin{figure}
 \includegraphics[width=\columnwidth]{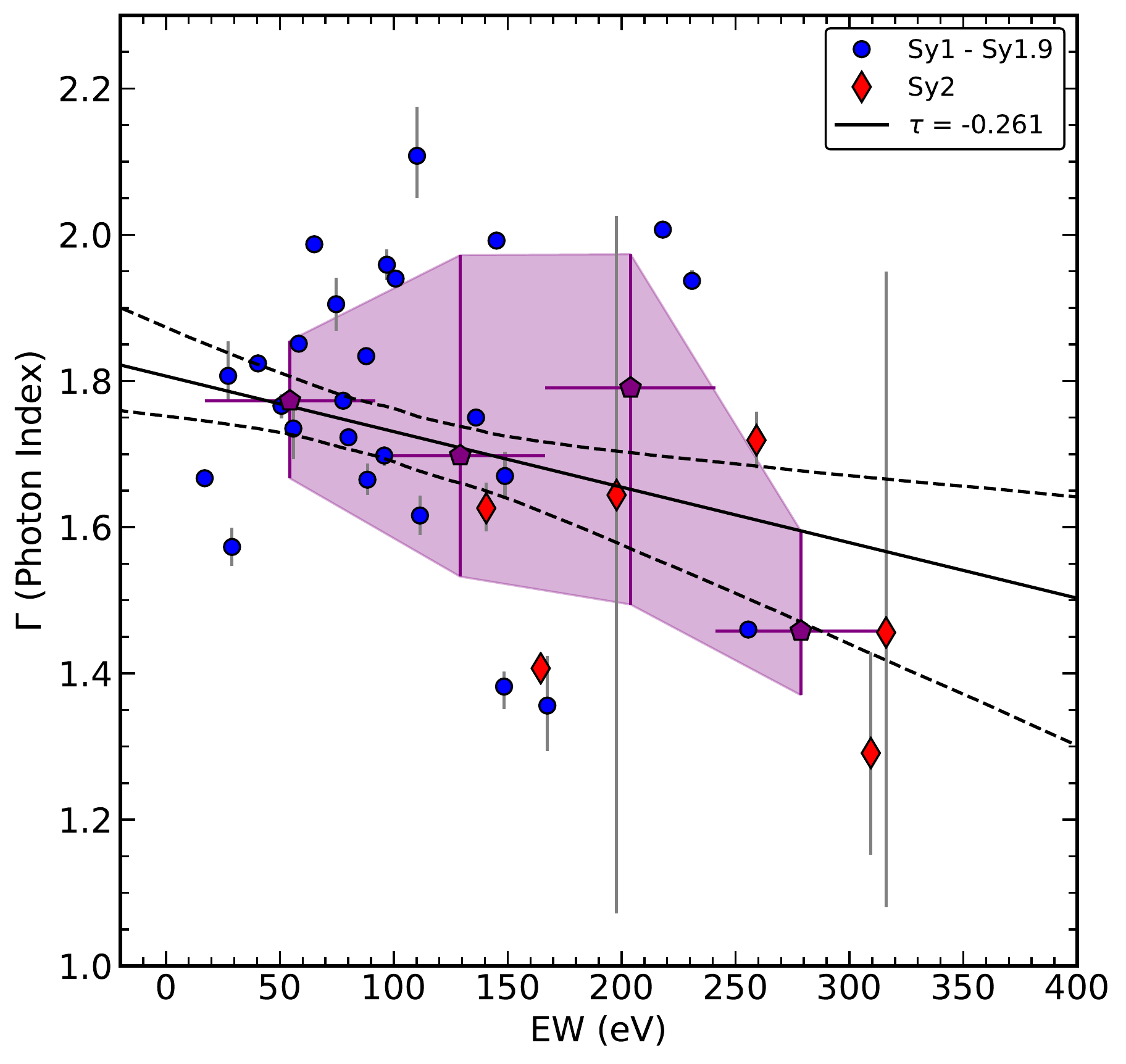}
 \caption{Photon index as compared to EW of the Fe \Ka line. The blue circles represent Seyfert 1 through Seyfert 1.9 classifications and the red diamonds are Seyfert 2 galaxies. The black solid line is the Theil-Sen estimator fit to the Seyfert 2 AGN and the dashed lines are plus/minus one sigma from our bootstrap runs. The purple pentagons are binned values of the photon index and the shading corresponds to 1$\sigma$ scatter on this binned point. Here we exclude the outlier NGC 1068, with an extremely large EW of 2875 eV.}
 \label{fig:ew_gamma}
\end{figure}

In Figure \ref{fig:ew_gamma} we compare the EW to the photon index. We find a weak, anti-correlation between the photon index and the Fe \Ka line EW, although this correlation is not recovered at the more stringent p-value of 0.005. Additionally, many of the AGN with large equivalent widths have large uncertainties on the photon index. It is worth noting that the photon index and bolometric luminosity are not themselves correlated.

\subsection{Coronal Parameter Contours}
In addition to obtaining constraints for the various coronal parameters individually and looking for correlations between them, we also investigated the covariance between the fundamental parameters. The combinations of parameters examined here are the folding energy/photon index, folding energy/reflection coefficient, and photon index/reflection coefficient. We show this for three of our AGN, to illustrate several different classifications of object and contour shapes. 

\begin{figure*}
  \includegraphics[width=0.33\textwidth]{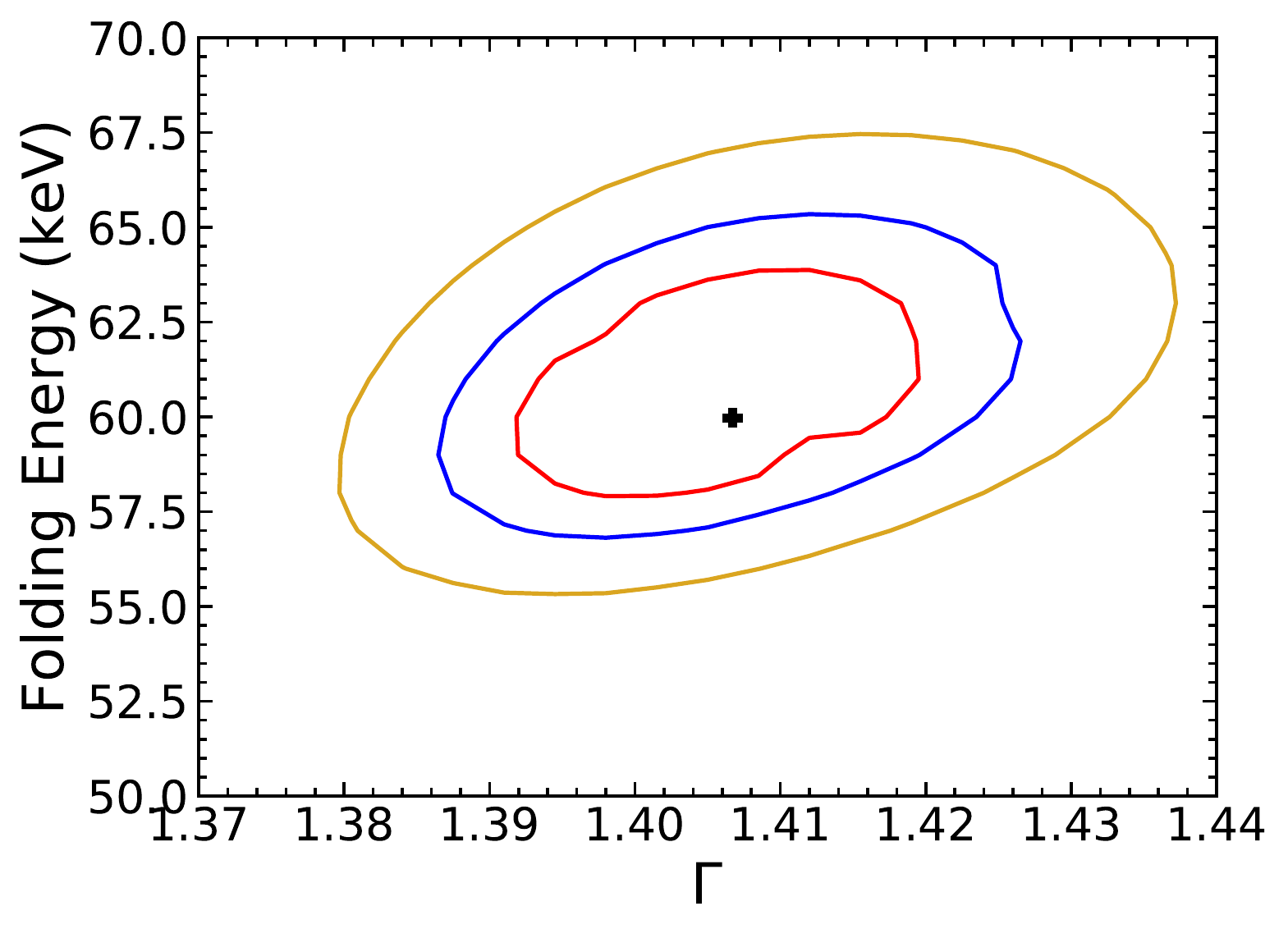}\hfill
 \includegraphics[width=0.33\textwidth]{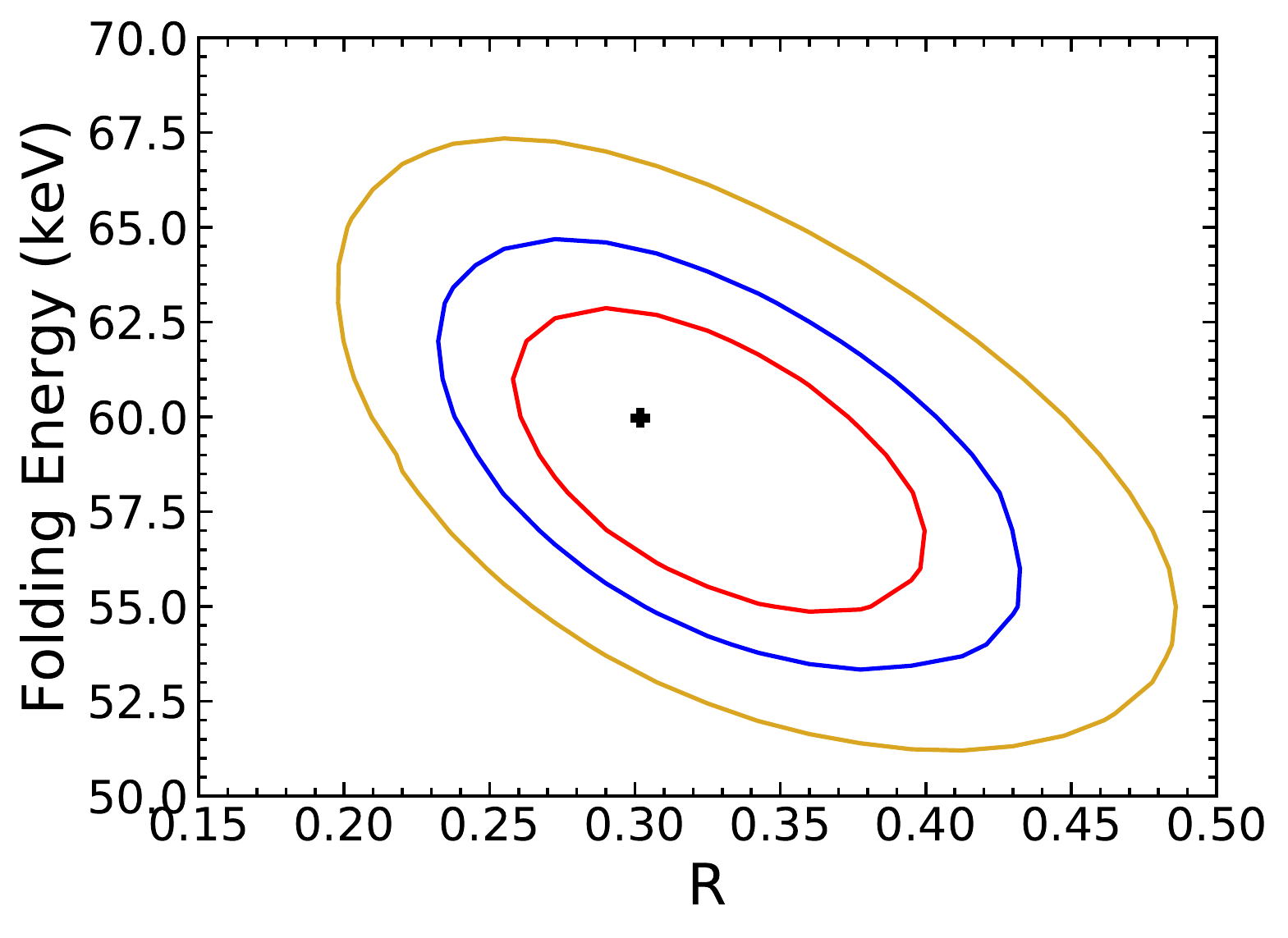}\hfill
 \includegraphics[width=0.33\textwidth]{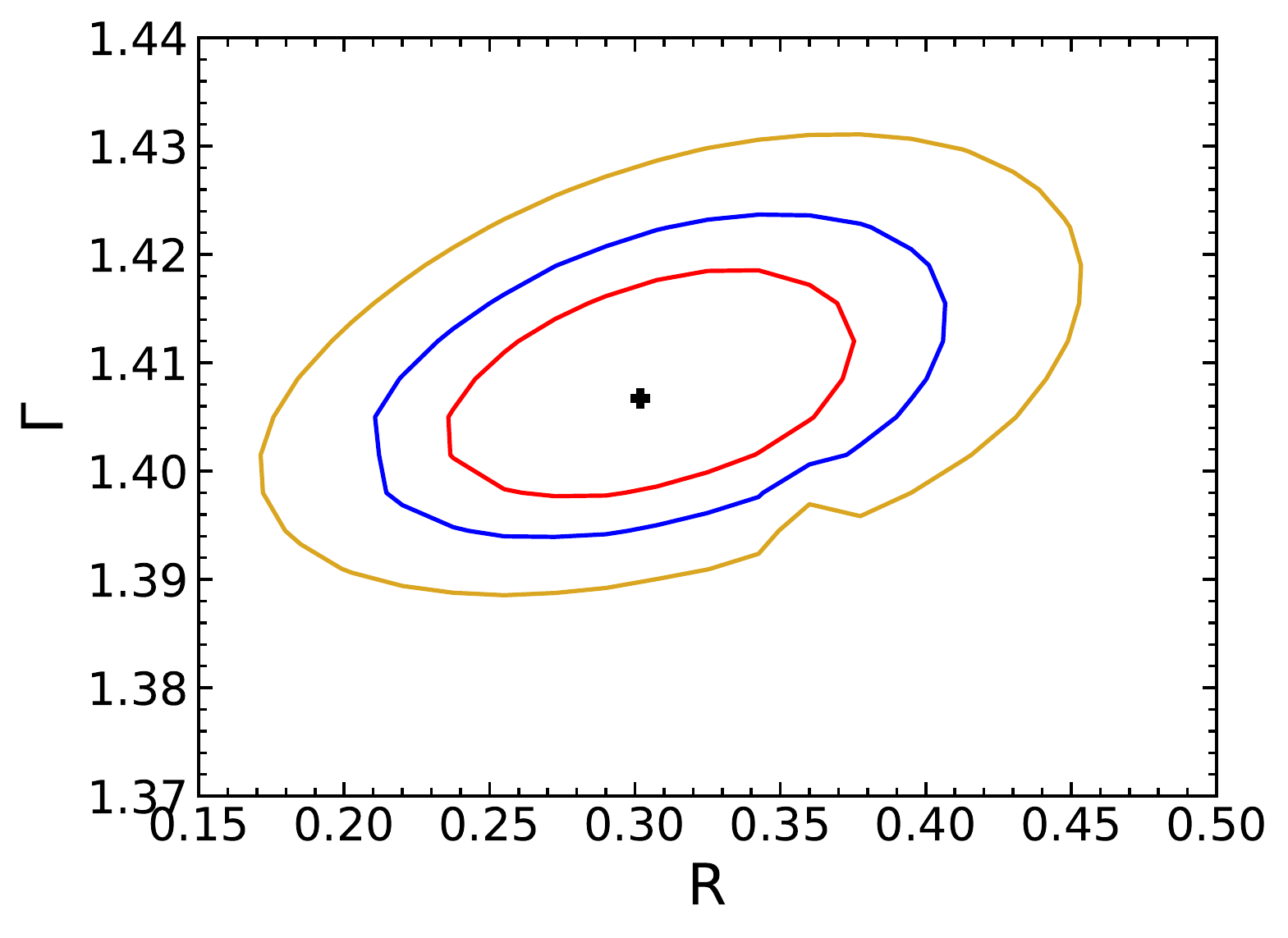}
 \caption{\textit{Left panel}: Contours for the combination of folding energy and photon index for NGC 3227. In each panel the best fit location is marked by the black plus, with one, two, and three sigma contours shown in red, blue, and gold respectively. \textit{Middle panel}: Contours for the combination of folding energy and reflection coefficient for NGC 3227. \textit{Right panel}: Contours for the combination of photon index and reflection coefficient for NGC 3227.}
 \label{fig:NGC3227_conts}
\end{figure*}

In Figure \ref{fig:NGC3227_conts} we plot the contours for the Seyfert 2 NGC 3227. For this object, the contours are well-behaved, appearing as ellipses with relatively smooth levels. There is only slight covariance between the sets of parameters. Additionally, the sizes of the error bars on this source are small, indicating a robust and good fit for this high S/N observation.

\begin{figure*}
  \includegraphics[width=0.33\textwidth]{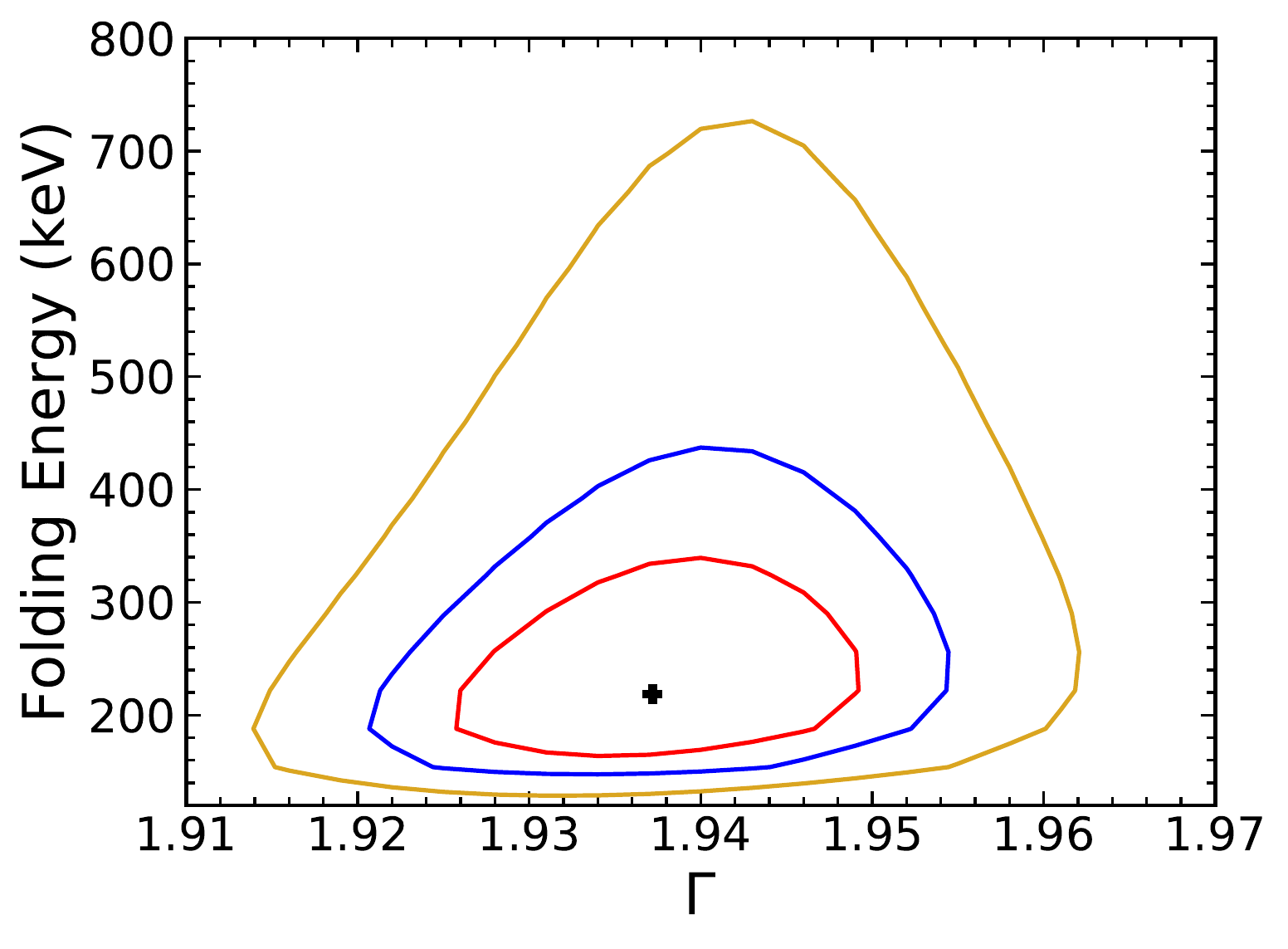}\hfill
 \includegraphics[width=0.33\textwidth]{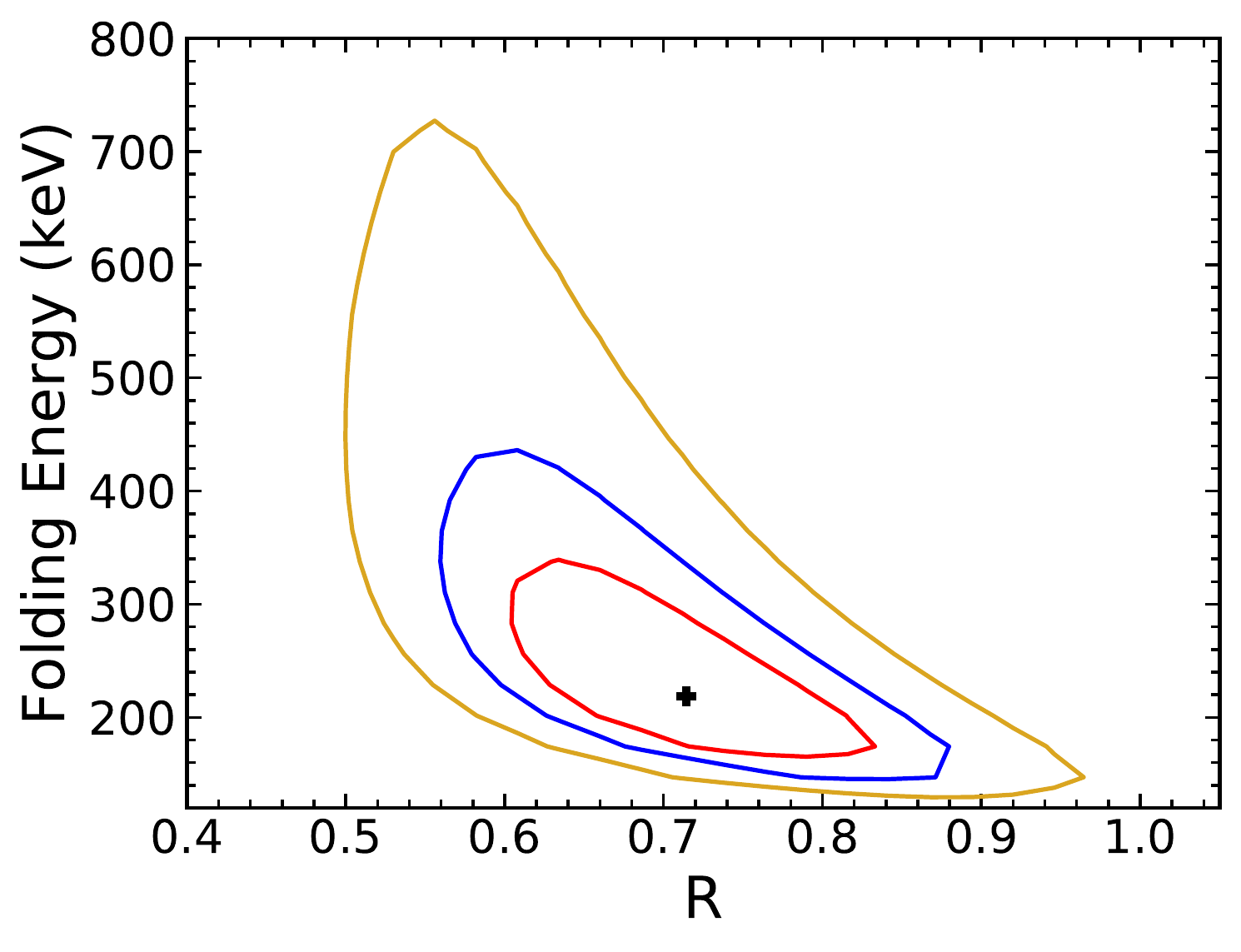}\hfill
 \includegraphics[width=0.33\textwidth]{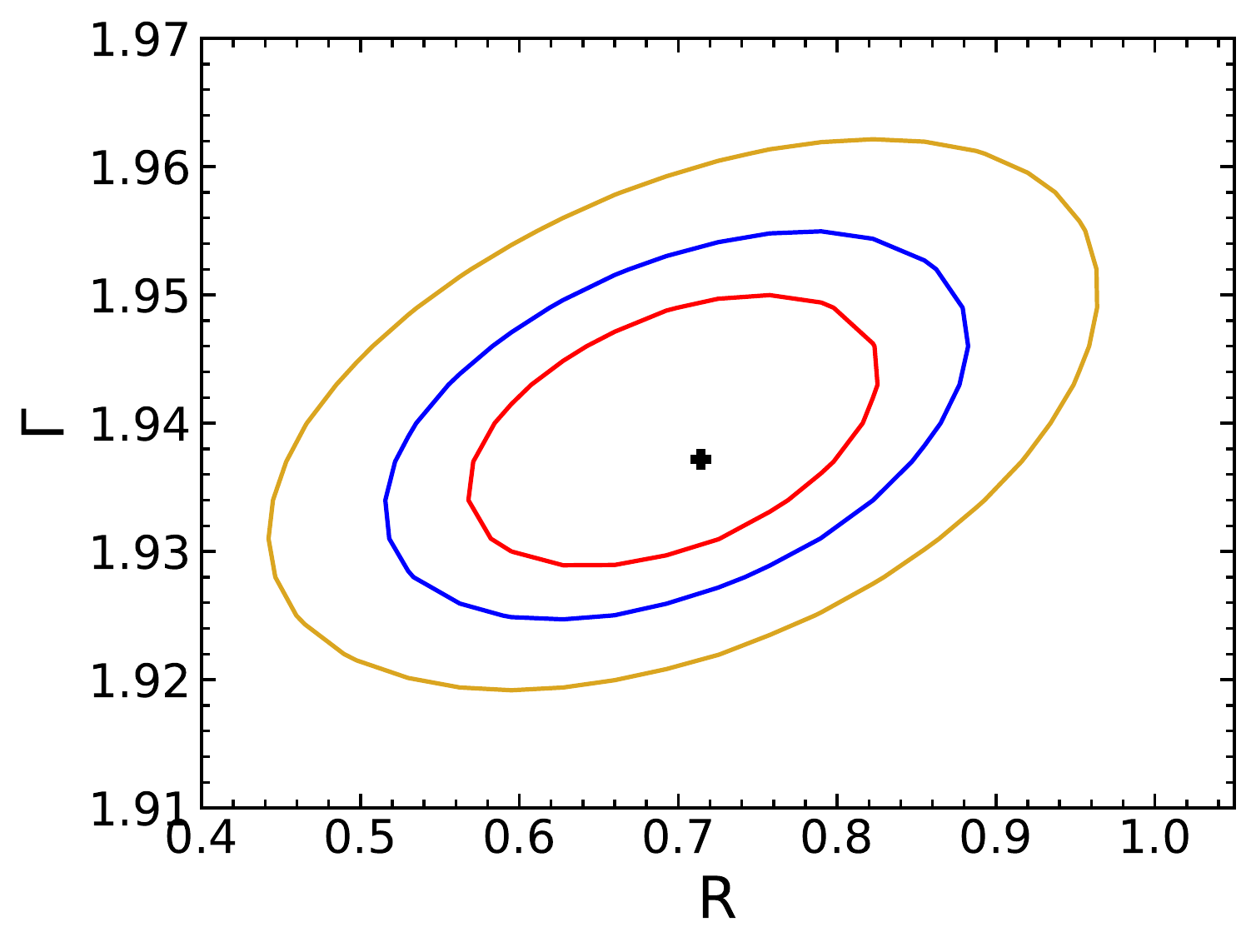}
 \caption{\textit{Left panel}: Contours for the combination of folding energy and photon index for NGC 7314. In each panel the best fit location is marked by the black plus, with one, two, and three sigma contours shown in red, blue, and gold respectively. \textit{Middle panel}: Contours for the combination of folding energy and reflection coefficient for NGC 7314. \textit{Right panel}: Contours for the combination of photon index and reflection coefficient for NGC 7314.}
 \label{fig:NGC7314_conts}
\end{figure*}

Next, in Figure \ref{fig:NGC7314_conts} we show the same plots but for the Seyfert 1 NGC 7314. Here there is noticeably more covariance between parameters, especially when one of them is the folding energy. Unlike NGC 3227, the error bars on folding energy are quite uneven, with the best-fit value lying towards the lower end of the confidence interval. Despite the misshapen contours, the folding energy is still constrained.

\begin{figure*}
  \includegraphics[width=0.33\textwidth]{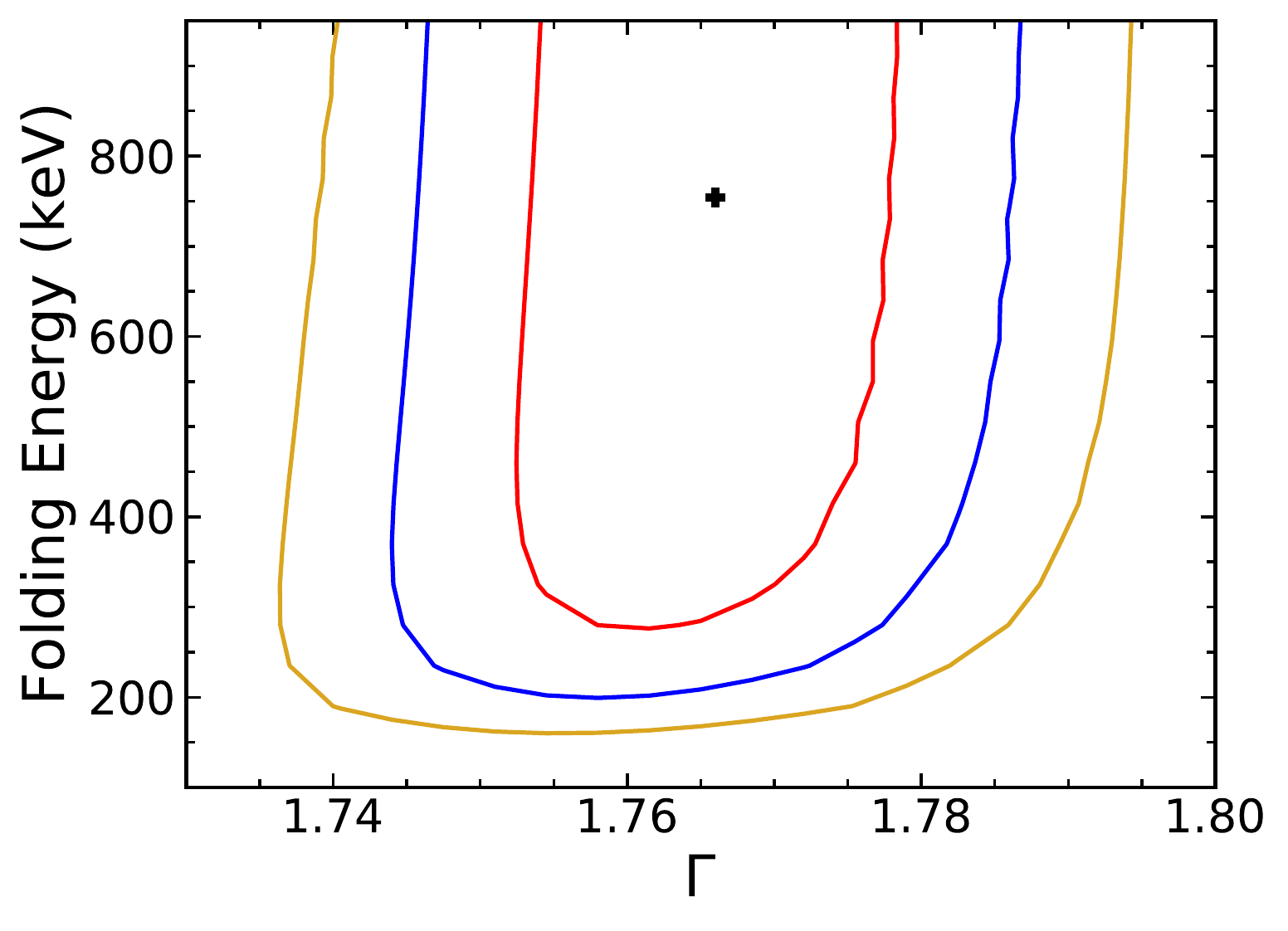}\hfill
 \includegraphics[width=0.33\textwidth]{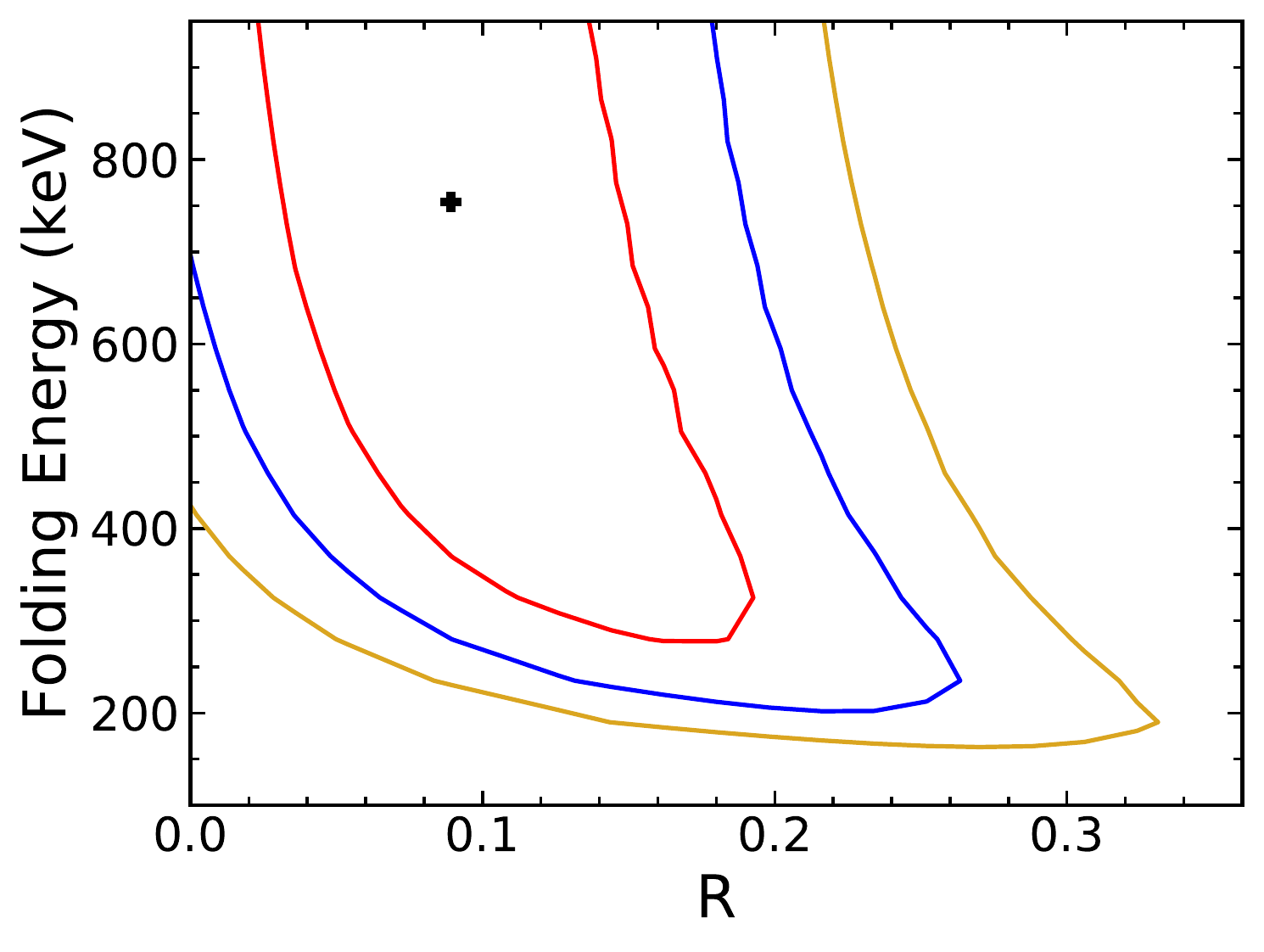}\hfill
 \includegraphics[width=0.33\textwidth]{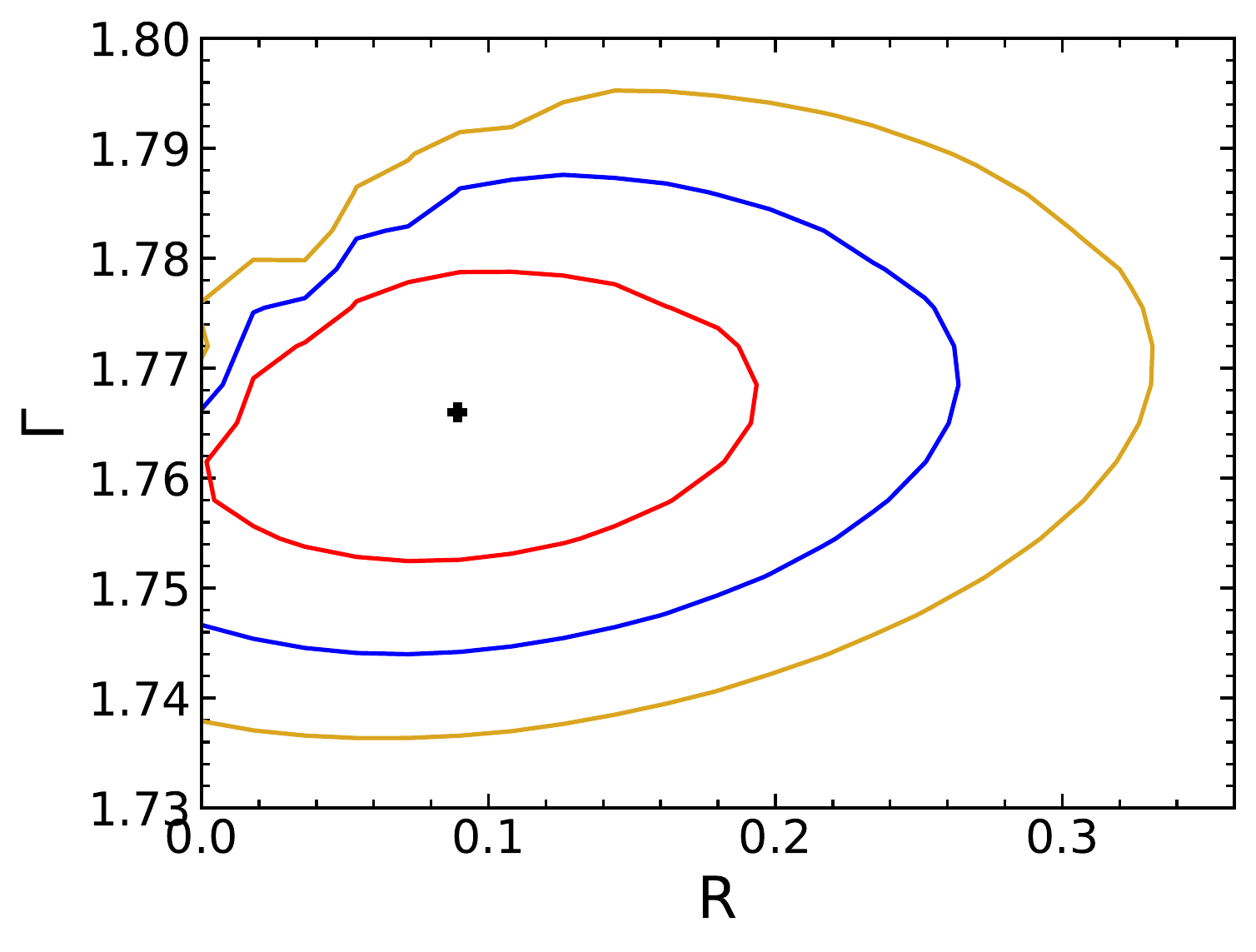}
 \caption{\textit{Left panel}: Contours for the combination of folding energy and photon index for Mrk 1501. In each panel the best fit location is marked by the black plus, with one, two, and three sigma contours shown in red, blue, and gold respectively. \textit{Middle panel}: Contours for the combination of folding energy and reflection coefficient for Mrk 1501. \textit{Right panel}: Contours for the combination of photon index and reflection coefficient for Mrk 1501.}
 \label{fig:Mrk1501_conts}
\end{figure*}

Finally, in Figure \ref{fig:Mrk1501_conts} we show the same plots for the Seyfert 1.2 Mrk 1501. For this AGN, we can only obtain a lower limit on the folding energy. Additionally, there are some jagged or sharp edges in the contours indicating lower quality constraints on the parameters.

\section{Discussion} \label{sec:discussion}

\subsection{Inclusion of Stacked BAT data}

When comparing our results to previous results \citep[e.g.,][]{ricci17, tortosa18, zhao20}, many using data from different instruments and with varying assumptions, we find that our results are generally consistent. The photon indices we obtain are usually within 10 percent of the values obtained in these studies. The folding energies are largely consistent, and oftentimes the limits we obtain are considerably smaller than in previous studies. In addition, for several objects, previous results could not measure a folding energy while our analysis yields at least lower limits for all our AGN. The reflection coefficients in our sample are also roughly consistent, and we find fewer upper limits than previous analyses. We should note though that in addition to using data from different instruments, previous studies often use slightly different models. Table \ref{tab:models} lists our choice of model for each object studied here. This may affect the best fit parameters we obtain as compared to results in the literature.

We additionally find that 29 out of our 33 objects are well-fit (reduced $\chi^2 < 1.3$) by our simple models. When increasing the cutoff to a more relaxed reduced $\chi^2$ of 1.5, we find that only one object, NGC 1068, which has strong emission lines, has a higher reduced $\chi^2$. Some of the objects did use a restricted energy range to avoid contamination by strong emission lines in the soft X-ray, but nonetheless, these low reduced chi-squared values indicate good fits in the energy range which determines the free parameters (folding energy, photon index, and reflection coefficient) we have focused on in this work.

The inclusion of the BAT data is crucial in constraining many of the fundamental parameters describing the X-ray emitting corona. To quantify this, we repeated our procedure for five AGN in our sample, with the Swift BAT data removed. We then compared the uncertainties from the fits with and without the BAT data. To test this over a range of different AGN types, we selected the Seyfert 1.5 Mrk 841, the Seyfert 1s NGC 7314 and NGC 4593, the Seyfert 2 Mrk 3, and the radio-loud Seyfert 1 3C 120 to re-fit. The main parameters of interest were the photon index, folding energy, and reflection coefficient. 

When fitting only the XMM-Newton and NuSTAR data, we found that the uncertainties on the photon index increased the least, by roughly 10\% as compared to when including the BAT data. This is to be expected, as significant information on the spectral slope is contained in the bandpasses covered by XMM-Newton and NuSTAR. All of the new photon indices were consistent with the values obtained in our original fits.

The reflection coefficient was the second most affected parameter, with an increase in the uncertainties by roughly 20\%. Again, we found that while the new errors were significantly larger, the values of the reflection coefficient without the BAT data were consistent with the values obtained from our full broadband fits. The one object with an upper-limit on the reflection coefficient (Mrk 3) still had an upper-limit without the BAT data, but the value of the limit was increased by 16\%.

As expected, the folding energy was most affected by the inclusion of the BAT data. On average, for these five objects, the uncertainties on folding energy increased by a factor of 1.4 for the lower error bar and 3.7 for the upper error bar as compared to including the BAT data. In two cases (NGC 7314 and Mrk 3), the best-fit folding energies were roughly 25\% lower, albeit still consistent within uncertainties. Similarly, for Mrk 841, the new folding energy was 60\% higher than before, but consistent within uncertainties. However, while with the BAT data, the folding energy was constrained for Mrk 841, without the BAT data, only a lower-limit was obtained. In the cases of Mrk 841, NGC 4593, and Mrk 3, the upper error bars on the folding energy in particular were multiple times those found when including the BAT data.

When we consider the fact that our approach yields results in rough agreement with previous results, and our reduced uncertainties, we have shown that the procedure of combining stacked BAT data with XMM-Newton and NuSTAR observations is valid. This is unsurprising as it is a natural extension of the earlier studies using stacked INTEGRAL data \citep{derosa08, ricci11}, the Rossi X-ray Timing Explorer \citep{rivers13}, and shorter BAT datasets \citep{kawamuro16}.

With our small error bars and improved constraints on parameters like the folding energy, we are able to continue the search for expected correlations in greater detail. With our sample of 33 AGN with small errors, correlations should be apparent. Some of these measured before, such as the X-ray Baldwin effect (see Fig. \ref{fig:ew_Lbol}) and the relationship between photon index and reflection coefficient (see Fig. \ref{fig:refl_gamma}) are indeed significant in our sample. However, correlations between fundamental corona parameters such as folding energy and photon index and the physical parameters of the system, like the black hole mass and Eddington ratio, are lacking even with the better constraints on all the free parameters. For example, a correlation between the folding energy and photon index is expected in thermal comptonization models from the analytic expression for the slope in optical depth -- electron temperature space. Our small error bars here make this lack of correlation more stark than in previous studies. Combining our analysis with the lack of expected correlations found in other recent observational studies \citep[e.g.,][]{ricci17, tortosa18} suggests that such correlations are indeed absent. This suggests that this set of free parameters might not be simply related to the physics of the sources as expected. It also prompts the question of what these lack of correlations mean in the context of our current understanding of the X-ray corona?

\subsection{Comparison with Black Hole Binaries}

We find that our results are generally consistent with studies on Comptonization in stellar mass black holes. When comparing our results on compactness and electron temperature, we find our AGN parameters are roughly consistent with those of Galactic black hole binaries (BHBs) \citep[c.f.][]{fabian15}. We also find that the values of electron temperature are roughly consistent with well-studied BHBs \citep{burke17, banerjee20}, although we find several AGN with higher temperatures. The range of Eddington ratios is similar between the BHBs and our AGN, although the distribution of BHB Eddington ratios may peak at lower values \citep{burke17, banerjee20}. Additionally the gaps in Eddington ratio seen for BHBs is not seen for our AGN. The reflection coefficients for our AGN on the lower end are consistent with well-studied BHBs \citep{burke17}, but we have several objects with significantly higher reflection coefficients, particularly those with the softest spectra (see Fig. \ref{fig:refl_gamma}). Finally, the EW of the Fe \Ka lines between the BHBs and AGN are similar \citep{burke17}.

\subsection{Detailed Measurements of AGN parameters}

In Figure \ref{fig:comp_theta}, we compared the electron temperature and compactness of our sources to theoretical expectations. We found that the majority of sources were below the lines from electron-electron pair annihilation and a slab geometry. However, our electron temperatures are calculated using the simple assumptions of \citet{petrucci01} and \citet{fabian12} that $\Theta = E_{fold} / 2m_{e}c^{2}$.  Additionally, when calculating the compactness parameter, we follow the assumption that the radius of the emitting region is 10 gravitational radii unless the source has had more detailed measurements \citep{fabian15}. This clearly introduces significant scatter into our compactness measurements. 

In the future, it will be imperative to constrain the sizes of AGN coronae directly through gravitational lensing \citep[e.g.,][]{jovanovic08} and X-ray timing analysis \citep[e.g,][]{mohan14}. This will not only be extremely valuable to understanding the structure of the X-ray corona, but also to placing the results of spectral studies such as ours in the appropriate context.

In this study we have also compared our best-fit parameters to physical parameters of the AGN such as black hole mass and Eddington ratio. As discussed previously, there is significant scatter in these values based on the fact that the masses are derived from several different methods. Recently there have been improvements in creating a consistent sample of black hole masses, particularly through reverberation mapping \citep[e.g.,][]{bentz15, fasnaugh17}. It will be important to obtain additional measurements of black hole masses in a uniform manner to allow for direct comparisons between physical and modeled parameters.

\section{Conclusion} \label{sec:conclusion}

In this work we have combined Swift BAT data of 33 AGN in the Swift BAT 105 month survey with simultaneous XMM-Newton and NuSTAR observations. We have used the simple \textsc{pexrav} model to obtain constraints on fundamental spectral shape parameters such as the photon index, folding energy, and reflection coefficient. We have also recorded soft excess, partial covering, and Fe \Ka line characteristics. We note that the use of an empirical model ignores much of the detailed physics driving the observed emission. Nonetheless, we have shown our fits to be acceptable, and the lack of specific choices on the underlying physics allows us to compare a broad sample of AGN uniformly. We find that the uncertainties on our best-fit parameters are significantly smaller than those in previously studies. In the case of the folding energies, the uncertainties are roughly half those obtained without the inclusion of the higher energy Swift BAT data in addition to NuSTAR. 

We have compared the properties of the Seyfert 1 - 1.9 and Seyfert 2 AGN in our sample and find that the Eddington ratio and photon index are the parameters for which their distributions are significantly different. We have recovered well-known relationships such as the one between reflection coefficient and photon index \citep{zdziarski99} and the X-ray Baldwin effect \citep{iwasawa93, nandra97}. 

Despite the small statistical uncertainties for most of the parameters for the bulk of the sample, we do not find strong correlations between coronal temperatures, compactness, black hole mass and Eddington ratios. While somewhat puzzling, this is consistent with previous work which had smaller samples and larger uncertainties. This indicates that the fundamental physics behind these Comptonization parameters is not yet understood. We hope that this work will stimulate physical models which can be compared to the data. We note that there is a wide range in effective folding energies which should be taken into account in modeling the contribution of AGN to the X-ray background and in calculating the effects of Comptonization on the continuum.

We are in the process of fitting Comptonization models such as \textsc{compPS} \citep{poutanen96} to the high S/N data and our preliminary findings show that this model can fit most of the sources -- however there is a strong correlation between the optical depth and electron temperature for many of the sources, making direct comparisons of the folding energy and electron temperature difficult. Alternatively the probability contours for the free parameters y (the Comptonization parameter) and T are well behaved allowing direct comparison.

In the late stages of preparing this manuscript, the Swift BAT 157 Month survey\footnote{\url{https://swift.gsfc.nasa.gov/results/bs157mon/}} was released. This expanded coverage will both increase the S/N of known AGN in our sample and add additional sources for which a similar analysis can be conducted. Additionally, while this study focused on AGN with simultaneous XMM-Newton and NuSTAR observations, for sources with low variability it may be possible to combine different epochs to increase S/N and the number of AGN for which such an analysis can be done. Such studies with increased sample sizes and data quality will be imperative to understanding if the lack of observed correlations is a result of observational constraints or a sign of missing physics in our understanding of AGN coronae.

\section*{Acknowledgements}
We thank the referee for helpful comments and suggestions that have improved the quality of this manuscript. We thank Benjamin Shappee and Michael Tucker for comments on the manuscript.

This research has made use of data and/or software provided by the High Energy Astrophysics Science Archive Research Center (HEASARC), which is a service of the Astrophysics Science Division at NASA/GSFC and the High Energy Astrophysics Division of the Smithsonian Astrophysical Observatory. This research has made use of the NASA/IPAC Extragalactic Database (NED), which is operated by the Jet Propulsion Laboratory, California Institute of Technology, under contract with the National Aeronautics and Space Administration. We acknowledge the use of public data from the Swift data archive. 

\section*{Data availability}
The data underlying this article are available in the article and in its online supplementary material.

\begin{landscape}
\begin{table} 
\renewcommand{\arraystretch}{1.3}
\caption{Spectral fit parameters of the objects in our sample, ordered by increasing right ascension. The references for black hole mass are as follows: (1) is \citet{afanasiev19}, (2) is \citet{koss17}, (3) is \citet{denney06}, (4) is \citet{lu19}, (5) is \citet{oliva99}, (6) uses the M-$\sigma$ relation \citep{gultekin09}, (7) is \citet{grier12}, (8) is \citet{bentz10}, (9) is \citet{bentz09}, (10) uses the NIR scaling of \citet{marconi03}, (11) is \citet{winter10}, (12) is \citet{peterson98}, (13) is \citet{peterson04}, (14) is \citet{u13}, (15) is \citet{davis19}, (16) is \citet{bentz06}, (17) is \citet{ho09}, (18) is \citet{walsh12}, (19) is \citet{mclure06}, (20) is \citet{collier98}, (21) is \citet{netzer90}, (22) is \citet{peterson02}, (23) is \citet{demarco13}, (24) is \citet{kollatschny14}, (25) is \citet{kaspi00}, (26) is \citet{santos97}.}
\label{tab:param}
\begin{tabular}{llllllllllll}
\hline
Object & Type & Redshift & log(M$_{\textrm{BH}}$) & log($\lambda_{e}$) & n$_\textrm{H}$ & $\Gamma$ & E$_{\textrm{fold}}$ & R & $\ell$ & $\chi^2$/d.o.f. & Ref.\\ & & & (M$_\odot$) &  & (10$^{22}$ cm$^{-2}$) &  & (keV) & & & &\\ 
\hline
Mrk1501 & S1.2 & 0.08934 & $8.07_{-0.17}^{+0.12}$ & $-0.43$ & $<2.90E-03$ & $1.77 \pm 0.02$ & $>240.5$ & $0.089_{-0.086}^{+0.131}$ & $9.5E+01 \pm 3.0E+00$ & 1.01 & 7\\ 
Mrk1148 & S1.5 & 0.06400 & $8.69 \pm 0.18$ & $-1.44$ & $<2.00E-03$ & $1.82 \pm 0.01$ & $100.6_{-9.1}^{+11.4}$ & $0.404_{-0.065}^{+0.064}$ & $1.0E+01_{-3.5E-01}^{+5.3E-01}$ & 1.01 & 1\\ 
Fairall9 & S1.2 & 0.04702 & $8.30_{-0.08}^{+0.12}$ & $-1.09$ & $<3.00E-03$ & $1.99 \pm 0.01$ & $271.4_{-120.0}^{+528.6}$ & $0.962_{-0.165}^{+0.207}$ & $1.1E+01_{-8.6E-02}^{+1.6E-01}$ & 1.15 & 13,26\\
Mrk359 & S1.5 & 0.01739 & $7.03 \pm 0.72$ & $-1.42$ & $<3.45E-03$ & $1.99 \pm 0.01$ & $98.3_{-17.1}^{+25.7}$ & $3.05 \pm 0.20$ & $1.6E+01_{-1.3E+00}^{+6.8E+00}$ & 0.90 & 2\\ 
NGC931 & S1 & 0.01665 & $7.64 \pm 0.40$ & $-1.34$ & $2.66E-01_{-5.00E-03}^{+5.70E-03}$ & $1.75 \pm 0.01$ & $141.8_{-27.1}^{+42.9}$ & $0.570_{-0.109}^{+0.117}$ & $1.6E+01_{-9.8E-01}^{+2.5E+00}$ & 1.21 & 23\\
NGC1052 & S2 & 0.00504 & $8.96 \pm 0.29$ & $-4.16$ & $1.40E+01 \pm 3.80E-01$ & $1.63_{-0.03}^{+0.04}$ & $374.7_{-211.1}^{+1522.0}$ & $<0.034$ & $3.5E-02_{-1.7E-03}^{+3.3E-03}$ & 1.15 & 2\\ 
NGC1068 & S2 & 0.00379 & $6.75 \pm 0.08$ & $-2.13$ & $7.40E+02_{-5.37E+01}^{+6.20E+01}$ & $1.21_{-0.07}^{+0.13}$ & $28.4_{-4.0}^{+7.7}$ & $< 0.002$ & $3.9E+00_{-6.6E-02}^{+7.9E-02}$ & 1.84 & 15\\
3C109 & S1.8 & 0.30560 & $8.30 \pm 0.40$ & $0.15$ & $4.69E-01_{-3.16E-02}^{+3.56E-02}$ & $1.62 \pm 0.03$ & $112.4_{-58.2}^{+62.1}$ & $<0.021$ & $3.0E+02_{-1.8E+01}^{+4.5E+01}$ & 1.03 & 19\\ 
3C120 & S1 & 0.03301 & $7.75 \pm 0.04$ & $-0.56$ & $7.60E-03 \pm 1.02E-03$ & $1.77 \pm 0.01$ & $158.0_{-6.9}^{+7.5}$ & $0.273_{-0.018}^{+0.017}$ & $7.9E+01_{-7.0E-01}^{+7.2E-01}$ & 1.04 & 7,12,13,24\\
Ark120 & S1 & 0.03271 & $8.07_{-0.06}^{+0.05}$ & $-1.03$ & $<1.72E-03$ & $2.01 \pm 0.01$ & $506.2_{-200.0}^{+813.8}$ & $0.985_{-0.121}^{+0.132}$ & $6.3E+01_{-1.9E+00}^{+1.7E+00}$ & 1.18 & 12,13\\
ESO362-18 & S1 & 0.01244 & $6.25 \pm 0.16$ & $-0.35$ & $<1.60E-03$ & $1.46 \pm 0.01$ & $88.8_{-24.6}^{+39.2}$ & $0.407_{-0.156}^{+0.209}$ & $3.4E+01_{-6.6E-01}^{+8.3E-01}$ & 1.42 & 10\\ 
Mrk3 & S2 & 0.01351 & $8.95 \pm 0.30$ & $-2.42$ & $5.71E+01_{-2.68E+00}^{+3.53E+00}$ & $1.29 \pm 0.14$ & $105.2_{-26.3}^{+10.8}$ & $<0.227$ & $1.3E+00_{-6.3E-02}^{+1.3E-01}$ & 1.06 & 2\\ 
IRAS09149-6206 & S1 & 0.05730 & $8.48 \pm 0.06$ & $-1.32$ & $1.16E+00_{-5.40E-02}^{+5.50E-02}$ & $1.91 \pm 0.04$ & $39.3_{-6.0}^{+9.1}$ & $2.04_{-0.39}^{+0.45}$ & $4.9E+00_{-3.4E-01}^{+1.0E+00}$ & 1.43 & 10\\
NGC3227 & S2 & 0.00386 & $7.18_{-0.88}^{+0.19}$ & $-2.00$ & $3.51E-02_{-2.43E-03}^{+9.83E-02}$ & $1.41 \pm 0.01$ & $60.0_{-4.4}^{+4.5}$ & $0.302_{-0.068}^{+0.080}$ & $4.6E+00_{-4.0E-01}^{+2.5E-01}$ & 1.19 & 2\\ 
NGC3998 & S1 & 0.00350 & $8.91_{-0.10}^{+0.11}$ & $-4.88$ & $9.61E-02_{-6.33E-02}^{+1.55E-01}$ & $1.81_{-0.04}^{+0.05}$ & $115.0_{-38.6}^{+128.2}$ & $<0.230$ & $8.0E-03_{-1.6E-04}^{+2.3E-04}$ & 0.98 & 18\\
NGC4151 & S1.5 & 0.00332 & $7.55 \pm 0.05$ & $-1.72$ & $1.03E+01_{-2.70E-01}^{+2.80E-01}$ & $1.67 \pm 0.02$ & $177.4_{-13.3}^{+16.8}$ & $0.494_{-0.043}^{+0.048}$ & $7.6E+00_{-7.8E-02}^{+9.4E-02}$ & 1.16 & 16\\
Mrk766 & S1 & 0.01293 & $6.82_{-0.06}^{+0.05}$ & $-1.19$ & $4.43E-02_{-1.07E-02}^{+9.50E-03}$ & $1.96 \pm 0.02$ & $124.5_{-40.2}^{+99.5}$ & $0.455_{-0.195}^{+0.225}$ & $7.9E+01 \pm 2.9E+00$ & 1.25 & 8,9\\ 
3C273 & S1 & 0.15834 & $8.84_{-0.08}^{+0.11}$ & $0.67$ & $<1.45E-03$ & $1.67 \pm 0.01$ & $700.2_{-122.7}^{+187.8}$ & $<0.034$ & $7.7E+02_{-1.2E+01}^{+2.3E+01}$ & 1.28 & 13,25\\ 
NGC4579 & S2 & 0.00506 & $7.77 \pm 0.11$ & $-3.60$ & $3.62E-02_{-2.02E-02}^{+2.28E-02}$ & $1.72 \pm 0.04$ & $67.4_{-15.5}^{+27.5}$ & $<0.195$ & $1.5E-01_{-3.4E-03}^{+4.4E-03}$ & 1.06 & 6,17\\ 
NGC4593 & S1 & 0.00900 & $6.88_{-0.08}^{+0.10}$ & $-1.01$ & $1.93E-02_{-1.93E-02}^{+2.27E-02}$ & $1.67 \pm 0.03$ & $223.3_{-64.8}^{+137.0}$ & $0.193_{-0.122}^{+0.135}$ & $3.8E+01_{-6.7E-01}^{+1.0E+00}$ & 1.03 & 3\\
NGC4785 & S2 & 0.01227 & $7.99 \pm 0.09$ & $-2.65$ & $3.74E+01_{-5.25E+00}^{+4.85E+00}$ & $1.46_{-0.38}^{+0.49}$ & $>53.0$ & $0.299_{-0.297}^{+0.797}$ & $9.9E-01_{-1.9E-02}^{+2.4E-02}$ & 1.01 & 5,6\\ 
Mrk273 & S2 & 0.03778 & $9.02 \pm 0.04$ & $-3.11$ & $4.18E+01_{-7.94E+00}^{+7.28E+00}$ & $1.64_{-0.57}^{+0.38}$ & $>36.8$ & $1.24_{-0.99}^{+1.82}$ & $3.0E-01_{-2.7E-03}^{+2.9E-03}$ & 0.94 & 14\\ 
Mrk841 & S1.5 & 0.03642 & $8.76 \pm 0.27$ & $-1.99$ & $<1.30E-02$ & $1.83 \pm 0.01$ & $139.1_{-49.4}^{+142.2}$ & $0.776_{-0.239}^{+0.282}$ & $3.3E+00_{-1.5E-01}^{+2.8E-01}$ & 1.07 & 1\\ 
Mrk1392 & S1.8 & 0.03614 & $6.88 \pm 0.14$ & $-0.40$ & $<1.10E-02$ & $2.11_{-0.06}^{+0.07}$ & $85.4_{-42.5}^{+414.6}$ & $4.68_{-2.05}^{+3.52}$ & $2.7E+01_{-5.2E-01}^{+6.5E-01}$ & 1.02 & 10\\ 
3C382 & S1 & 0.05787 & $8.01_{-0.05}^{+0.09}$ & $-0.33$ & $<3.00E-03$ & $1.72 \pm 0.01$ & $163.4_{-33.8}^{+53.7}$ & $0.204_{-0.095}^{+0.103}$ & $1.2E+02_{-1.3E+00}^{+2.7E+00}$ & 1.06 & 4\\ 
SwiftJ2127.4+5654 & NLS1 & 0.01470 & $6.34 \pm 0.19$ & $-0.38$ & $5.92E-03_{-1.76E-03}^{+1.74E-03}$ & $1.85 \pm 0.01$ & $39.0_{-0.7}^{+0.8}$ & $1.83_{-0.040}^{+0.039}$ & $3.6E+01_{-2.2E+00}^{+1.1E+01}$ & 1.30 & 10\\ 
IIZw171 & S1 & 0.07000 & $7.33 \pm 0.17$ & $-0.27$ & $<1.10E-02$ & $1.74 \pm 0.04$ & $60.3_{-15.3}^{+28.9}$ & $<0.494$ & $3.1E+01_{-5.2E-01}^{+6.3E-01}$ & 1.13 & 10\\ 
NGC7314 & S1 & 0.00476 & $7.24 \pm 0.56$ & $-2.18$ & $7.64E-01_{-8.80E-03}^{+9.00E-03}$ & $1.94 \pm 0.01$ & $218.5_{-62.4}^{+138.3}$ & $0.715_{-0.121}^{+0.130}$ & $3.1E+00_{-2.3E-01}^{+8.2E-01}$ & 1.05 & 2\\ 
Mrk915 & S1 & 0.02411 & $7.76 \pm 0.37$ & $-1.41$ & $1.36E+00_{-3.40E-02}^{+3.30E-02}$ & $1.38_{-0.03}^{+0.02}$ & $57.9_{-7.4}^{+11.2}$ & $<0.041$ & $1.4E+01_{-7.8E-01}^{+1.8E+00}$ & 1.05 & 2\\
MR2251-178 & S1.5 & 0.06398 & $7.82 \pm 0.15$ & $0.05$ & $1.42E-01_{-5.00E-02}^{+4.85E-02}$ & $1.57 \pm 0.03$ & $58.0_{-7.5}^{+9.5}$ & $<0.011$ & $5.4E+01_{-1.3E+00}^{+1.8E+00}$ & 1.11 & 10\\ 
NGC7469 & S1 & 0.01632 & $6.96 \pm 0.05$ & $-0.61$ & $<1.90E-03$ & $1.94 \pm 0.01$ & $112.8_{-21.9}^{+32.8}$ & $2.04_{-0.24}^{+0.26}$ &$8.7E+01_{-9.4E-01}^{+1.0E+00}$ & 1.17 & 13,20\\
Mrk926 & S1.5 & 0.04686 & $8.36 \pm 0.02$ & $-0.75$ & $<4.65E-03$ & $1.70 \pm 0.01$ & $172.8_{-26.4}^{+36.2}$ & $<0.051$ & $4.6E+01_{-2.1E-01}^{+2.2E-01}$ & 1.01 & 11\\
NGC7582 & S1 & 0.00525 & $7.67_{-0.08}^{+0.09}$ & $-2.37$ & $2.55E+01_{-6.10E-01}^{+6.30E-01}$ & $1.36_{-0.06}^{+0.07}$ & $72.8_{-9.9}^{+13.9}$ & $0.518_{-0.135}^{+0.174}$ & $1.9E+00_{-3.2E-02}^{+4.4E-02}$ & 1.08 & 15\\ 
\hline
\end{tabular}
\end{table} 
\end{landscape}

\begin{landscape}
\begin{table} 
\renewcommand{\arraystretch}{1.3}
\caption{Soft X-ray emission and Fe \Ka line parameters for the objects in our sample.}
\label{tab:soft_fe}
\begin{tabular}{llllllll}
\hline
Object & kT & log(L$_{BB}$) & Covering Fraction & Fe K$\alpha$ Energy & $\sigma$ & Fe K$\alpha$ Normalization & Fe K$\alpha$ EW\\ & (keV) & (erg s$^{-1}$) & & (keV) & (keV) & (photons cm$^{-2}$ s$^{-1}$) & (eV)\\ 
\hline
Mrk1501 & $2.03E-01_{-8.80E-03}^{+8.30E-03}$ & $43.5 \pm 0.1$ & --- & $6.33_{-0.08}^{+0.19}$ & $1.7E-01_{-1.3E-01}^{+2.8E-01}$ & $1.0E-05_{-3.7E-06}^{+6.4E-06}$ & 50.7\\ 
Mrk1148 & $1.41E-01 \pm 2.40E-03$ & $43.8 \pm 0.1$ & --- & $6.40$ & $1.0E-01$ & $9.1E-06 \pm 4.1E-06$ & 40.4\\ 
Fairall9 & $1.60E-01_{-3.00E-03}^{+4.00E-03}$ & $43.5 \pm 0.1$ & --- & $6.39 \pm 0.02$ & $1.9E-01_{-5.4E-02}^{+6.4E-02}$ & $3.8E-05_{-5.0E-06}^{+5.7E-06}$ & 145.1\\
Mrk359 & $1.64E-01_{-4.30E-03}^{+4.40E-03}$ & $42.0 \pm 0.1$ & --- & $6.42_{-0.11}^{+0.05}$ & $<1.0E-01$ & $4.3E-06 \pm 1.2E-06$ & 65.1\\ 
NGC931 & $6.66E-02_{-1.15E-03}^{+1.18E-03}$ & $43.7 \pm 0.1$ & --- & $6.44 \pm 0.2$ & $1.4E-01_{-2.9E-02}^{+3.6E-02}$ & $3.8E-05_{-4.7E-06}^{+4.1E-06}$ & 136.1\\ 
NGC1052 & --- & --- & $0.93 \pm 0.01$ & $6.40 \pm 0.2$ & $5.8E-02_{-3.7E-02}^{+2.8E-02}$ & $1.4E-05_{-1.9E-06}^{+2.0E-06}$ & 140.6\\ 
NGC1068 & --- & --- & $0.88_{-0.01}^{+0.04}$ & $6.57 \pm 0.01$ & $2.8E-01 \pm 1.3E-02$ & $9.3E-04_{-2.2E-04}^{+1.3E-04}$ & 2875\\ 
3C109 & --- & --- & $0.98 \pm 0.02$ & $6.60_{-0.12}^{+0.11}$ & $4.1E-01_{-1.6E-01}^{+2.4E-01}$ & $1.3E-05_{-3.8E-06}^{+5.3E-06}$ & 111.5\\ 
3C120 & $2.50E-01_{-3.20E-03}^{+2.90E-03}$ & $43.4\pm 0.1$ & --- & $6.44 \pm 0.03$ & $1.7E-01_{-4.2E-02}^{+5.6E-02}$ & $4.1E-05_{-4.2E-06}^{+3.8E-06}$ & 77.8\\ 
Ark120 & $1.54E-01_{-2.80E-03}^{+2.60E-03}$ & $43.5\pm 0.1$ & --- & $6.50 \pm 0.03$ & $4.1E-01_{-6.8E-02}^{+8.7E-02}$ & $8.9E-05_{-9.3E-06}^{+1.1E-05}$ & 218.1\\ 
ESO362-18 & $1.13E-01_{-1.80E-03}^{+1.70E-03}$ & $41.7 \pm 0.1$ & -- & $6.39 \pm 0.01$ & $7.2E-02 \pm 1.3E-02$ & $2.4E-05_{-1.3E-06}^{+1.4E-06}$ & 255.6\\ 
Mrk3 & --- & --- & $0.94 \pm 0.01$ & $6.44 \pm 0.03$ & $<7.1E-02$ & $9.1E-05_{-1.4E-05}^{+2.0E-05}$ & 309.4\\ 
IRAS09149-6206 & $1.26E-01_{-2.80E-03}^{+2.90E-03}$ & $45.1\pm 0.1$ & --- & $6.40$ & $1.1E-01_{-3.8E-02}^{+3.6E-02}$ & $1.5E-05 \pm 2.6E-06$ & 74.7\\
NGC3227 & --- & --- & $0.98_{-0.02}^{+0.63}$ & $6.41_{-0.01}^{+0.02}$ & $1.3E-01_{-2.3E-02}^{+2.8E-02}$ & $5.9E-05_{-4.8E-06}^{+5.2E-06}$ & 164.5\\
NGC3998 & --- & --- & $0.49_{-0.51}^{+0.21}$ & $<6.42$ & $<1.7E-05$ & $1.9E-06_{-1.5E-06}^{+1.7E-06}$ & 27.3\\ 
NGC4151 & --- & --- & $0.93 \pm 0.01$ & $6.35_{-0.01}^{+0.02}$ & $6.8E-02_{-2.8E-02}^{+1.8E-02}$ & $1.5E-04_{-1.1E-05}^{+1.0E-05}$ & 88.4\\ 
Mrk766 & $7.50E-02_{-1.70E-03}^{+1.60E-03}$ & $43.0\pm 0.1$ & --- & $6.55_{-0.14}^{+0.15}$ & $3.2E-01_{-1.5E-01}^{+1.8E-01}$ & $1.4E-05_{-4.9E-06}^{+5.5E-06}$ & 97.0\\ 
3C273 & $1.46E-01_{-2.80E-03}^{+2.70E-03}$ & $44.7\pm 0.1$ & --- & $6.40$ & $1.0E-01$ & $1.5E-05_{-4.0E-06}^{+4.1E-06}$ & 17.0\\ 
NGC4579 & $1.54E-01_{-1.00E-02}^{+1.30E-02}$ & $41.0_{-0.1}^{+0.2}$ & --- & $6.56 \pm 0.06$ & $3.3E-01_{-6.7E-02}^{+8.2E-02}$ & $1.6E-05_{-2.5E-06}^{+2.7E-06}$ & 259.2\\ 
NGC4593 & $8.27E-02_{-6.22E-03}^{+5.51E-03}$ & $42.4\pm 0.1$ & --- & $6.43 \pm 0.04$ & $1.0E-01_{-6.7E-02}^{+9.5E-02}$ & $3.6E-05_{-7.5E-06}^{+8.7E-06}$ & 148.8\\
NGC4785 & --- & --- & --- & $6.35 \pm 0.06$ & $1.4E-01_{-8.8E-02}^{+9.3E-02}$ & $9.5E-06_{-2.5E-06}^{+3.3E-06}$ & 316.1\\ 
Mrk273 & --- & --- & $0.95_{-0.02}^{+0.06}$ & $6.37_{-0.21}^{+0.14}$ & $<5.2E-01$ & $6.4E-06_{-3.8E-06}^{+7.8E-06}$ & 197.8\\ 
Mrk841 & $8.67E-02_{-2.15E-03}^{+2.13E-03}$ & $43.5\pm 0.1$ & --- & $6.42_{-0.07}^{+0.06}$ & $1.5E-01_{-7.5E-02}^{+1.3E-01}$ & $1.3E-05_{-3.5E-06}^{+4.5E-06}$ & 87.9\\ 
Mrk1392 & $7.91E-02_{-1.41E-02}^{+1.29E-02}$ & $43.1_{-0.1}^{+0.4}$ & --- & $6.39_{-0.11}^{+0.15}$ & $1.0E-01$ & $6.8E-06_{-3.6E-06}^{+3.4E-06}$ & 110.2\\ 
3C382 & $1.17E-01_{-5.30E-03}^{+5.20E-03}$ & $43.7\pm 0.1$ & --- & $6.44 \pm 0.04$ & $1.1E-01_{-5.7E-02}^{+1.0E-01}$ & $3.2E-05_{-6.6E-06}^{+8.7E-06}$ & 80.1\\ 
SwiftJ2127.4+5654 & $2.66E-01 \pm 4.50E-03$ & $42.3\pm 0.1$ & --- & $6.41_{-0.04}^{+0.06}$ & $2.3E-01_{-1.8E-01}^{+2.9E-01}$ & $2.0E-05_{-2.6E-06}^{+2.6E-06}$ & 58.3\\ 
IIZw171 & $2.21E-01_{-1.12E-02}^{+1.02E-02}$ & $43.3\pm 0.1$ & --- & $6.36_{-0.05}^{+0.19}$ & $<8.9E-01$ & $5.5E-06_{-1.6E-06}^{+1.4E-05}$ & 55.9\\ 
NGC7314 & $6.44E-02_{-1.41E-03}^{+1.40E-03}$ & $43.4\pm 0.1$ & --- & $6.53 \pm 0.05$ & $5.1E-01_{-7.6E-02}^{+9.4E-02}$ & $7.3E-05_{-8.4E-06}^{+9.5E-06}$ & 230.9\\
Mrk915 & --- & --- & $0.85 \pm 0.01$ & $6.41_{-0.01}^{+0.02}$ & $1.1E-01_{-2.7E-02}^{+2.8E-02}$ & $1.5E-05_{-1.6E-06}^{+1.5E-06}$ & 148.4\\
MR2251-178 & $1.84E-01_{-9.70E-03}^{+1.18E-02}$ & $44.1\pm 0.2$ & --- & $6.40$ & $>1.7E-02$ & $1.9E-05_{-6.8E-06}^{+7.8E-04}$ & 29.0\\
NGC7469 & $1.00E-01 \pm 1.70E-03$ & $42.8\pm 0.1$ & --- & $6.43 \pm 0.01$ & $6.8E-02_{-2.3E-02}^{+2.4E-02}$ & $3.3E-05_{-3.8E-06}^{+3.7E-06}$ & 100.8\\ 
Mrk926 & $1.89E-01_{-7.40E-03}^{+7.00E-03}$ & $43.6 \pm 0.1$ & --- & $6.43 \pm 0.06$ & $3.3E-01_{-6.0E-02}^{+1.1E-01}$ & $5.5E-05_{-8.4E-06}^{+5.6E-06}$ & 95.8\\ 
NGC7582 & --- & --- & --- & $6.40 \pm 0.01$ & $5.7E-02_{-1.9E-02}^{+1.6E-02}$ & $3.4E-05_{-2.7E-06}^{+3.2E-06}$ & 167.4\\ 
\hline 
\end{tabular}
\end{table} 
\end{landscape} 

\begin{table*} 
\caption{Model parameters used in the fits for our various objects. Each object fit includes an additional TBabs component frozen to the Galactic value.}
\label{tab:models}
\begin{tabular}{lcccccc}
\hline
Object & TBabs & TBpcf & zbbody & zgauss & pexrav & other\\ 
\hline
Mrk1501 & \checkmark & & \checkmark & \checkmark & \checkmark \\ 
Mrk1148 & \checkmark & & \checkmark & \checkmark & \checkmark \\ 
Fairall9 & \checkmark & & \checkmark & \checkmark & \checkmark \\  
Mrk359 & \checkmark & & \checkmark & \checkmark & \checkmark \\
NGC931 & \checkmark & & \checkmark & \checkmark & \checkmark \\
NGC1052 & & \checkmark & & \checkmark & \checkmark \\ 
NGC1068 & & \checkmark & & \checkmark & \checkmark \\ 
3C109 & & \checkmark & & \checkmark & \checkmark \\ 
3C120 & \checkmark & & \checkmark & \checkmark & \checkmark \\ 
Ark120 & \checkmark & & \checkmark & \checkmark & \checkmark \\
ESO362-18 & \checkmark & & \checkmark & \checkmark & \checkmark \\ 
Mrk3 & & \checkmark & & \checkmark & \checkmark \\ 
IRAS09149-6206 & \checkmark & & \checkmark & \checkmark & \checkmark \\ 
NGC3227 & & \checkmark & & \checkmark & \checkmark \\ 
NGC3998 & & \checkmark & & \checkmark & \checkmark \\ 
NGC4151 & & \checkmark & & \checkmark & \checkmark \\ 
Mrk766 & \checkmark & & \checkmark & \checkmark & \checkmark \\ 
3C273 & \checkmark & & \checkmark & \checkmark & \checkmark \\ 
NGC4579 & \checkmark & & \checkmark & \checkmark & \checkmark \\ 
NGC4593 & \checkmark & & \checkmark & \checkmark & \checkmark \\
NGC4785 & & \checkmark & & \checkmark & \checkmark \\ 
Mrk273 & & \checkmark & & \checkmark & \checkmark & \checkmark \\ 
Mrk841 & \checkmark & & \checkmark & \checkmark & \checkmark \\
Mrk1392 & \checkmark & & \checkmark & \checkmark & \checkmark \\
3C382 & \checkmark & & \checkmark & \checkmark & \checkmark \\ 
SwiftJ2127.4+5654 & \checkmark & & \checkmark & \checkmark & \checkmark \\
IIZw171 & \checkmark & & \checkmark & \checkmark & \checkmark \\
NGC7314 & \checkmark & & \checkmark & \checkmark & \checkmark \\
Mrk915 & & \checkmark & & \checkmark & \checkmark \\ 
MR2251-178 & & \checkmark & & \checkmark & \checkmark \\  
NGC7469 & \checkmark & & \checkmark & \checkmark & \checkmark \\
Mrk926 & \checkmark & & \checkmark & \checkmark & \checkmark \\
NGC7582 & & \checkmark & & \checkmark & \checkmark \\ 
\hline
\end{tabular}
\end{table*} 

\bibliographystyle{mnras}
\bibliography{biblio}

\begin{thebibliography}{}
\makeatletter
\relax
\def\mn@urlcharsother{\let\do\@makeother \do\$\do\&\do\#\do\^\do\_\do\%\do\~}
\def\mn@doi{\begingroup\mn@urlcharsother \@ifnextchar [ {\mn@doi@}
  {\mn@doi@[]}}
\def\mn@doi@[#1]#2{\def\@tempa{#1}\ifx\@tempa\@empty \href
  {http://dx.doi.org/#2} {doi:#2}\else \href {http://dx.doi.org/#2} {#1}\fi
  \endgroup}
\def\mn@eprint#1#2{\mn@eprint@#1:#2::\@nil}
\def\mn@eprint@arXiv#1{\href {http://arxiv.org/abs/#1} {{\tt arXiv:#1}}}
\def\mn@eprint@dblp#1{\href {http://dblp.uni-trier.de/rec/bibtex/#1.xml}
  {dblp:#1}}
\def\mn@eprint@#1:#2:#3:#4\@nil{\def\@tempa {#1}\def\@tempb {#2}\def\@tempc
  {#3}\ifx \@tempc \@empty \let \@tempc \@tempb \let \@tempb \@tempa \fi \ifx
  \@tempb \@empty \def\@tempb {arXiv}\fi \@ifundefined
  {mn@eprint@\@tempb}{\@tempb:\@tempc}{\expandafter \expandafter \csname
  mn@eprint@\@tempb\endcsname \expandafter{\@tempc}}}

\bibitem[\protect\citeauthoryear{{Abarr} \& {Krawczynski}}{{Abarr} \&
  {Krawczynski}}{2021}]{abarr21}
{Abarr} Q.,  {Krawczynski} H.,  2021, \mn@doi [\apj]
  {10.3847/1538-4357/abc826}, \href
  {https://ui.adsabs.harvard.edu/abs/2021ApJ...906...28A} {906, 28}

\bibitem[\protect\citeauthoryear{{Acero} et~al.,}{{Acero}
  et~al.}{2015}]{fermi15}
{Acero} F.,  et~al., 2015, \mn@doi [\apjs] {10.1088/0067-0049/218/2/23}, \href
  {https://ui.adsabs.harvard.edu/abs/2015ApJS..218...23A} {218, 23}

\bibitem[\protect\citeauthoryear{{Afanasiev}, {Popovi{\'c}}  \&
  {Shapovalova}}{{Afanasiev} et~al.}{2019}]{afanasiev19}
{Afanasiev} V.~L.,  {Popovi{\'c}} L.~{\v{C}}.,   {Shapovalova} A.~I.,  2019,
  \mn@doi [\mnras] {10.1093/mnras/sty2995}, \href
  {https://ui.adsabs.harvard.edu/abs/2019MNRAS.482.4985A} {482, 4985}

\bibitem[\protect\citeauthoryear{{Antonucci}}{{Antonucci}}{1993}]{antonucci93}
{Antonucci} R.,  1993, \mn@doi [\araa] {10.1146/annurev.aa.31.090193.002353},
  \href {https://ui.adsabs.harvard.edu/abs/1993ARA&A..31..473A} {31, 473}

\bibitem[\protect\citeauthoryear{{Arnaud}}{{Arnaud}}{1996}]{arnaud96}
{Arnaud} K.~A.,  1996, {XSPEC: The First Ten Years}.
p.~17

\bibitem[\protect\citeauthoryear{{Ballantyne}}{{Ballantyne}}{2020}]{ballantyne20}
{Ballantyne} D.~R.,  2020, \mn@doi [\mnras] {10.1093/mnras/stz3294}, \href
  {https://ui.adsabs.harvard.edu/abs/2020MNRAS.491.3553B} {491, 3553}

\bibitem[\protect\citeauthoryear{{Balokovi{\'c}} et~al.,}{{Balokovi{\'c}}
  et~al.}{2015}]{balokovic15}
{Balokovi{\'c}} M.,  et~al., 2015, \mn@doi [\apj] {10.1088/0004-637X/800/1/62},
  \href {https://ui.adsabs.harvard.edu/abs/2015ApJ...800...62B} {800, 62}

\bibitem[\protect\citeauthoryear{{Balokovi{\'c}} et~al.,}{{Balokovi{\'c}}
  et~al.}{2018}]{balokovic18}
{Balokovi{\'c}} M.,  et~al., 2018, \mn@doi [\apj] {10.3847/1538-4357/aaa7eb},
  \href {https://ui.adsabs.harvard.edu/abs/2018ApJ...854...42B} {854, 42}

\bibitem[\protect\citeauthoryear{{Bambi}, {C{\'a}rdenas-Avenda{\~n}o},
  {Dauser}, {Garc{\'\i}a}  \& {Nampalliwar}}{{Bambi} et~al.}{2017}]{bambi17}
{Bambi} C.,  {C{\'a}rdenas-Avenda{\~n}o} A.,  {Dauser} T.,  {Garc{\'\i}a}
  J.~A.,   {Nampalliwar} S.,  2017, \mn@doi [\apj] {10.3847/1538-4357/aa74c0},
  \href {https://ui.adsabs.harvard.edu/abs/2017ApJ...842...76B} {842, 76}

\bibitem[\protect\citeauthoryear{{Banerjee}, {Gilfanov}, {Bhattacharyya}  \&
  {Sunyaev}}{{Banerjee} et~al.}{2020}]{banerjee20}
{Banerjee} S.,  {Gilfanov} M.,  {Bhattacharyya} S.,   {Sunyaev} R.,  2020,
  \mn@doi [\mnras] {10.1093/mnras/staa2788}, \href
  {https://ui.adsabs.harvard.edu/abs/2020MNRAS.498.5353B} {498, 5353}

\bibitem[\protect\citeauthoryear{{Barthelmy} et~al.,}{{Barthelmy}
  et~al.}{2005}]{barthelmy05}
{Barthelmy} S.~D.,  et~al., 2005, \mn@doi [\ssr] {10.1007/s11214-005-5096-3},
  \href {https://ui.adsabs.harvard.edu/abs/2005SSRv..120..143B} {120, 143}

\bibitem[\protect\citeauthoryear{{Beloborodov}}{{Beloborodov}}{1999}]{beloborodov99}
{Beloborodov} A.~M.,  1999, in {Poutanen} J.,  {Svensson} R.,  eds,
  Astronomical Society of the Pacific Conference Series Vol. 161, High Energy
  Processes in Accreting Black Holes. p.~295 (\mn@eprint {arXiv}
  {astro-ph/9901108})

\bibitem[\protect\citeauthoryear{{Bentz} \& {Katz}}{{Bentz} \&
  {Katz}}{2015}]{bentz15}
{Bentz} M.~C.,  {Katz} S.,  2015, \mn@doi [\pasp] {10.1086/679601}, \href
  {https://ui.adsabs.harvard.edu/abs/2015PASP..127...67B} {127, 67}

\bibitem[\protect\citeauthoryear{{Bentz} et~al.,}{{Bentz}
  et~al.}{2006}]{bentz06}
{Bentz} M.~C.,  et~al., 2006, \mn@doi [\apj] {10.1086/507417}, \href
  {https://ui.adsabs.harvard.edu/abs/2006ApJ...651..775B} {651, 775}

\bibitem[\protect\citeauthoryear{{Bentz} et~al.,}{{Bentz}
  et~al.}{2009}]{bentz09}
{Bentz} M.~C.,  et~al., 2009, \mn@doi [\apj] {10.1088/0004-637X/705/1/199},
  \href {https://ui.adsabs.harvard.edu/abs/2009ApJ...705..199B} {705, 199}

\bibitem[\protect\citeauthoryear{{Bentz} et~al.,}{{Bentz}
  et~al.}{2010}]{bentz10}
{Bentz} M.~C.,  et~al., 2010, \mn@doi [\apj] {10.1088/0004-637X/716/2/993},
  \href {https://ui.adsabs.harvard.edu/abs/2010ApJ...716..993B} {716, 993}

\bibitem[\protect\citeauthoryear{{Best}, {Kauffmann}, {Heckman}, {Brinchmann},
  {Charlot}, {Ivezi{\'c}}  \& {White}}{{Best} et~al.}{2005}]{best05}
{Best} P.~N.,  {Kauffmann} G.,  {Heckman} T.~M.,  {Brinchmann} J.,  {Charlot}
  S.,  {Ivezi{\'c}} {\v{Z}}.,   {White} S.~D.~M.,  2005, \mn@doi [\mnras]
  {10.1111/j.1365-2966.2005.09192.x}, \href
  {https://ui.adsabs.harvard.edu/abs/2005MNRAS.362...25B} {362, 25}

\bibitem[\protect\citeauthoryear{{Boissay}, {Ricci}  \& {Paltani}}{{Boissay}
  et~al.}{2016}]{boissay16}
{Boissay} R.,  {Ricci} C.,   {Paltani} S.,  2016, \mn@doi [\aap]
  {10.1051/0004-6361/201526982}, \href
  {https://ui.adsabs.harvard.edu/abs/2016A&A...588A..70B} {588, A70}

\bibitem[\protect\citeauthoryear{{Boroson} \& {Green}}{{Boroson} \&
  {Green}}{1992}]{boroson92}
{Boroson} T.~A.,  {Green} R.~F.,  1992, \mn@doi [\apjs] {10.1086/191661}, \href
  {https://ui.adsabs.harvard.edu/abs/1992ApJS...80..109B} {80, 109}

\bibitem[\protect\citeauthoryear{{Brenneman} et~al.,}{{Brenneman}
  et~al.}{2014}]{brenneman14}
{Brenneman} L.~W.,  et~al., 2014, \mn@doi [\apj] {10.1088/0004-637X/788/1/61},
  \href {https://ui.adsabs.harvard.edu/abs/2014ApJ...788...61B} {788, 61}

\bibitem[\protect\citeauthoryear{{Burke}, {Gilfanov}  \& {Sunyaev}}{{Burke}
  et~al.}{2017}]{burke17}
{Burke} M.~J.,  {Gilfanov} M.,   {Sunyaev} R.,  2017, \mn@doi [\mnras]
  {10.1093/mnras/stw2514}, \href
  {https://ui.adsabs.harvard.edu/abs/2017MNRAS.466..194B} {466, 194}

\bibitem[\protect\citeauthoryear{{Burtscher} et~al.,}{{Burtscher}
  et~al.}{2015}]{burtscher15}
{Burtscher} L.,  et~al., 2015, \mn@doi [\aap] {10.1051/0004-6361/201525817},
  \href {https://ui.adsabs.harvard.edu/abs/2015A&A...578A..47B} {578, A47}

\bibitem[\protect\citeauthoryear{{Chalise}, {Lohfink}, {Kara}  \&
  {Fabian}}{{Chalise} et~al.}{2020}]{chalise20}
{Chalise} S.,  {Lohfink} A.~M.,  {Kara} E.,   {Fabian} A.~C.,  2020, \mn@doi
  [\apj] {10.3847/1538-4357/ab94a2}, \href
  {https://ui.adsabs.harvard.edu/abs/2020ApJ...897...47C} {897, 47}

\bibitem[\protect\citeauthoryear{{Chiang}, {Reynolds}, {Blaes}, {Nowak},
  {Murray}, {Madejski}, {Marshall}  \& {Magdziarz}}{{Chiang}
  et~al.}{2000}]{chiang00}
{Chiang} J.,  {Reynolds} C.~S.,  {Blaes} O.~M.,  {Nowak} M.~A.,  {Murray} N.,
  {Madejski} G.,  {Marshall} H.~L.,   {Magdziarz} P.,  2000, \mn@doi [\apj]
  {10.1086/308178}, \href
  {https://ui.adsabs.harvard.edu/abs/2000ApJ...528..292C} {528, 292}

\bibitem[\protect\citeauthoryear{{Collier} et~al.,}{{Collier}
  et~al.}{1998}]{collier98}
{Collier} S.~J.,  et~al., 1998, \mn@doi [\apj] {10.1086/305720}, \href
  {https://ui.adsabs.harvard.edu/abs/1998ApJ...500..162C} {500, 162}

\bibitem[\protect\citeauthoryear{{Crummy}, {Fabian}, {Gallo}  \&
  {Ross}}{{Crummy} et~al.}{2006}]{crummy06}
{Crummy} J.,  {Fabian} A.~C.,  {Gallo} L.,   {Ross} R.~R.,  2006, \mn@doi
  [\mnras] {10.1111/j.1365-2966.2005.09844.x}, \href
  {https://ui.adsabs.harvard.edu/abs/2006MNRAS.365.1067C} {365, 1067}

\bibitem[\protect\citeauthoryear{{Dadina}}{{Dadina}}{2007}]{dadina07}
{Dadina} M.,  2007, \mn@doi [\aap] {10.1051/0004-6361:20065734}, \href
  {https://ui.adsabs.harvard.edu/abs/2007A&A...461.1209D} {461, 1209}

\bibitem[\protect\citeauthoryear{{Davis}, {Graham}  \& {Combes}}{{Davis}
  et~al.}{2019}]{davis19}
{Davis} B.~L.,  {Graham} A.~W.,   {Combes} F.,  2019, \mn@doi [\apj]
  {10.3847/1538-4357/ab1aa4}, \href
  {https://ui.adsabs.harvard.edu/abs/2019ApJ...877...64D} {877, 64}

\bibitem[\protect\citeauthoryear{{De Marco}, {Ponti}, {Cappi}, {Dadina},
  {Uttley}, {Cackett}, {Fabian}  \& {Miniutti}}{{De Marco}
  et~al.}{2013}]{demarco13}
{De Marco} B.,  {Ponti} G.,  {Cappi} M.,  {Dadina} M.,  {Uttley} P.,  {Cackett}
  E.~M.,  {Fabian} A.~C.,   {Miniutti} G.,  2013, \mn@doi [\mnras]
  {10.1093/mnras/stt339}, \href
  {https://ui.adsabs.harvard.edu/abs/2013MNRAS.431.2441D} {431, 2441}

\bibitem[\protect\citeauthoryear{{Denney} et~al.,}{{Denney}
  et~al.}{2006}]{denney06}
{Denney} K.~D.,  et~al., 2006, \mn@doi [\apj] {10.1086/508533}, \href
  {https://ui.adsabs.harvard.edu/abs/2006ApJ...653..152D} {653, 152}

\bibitem[\protect\citeauthoryear{{Denney} et~al.,}{{Denney}
  et~al.}{2014}]{denney14}
{Denney} K.~D.,  et~al., 2014, \mn@doi [\apj] {10.1088/0004-637X/796/2/134},
  \href {https://ui.adsabs.harvard.edu/abs/2014ApJ...796..134D} {796, 134}

\bibitem[\protect\citeauthoryear{{Done} \& {Nayakshin}}{{Done} \&
  {Nayakshin}}{2007}]{done07a}
{Done} C.,  {Nayakshin} S.,  2007, \mn@doi [\mnras]
  {10.1111/j.1745-3933.2007.00303.x}, \href
  {https://ui.adsabs.harvard.edu/abs/2007MNRAS.377L..59D} {377, L59}

\bibitem[\protect\citeauthoryear{{Done}, {Gierli{\'n}ski}  \& {Kubota}}{{Done}
  et~al.}{2007}]{done07b}
{Done} C.,  {Gierli{\'n}ski} M.,   {Kubota} A.,  2007, \mn@doi [\aapr]
  {10.1007/s00159-007-0006-1}, \href
  {https://ui.adsabs.harvard.edu/abs/2007A&ARv..15....1D} {15, 1}

\bibitem[\protect\citeauthoryear{{Drake} et~al.,}{{Drake}
  et~al.}{2009}]{drake09}
{Drake} A.~J.,  et~al., 2009, \mn@doi [\apj] {10.1088/0004-637X/696/1/870},
  \href {https://ui.adsabs.harvard.edu/abs/2009ApJ...696..870D} {696, 870}

\bibitem[\protect\citeauthoryear{{Elvis}}{{Elvis}}{2000}]{elvis00}
{Elvis} M.,  2000, \mn@doi [\apj] {10.1086/317778}, \href
  {https://ui.adsabs.harvard.edu/abs/2000ApJ...545...63E} {545, 63}

\bibitem[\protect\citeauthoryear{{Elvis}, {Maccacaro}, {Wilson}, {Ward},
  {Penston}, {Fosbury}  \& {Perola}}{{Elvis} et~al.}{1978}]{elvis78}
{Elvis} M.,  {Maccacaro} T.,  {Wilson} A.~S.,  {Ward} M.~J.,  {Penston} M.~V.,
  {Fosbury} R.~A.~E.,   {Perola} G.~C.,  1978, \mn@doi [\mnras]
  {10.1093/mnras/183.2.129}, \href
  {https://ui.adsabs.harvard.edu/abs/1978MNRAS.183..129E} {183, 129}

\bibitem[\protect\citeauthoryear{{Ezhikode}, {Dewangan}, {Misra}  \&
  {Philip}}{{Ezhikode} et~al.}{2020}]{ezhikode20}
{Ezhikode} S.~H.,  {Dewangan} G.~C.,  {Misra} R.,   {Philip} N.~S.,  2020,
  \mn@doi [\mnras] {10.1093/mnras/staa1288}, \href
  {https://ui.adsabs.harvard.edu/abs/2020MNRAS.495.3373E} {495, 3373}

\bibitem[\protect\citeauthoryear{{Fabian}}{{Fabian}}{2012}]{fabian12}
{Fabian} A.~C.,  2012, \mn@doi [\araa] {10.1146/annurev-astro-081811-125521},
  \href {https://ui.adsabs.harvard.edu/abs/2012ARA&A..50..455F} {50, 455}

\bibitem[\protect\citeauthoryear{{Fabian}, {Iwasawa}, {Reynolds}  \&
  {Young}}{{Fabian} et~al.}{2000}]{fabian00}
{Fabian} A.~C.,  {Iwasawa} K.,  {Reynolds} C.~S.,   {Young} A.~J.,  2000,
  \mn@doi [\pasp] {10.1086/316610}, \href
  {https://ui.adsabs.harvard.edu/abs/2000PASP..112.1145F} {112, 1145}

\bibitem[\protect\citeauthoryear{{Fabian}, {Miniutti}, {Iwasawa}  \&
  {Ross}}{{Fabian} et~al.}{2005}]{fabian05}
{Fabian} A.~C.,  {Miniutti} G.,  {Iwasawa} K.,   {Ross} R.~R.,  2005, \mn@doi
  [\mnras] {10.1111/j.1365-2966.2005.09148.x}, \href
  {https://ui.adsabs.harvard.edu/abs/2005MNRAS.361..795F} {361, 795}

\bibitem[\protect\citeauthoryear{{Fabian}, {Lohfink}, {Kara}, {Parker},
  {Vasudevan}  \& {Reynolds}}{{Fabian} et~al.}{2015}]{fabian15}
{Fabian} A.~C.,  {Lohfink} A.,  {Kara} E.,  {Parker} M.~L.,  {Vasudevan} R.,
  {Reynolds} C.~S.,  2015, \mn@doi [\mnras] {10.1093/mnras/stv1218}, \href
  {https://ui.adsabs.harvard.edu/abs/2015MNRAS.451.4375F} {451, 4375}

\bibitem[\protect\citeauthoryear{{Fausnaugh} et~al.,}{{Fausnaugh}
  et~al.}{2017}]{fasnaugh17}
{Fausnaugh} M.~M.,  et~al., 2017, \mn@doi [\apj] {10.3847/1538-4357/aa6d52},
  \href {https://ui.adsabs.harvard.edu/abs/2017ApJ...840...97F} {840, 97}

\bibitem[\protect\citeauthoryear{{Garc{\'\i}a} et~al.,}{{Garc{\'\i}a}
  et~al.}{2019}]{garcia19}
{Garc{\'\i}a} J.~A.,  et~al., 2019, \mn@doi [\apj] {10.3847/1538-4357/aaf739},
  \href {https://ui.adsabs.harvard.edu/abs/2019ApJ...871...88G} {871, 88}

\bibitem[\protect\citeauthoryear{{Gehrels} et~al.,}{{Gehrels}
  et~al.}{2004}]{gehrels04}
{Gehrels} N.,  et~al., 2004, \mn@doi [\apj] {10.1086/422091}, \href
  {https://ui.adsabs.harvard.edu/abs/2004ApJ...611.1005G} {611, 1005}

\bibitem[\protect\citeauthoryear{{George} \& {Fabian}}{{George} \&
  {Fabian}}{1991}]{george91}
{George} I.~M.,  {Fabian} A.~C.,  1991, \mn@doi [\mnras]
  {10.1093/mnras/249.2.352}, \href
  {https://ui.adsabs.harvard.edu/abs/1991MNRAS.249..352G} {249, 352}

\bibitem[\protect\citeauthoryear{{Ghisellini}, {Haardt}  \&
  {Fabian}}{{Ghisellini} et~al.}{1993}]{ghisellini93}
{Ghisellini} G.,  {Haardt} F.,   {Fabian} A.~C.,  1993, \mn@doi [\mnras]
  {10.1093/mnras/263.1.L9}, \href
  {https://ui.adsabs.harvard.edu/abs/1993MNRAS.263L...9G} {263, L9}

\bibitem[\protect\citeauthoryear{{Gierli{\'n}ski} \& {Done}}{{Gierli{\'n}ski}
  \& {Done}}{2004}]{gierlinski04}
{Gierli{\'n}ski} M.,  {Done} C.,  2004, \mn@doi [\mnras]
  {10.1111/j.1365-2966.2004.07687.x}, \href
  {https://ui.adsabs.harvard.edu/abs/2004MNRAS.349L...7G} {349, L7}

\bibitem[\protect\citeauthoryear{{Gilli}, {Comastri}  \& {Hasinger}}{{Gilli}
  et~al.}{2007}]{gilli07}
{Gilli} R.,  {Comastri} A.,   {Hasinger} G.,  2007, \mn@doi [\aap]
  {10.1051/0004-6361:20066334}, \href
  {https://ui.adsabs.harvard.edu/abs/2007A&A...463...79G} {463, 79}

\bibitem[\protect\citeauthoryear{{Grier} et~al.,}{{Grier}
  et~al.}{2012}]{grier12}
{Grier} C.~J.,  et~al., 2012, \mn@doi [\apj] {10.1088/0004-637X/755/1/60},
  \href {https://ui.adsabs.harvard.edu/abs/2012ApJ...755...60G} {755, 60}

\bibitem[\protect\citeauthoryear{{Grupe}, {Komossa}, {Gallo}, {Longinotti},
  {Fabian}, {Pradhan}, {Gruberbauer}  \& {Xu}}{{Grupe} et~al.}{2012}]{grupe12}
{Grupe} D.,  {Komossa} S.,  {Gallo} L.~C.,  {Longinotti} A.~L.,  {Fabian}
  A.~C.,  {Pradhan} A.~K.,  {Gruberbauer} M.,   {Xu} D.,  2012, \mn@doi [\apjs]
  {10.1088/0067-0049/199/2/28}, \href
  {https://ui.adsabs.harvard.edu/abs/2012ApJS..199...28G} {199, 28}

\bibitem[\protect\citeauthoryear{{Guilbert}, {Fabian}  \& {Rees}}{{Guilbert}
  et~al.}{1983}]{guilbert83}
{Guilbert} P.~W.,  {Fabian} A.~C.,   {Rees} M.~J.,  1983, \mn@doi [\mnras]
  {10.1093/mnras/205.3.593}, \href
  {https://ui.adsabs.harvard.edu/abs/1983MNRAS.205..593G} {205, 593}

\bibitem[\protect\citeauthoryear{{G{\"u}ltekin} et~al.,}{{G{\"u}ltekin}
  et~al.}{2009}]{gultekin09}
{G{\"u}ltekin} K.,  et~al., 2009, \mn@doi [\apj] {10.1088/0004-637X/698/1/198},
  \href {https://ui.adsabs.harvard.edu/abs/2009ApJ...698..198G} {698, 198}

\bibitem[\protect\citeauthoryear{{Guo} et~al.,}{{Guo} et~al.}{2020}]{guo20}
{Guo} H.,  et~al., 2020, arXiv e-prints, \href
  {https://ui.adsabs.harvard.edu/abs/2020arXiv200608645G} {p. arXiv:2006.08645}

\bibitem[\protect\citeauthoryear{{HI4PI Collaboration} et~al.,}{{HI4PI
  Collaboration} et~al.}{2016}]{hip4i16}
{HI4PI Collaboration} et~al., 2016, \mn@doi [\aap]
  {10.1051/0004-6361/201629178}, \href
  {https://ui.adsabs.harvard.edu/abs/2016A&A...594A.116H} {594, A116}

\bibitem[\protect\citeauthoryear{{Haardt} \& {Madau}}{{Haardt} \&
  {Madau}}{1996}]{haardt96}
{Haardt} F.,  {Madau} P.,  1996, \mn@doi [\apj] {10.1086/177035}, \href
  {https://ui.adsabs.harvard.edu/abs/1996ApJ...461...20H} {461, 20}

\bibitem[\protect\citeauthoryear{{Haardt} \& {Maraschi}}{{Haardt} \&
  {Maraschi}}{1991}]{haardt91}
{Haardt} F.,  {Maraschi} L.,  1991, \mn@doi [\apjl] {10.1086/186171}, \href
  {https://ui.adsabs.harvard.edu/abs/1991ApJ...380L..51H} {380, L51}

\bibitem[\protect\citeauthoryear{{Haardt} \& {Maraschi}}{{Haardt} \&
  {Maraschi}}{1993}]{haardt93}
{Haardt} F.,  {Maraschi} L.,  1993, \mn@doi [\apj] {10.1086/173020}, \href
  {https://ui.adsabs.harvard.edu/abs/1993ApJ...413..507H} {413, 507}

\bibitem[\protect\citeauthoryear{{Harrison} et~al.,}{{Harrison}
  et~al.}{2013}]{harrison13}
{Harrison} F.~A.,  et~al., 2013, \mn@doi [\apj] {10.1088/0004-637X/770/2/103},
  \href {https://ui.adsabs.harvard.edu/abs/2013ApJ...770..103H} {770, 103}

\bibitem[\protect\citeauthoryear{{Hartman} et~al.,}{{Hartman}
  et~al.}{1999}]{hartman99}
{Hartman} R.~C.,  et~al., 1999, \mn@doi [\apjs] {10.1086/313231}, \href
  {https://ui.adsabs.harvard.edu/abs/1999ApJS..123...79H} {123, 79}

\bibitem[\protect\citeauthoryear{{Helou}, {Madore}, {Schmitz}, {Bicay}, {Wu}
  \& {Bennett}}{{Helou} et~al.}{1991}]{helou91}
{Helou} G.,  {Madore} B.~F.,  {Schmitz} M.,  {Bicay} M.~D.,  {Wu} X.,
  {Bennett} J.,  1991, {The NASA/IPAC extragalactic database.}.
pp 89--106, \mn@doi{10.1007/978-94-011-3250-3_10}

\bibitem[\protect\citeauthoryear{{Henri} \& {Petrucci}}{{Henri} \&
  {Petrucci}}{1997}]{henri97}
{Henri} G.,  {Petrucci} P.~O.,  1997, \aap, \href
  {https://ui.adsabs.harvard.edu/abs/1997A&A...326...87H} {326, 87}

\bibitem[\protect\citeauthoryear{{Ho}, {Greene}, {Filippenko}  \&
  {Sargent}}{{Ho} et~al.}{2009}]{ho09}
{Ho} L.~C.,  {Greene} J.~E.,  {Filippenko} A.~V.,   {Sargent} W. L.~W.,  2009,
  \mn@doi [\apjs] {10.1088/0067-0049/183/1/1}, \href
  {https://ui.adsabs.harvard.edu/abs/2009ApJS..183....1H} {183, 1}

\bibitem[\protect\citeauthoryear{{Huchra}, {Davis}, {Latham}  \&
  {Tonry}}{{Huchra} et~al.}{1983}]{huchra83}
{Huchra} J.,  {Davis} M.,  {Latham} D.,   {Tonry} J.,  1983, \mn@doi [\apjs]
  {10.1086/190860}, \href
  {https://ui.adsabs.harvard.edu/abs/1983ApJS...52...89H} {52, 89}

\bibitem[\protect\citeauthoryear{{Iwasawa} \& {Taniguchi}}{{Iwasawa} \&
  {Taniguchi}}{1993}]{iwasawa93}
{Iwasawa} K.,  {Taniguchi} Y.,  1993, \mn@doi [\apjl] {10.1086/186948}, \href
  {https://ui.adsabs.harvard.edu/abs/1993ApJ...413L..15I} {413, L15}

\bibitem[\protect\citeauthoryear{{Iwasawa}, {Fabian}, {Kara}, {Reynolds},
  {Miniutti}  \& {Tombesi}}{{Iwasawa} et~al.}{2016}]{iwasawa16}
{Iwasawa} K.,  {Fabian} A.~C.,  {Kara} E.,  {Reynolds} C.~S.,  {Miniutti} G.,
  {Tombesi} F.,  2016, \mn@doi [\aap] {10.1051/0004-6361/201528030}, \href
  {https://ui.adsabs.harvard.edu/abs/2016A&A...592A..98I} {592, A98}

\bibitem[\protect\citeauthoryear{{Jansen} et~al.,}{{Jansen}
  et~al.}{2001}]{jansen01}
{Jansen} F.,  et~al., 2001, \mn@doi [\aap] {10.1051/0004-6361:20000036}, \href
  {https://ui.adsabs.harvard.edu/abs/2001A&A...365L...1J} {365, L1}

\bibitem[\protect\citeauthoryear{{Jiang} et~al.,}{{Jiang}
  et~al.}{2018}]{jiang18}
{Jiang} J.,  et~al., 2018, \mn@doi [\mnras] {10.1093/mnras/sty836}, \href
  {https://ui.adsabs.harvard.edu/abs/2018MNRAS.477.3711J} {477, 3711}

\bibitem[\protect\citeauthoryear{{Jiang}, {Walton}, {Fabian}  \&
  {Parker}}{{Jiang} et~al.}{2019}]{jiang19}
{Jiang} J.,  {Walton} D.~J.,  {Fabian} A.~C.,   {Parker} M.~L.,  2019, \mn@doi
  [\mnras] {10.1093/mnras/sty3228}, \href
  {https://ui.adsabs.harvard.edu/abs/2019MNRAS.483.2958J} {483, 2958}

\bibitem[\protect\citeauthoryear{{Jin}, {Ward}, {Done}  \& {Gelbord}}{{Jin}
  et~al.}{2012}]{jin12}
{Jin} C.,  {Ward} M.,  {Done} C.,   {Gelbord} J.,  2012, \mn@doi [\mnras]
  {10.1111/j.1365-2966.2011.19805.x}, \href
  {https://ui.adsabs.harvard.edu/abs/2012MNRAS.420.1825J} {420, 1825}

\bibitem[\protect\citeauthoryear{{Jones}, {Brenneman}, {Civano}, {Lanzuisi}  \&
  {Marchesi}}{{Jones} et~al.}{2020}]{jones20}
{Jones} M.,  {Brenneman} L.,  {Civano} F.,  {Lanzuisi} G.,   {Marchesi} S.,
  2020, arXiv e-prints, \href
  {https://ui.adsabs.harvard.edu/abs/2020arXiv200808588J} {p. arXiv:2008.08588}

\bibitem[\protect\citeauthoryear{{Jovanovi{\'c}}, {Zakharov}, {Popovi{\'c}}  \&
  {Petrovi{\'c}}}{{Jovanovi{\'c}} et~al.}{2008}]{jovanovic08}
{Jovanovi{\'c}} P.,  {Zakharov} A.~F.,  {Popovi{\'c}} L.~{\v{C}}.,
  {Petrovi{\'c}} T.,  2008, \mn@doi [\mnras]
  {10.1111/j.1365-2966.2008.13036.x}, \href
  {https://ui.adsabs.harvard.edu/abs/2008MNRAS.386..397J} {386, 397}

\bibitem[\protect\citeauthoryear{{Kamraj}, {Harrison}, {Balokovi{\'c}},
  {Lohfink}  \& {Brightman}}{{Kamraj} et~al.}{2018}]{kamraj18}
{Kamraj} N.,  {Harrison} F.~A.,  {Balokovi{\'c}} M.,  {Lohfink} A.,
  {Brightman} M.,  2018, \mn@doi [\apj] {10.3847/1538-4357/aadd0d}, \href
  {https://ui.adsabs.harvard.edu/abs/2018ApJ...866..124K} {866, 124}

\bibitem[\protect\citeauthoryear{{Kang}, {Wang}  \& {Kang}}{{Kang}
  et~al.}{2021}]{kang21}
{Kang} J.-L.,  {Wang} J.-X.,   {Kang} W.-Y.,  2021, \mn@doi [\mnras]
  {10.1093/mnras/stab039}, \href
  {https://ui.adsabs.harvard.edu/abs/2021MNRAS.tmp...68K} {}

\bibitem[\protect\citeauthoryear{{Kaspi}, {Smith}, {Netzer}, {Maoz}, {Jannuzi}
  \& {Giveon}}{{Kaspi} et~al.}{2000}]{kaspi00}
{Kaspi} S.,  {Smith} P.~S.,  {Netzer} H.,  {Maoz} D.,  {Jannuzi} B.~T.,
  {Giveon} U.,  2000, \mn@doi [\apj] {10.1086/308704}, \href
  {https://ui.adsabs.harvard.edu/abs/2000ApJ...533..631K} {533, 631}

\bibitem[\protect\citeauthoryear{{Kawamuro}, {Ueda}, {Tazaki}, {Ricci}  \&
  {Terashima}}{{Kawamuro} et~al.}{2016}]{kawamuro16}
{Kawamuro} T.,  {Ueda} Y.,  {Tazaki} F.,  {Ricci} C.,   {Terashima} Y.,  2016,
  \mn@doi [\apjs] {10.3847/0067-0049/225/1/14}, \href
  {https://ui.adsabs.harvard.edu/abs/2016ApJS..225...14K} {225, 14}

\bibitem[\protect\citeauthoryear{{Kewley}, {Groves}, {Kauffmann}  \&
  {Heckman}}{{Kewley} et~al.}{2006}]{kewley06}
{Kewley} L.~J.,  {Groves} B.,  {Kauffmann} G.,   {Heckman} T.,  2006, \mn@doi
  [\mnras] {10.1111/j.1365-2966.2006.10859.x}, \href
  {https://ui.adsabs.harvard.edu/abs/2006MNRAS.372..961K} {372, 961}

\bibitem[\protect\citeauthoryear{{Kollatschny}, {Ulbrich}, {Zetzl}, {Kaspi}  \&
  {Haas}}{{Kollatschny} et~al.}{2014}]{kollatschny14}
{Kollatschny} W.,  {Ulbrich} K.,  {Zetzl} M.,  {Kaspi} S.,   {Haas} M.,  2014,
  \mn@doi [\aap] {10.1051/0004-6361/201423901}, \href
  {https://ui.adsabs.harvard.edu/abs/2014A&A...566A.106K} {566, A106}

\bibitem[\protect\citeauthoryear{{Koss} et~al.,}{{Koss} et~al.}{2017}]{koss17}
{Koss} M.,  et~al., 2017, \mn@doi [\apj] {10.3847/1538-4357/aa8ec9}, \href
  {https://ui.adsabs.harvard.edu/abs/2017ApJ...850...74K} {850, 74}

\bibitem[\protect\citeauthoryear{{Lanz} et~al.,}{{Lanz} et~al.}{2019}]{lanz19}
{Lanz} L.,  et~al., 2019, \mn@doi [\apj] {10.3847/1538-4357/aaee6c}, \href
  {https://ui.adsabs.harvard.edu/abs/2019ApJ...870...26L} {870, 26}

\bibitem[\protect\citeauthoryear{{Laor}}{{Laor}}{1991}]{laor91}
{Laor} A.,  1991, \mn@doi [\apj] {10.1086/170257}, \href
  {https://ui.adsabs.harvard.edu/abs/1991ApJ...376...90L} {376, 90}

\bibitem[\protect\citeauthoryear{{Lu} et~al.,}{{Lu} et~al.}{2019}]{lu19}
{Lu} K.-X.,  et~al., 2019, \mn@doi [\apj] {10.3847/1538-4357/ab16e8}, \href
  {https://ui.adsabs.harvard.edu/abs/2019ApJ...877...23L} {877, 23}

\bibitem[\protect\citeauthoryear{{Lubi{\'n}ski} et~al.,}{{Lubi{\'n}ski}
  et~al.}{2016}]{lubinski16}
{Lubi{\'n}ski} P.,  et~al., 2016, \mn@doi [\mnras] {10.1093/mnras/stw454},
  \href {https://ui.adsabs.harvard.edu/abs/2016MNRAS.458.2454L} {458, 2454}

\bibitem[\protect\citeauthoryear{{Magdziarz} \& {Zdziarski}}{{Magdziarz} \&
  {Zdziarski}}{1995}]{magdziarz95}
{Magdziarz} P.,  {Zdziarski} A.~A.,  1995, \mn@doi [\mnras]
  {10.1093/mnras/273.3.837}, \href
  {https://ui.adsabs.harvard.edu/abs/1995MNRAS.273..837M} {273, 837}

\bibitem[\protect\citeauthoryear{{Mallick} et~al.,}{{Mallick}
  et~al.}{2018}]{mallick18}
{Mallick} L.,  et~al., 2018, \mn@doi [\mnras] {10.1093/mnras/sty1487}, \href
  {https://ui.adsabs.harvard.edu/abs/2018MNRAS.479..615M} {479, 615}

\bibitem[\protect\citeauthoryear{{Mao} et~al.,}{{Mao} et~al.}{2019}]{mao19}
{Mao} J.,  et~al., 2019, \mn@doi [\aap] {10.1051/0004-6361/201833191}, \href
  {https://ui.adsabs.harvard.edu/abs/2019A&A...621A..99M} {621, A99}

\bibitem[\protect\citeauthoryear{{Marconi} \& {Hunt}}{{Marconi} \&
  {Hunt}}{2003}]{marconi03}
{Marconi} A.,  {Hunt} L.~K.,  2003, \mn@doi [\apjl] {10.1086/375804}, \href
  {https://ui.adsabs.harvard.edu/abs/2003ApJ...589L..21M} {589, L21}

\bibitem[\protect\citeauthoryear{{Marinucci}, {Bianchi}, {Nicastro}, {Matt}  \&
  {Goulding}}{{Marinucci} et~al.}{2012}]{marinucci12}
{Marinucci} A.,  {Bianchi} S.,  {Nicastro} F.,  {Matt} G.,   {Goulding} A.~D.,
  2012, \mn@doi [\apj] {10.1088/0004-637X/748/2/130}, \href
  {https://ui.adsabs.harvard.edu/abs/2012ApJ...748..130M} {748, 130}

\bibitem[\protect\citeauthoryear{{Marinucci} et~al.,}{{Marinucci}
  et~al.}{2014a}]{marinucci14b}
{Marinucci} A.,  et~al., 2014a, \mn@doi [\mnras] {10.1093/mnras/stu404}, \href
  {https://ui.adsabs.harvard.edu/abs/2014MNRAS.440.2347M} {440, 2347}

\bibitem[\protect\citeauthoryear{{Marinucci} et~al.,}{{Marinucci}
  et~al.}{2014b}]{marinucci14a}
{Marinucci} A.,  et~al., 2014b, \mn@doi [\apj] {10.1088/0004-637X/787/1/83},
  \href {https://ui.adsabs.harvard.edu/abs/2014ApJ...787...83M} {787, 83}

\bibitem[\protect\citeauthoryear{{Marinucci} et~al.,}{{Marinucci}
  et~al.}{2016}]{marinucci16}
{Marinucci} A.,  et~al., 2016, \mn@doi [\mnras] {10.1093/mnrasl/slv178}, \href
  {https://ui.adsabs.harvard.edu/abs/2016MNRAS.456L..94M} {456, L94}

\bibitem[\protect\citeauthoryear{{Markoff}, {Nowak}  \& {Wilms}}{{Markoff}
  et~al.}{2005}]{markoff05}
{Markoff} S.,  {Nowak} M.~A.,   {Wilms} J.,  2005, \mn@doi [\apj]
  {10.1086/497628}, \href
  {https://ui.adsabs.harvard.edu/abs/2005ApJ...635.1203M} {635, 1203}

\bibitem[\protect\citeauthoryear{{Martocchia} \& {Matt}}{{Martocchia} \&
  {Matt}}{1996}]{martocchia96}
{Martocchia} A.,  {Matt} G.,  1996, \mn@doi [\mnras] {10.1093/mnras/282.4.L53},
  \href {https://ui.adsabs.harvard.edu/abs/1996MNRAS.282L..53M} {282, L53}

\bibitem[\protect\citeauthoryear{Massey}{Massey}{1951}]{massey51}
Massey F.~J.,  1951, \mn@doi [Journal of the American Statistical Association]
  {10.1080/01621459.1951.10500769}, 46, 68

\bibitem[\protect\citeauthoryear{{Matt}, {Perola}  \& {Piro}}{{Matt}
  et~al.}{1991}]{matt91}
{Matt} G.,  {Perola} G.~C.,   {Piro} L.,  1991, \aap, \href
  {https://ui.adsabs.harvard.edu/abs/1991A&A...247...25M} {247, 25}

\bibitem[\protect\citeauthoryear{{Matt} et~al.,}{{Matt} et~al.}{2015}]{matt15}
{Matt} G.,  et~al., 2015, \mn@doi [\mnras] {10.1093/mnras/stu2653}, \href
  {https://ui.adsabs.harvard.edu/abs/2015MNRAS.447.3029M} {447, 3029}

\bibitem[\protect\citeauthoryear{{McLure}, {Jarvis}, {Targett}, {Dunlop}  \&
  {Best}}{{McLure} et~al.}{2006}]{mclure06}
{McLure} R.~J.,  {Jarvis} M.~J.,  {Targett} T.~A.,  {Dunlop} J.~S.,   {Best}
  P.~N.,  2006, \mn@doi [\mnras] {10.1111/j.1365-2966.2006.10228.x}, \href
  {https://ui.adsabs.harvard.edu/abs/2006MNRAS.368.1395M} {368, 1395}

\bibitem[\protect\citeauthoryear{{Merloni}, {Heinz}  \& {di Matteo}}{{Merloni}
  et~al.}{2003}]{merloni03}
{Merloni} A.,  {Heinz} S.,   {di Matteo} T.,  2003, \mn@doi [\mnras]
  {10.1046/j.1365-2966.2003.07017.x}, \href
  {https://ui.adsabs.harvard.edu/abs/2003MNRAS.345.1057M} {345, 1057}

\bibitem[\protect\citeauthoryear{{Middei}, {Bianchi}, {Marinucci}, {Matt},
  {Petrucci}, {Tamborra}  \& {Tortosa}}{{Middei} et~al.}{2019}]{middei19}
{Middei} R.,  {Bianchi} S.,  {Marinucci} A.,  {Matt} G.,  {Petrucci} P.~O.,
  {Tamborra} F.,   {Tortosa} A.,  2019, \mn@doi [\aap]
  {10.1051/0004-6361/201935881}, \href
  {https://ui.adsabs.harvard.edu/abs/2019A&A...630A.131M} {630, A131}

\bibitem[\protect\citeauthoryear{{Mitsuda} et~al.,}{{Mitsuda}
  et~al.}{2007}]{mitsuda07}
{Mitsuda} K.,  et~al., 2007, \mn@doi [\pasj] {10.1093/pasj/59.sp1.S1}, \href
  {https://ui.adsabs.harvard.edu/abs/2007PASJ...59S...1M} {59, S1}

\bibitem[\protect\citeauthoryear{{Mohan} \& {Mangalam}}{{Mohan} \&
  {Mangalam}}{2014}]{mohan14}
{Mohan} P.,  {Mangalam} A.,  2014, \mn@doi [\apj] {10.1088/0004-637X/791/2/74},
  \href {https://ui.adsabs.harvard.edu/abs/2014ApJ...791...74M} {791, 74}

\bibitem[\protect\citeauthoryear{{Molina}, {Bassani}, {Malizia}, {Stephen},
  {Bird}, {Bazzano}  \& {Ubertini}}{{Molina} et~al.}{2013}]{molina13}
{Molina} M.,  {Bassani} L.,  {Malizia} A.,  {Stephen} J.~B.,  {Bird} A.~J.,
  {Bazzano} A.,   {Ubertini} P.,  2013, \mn@doi [\mnras]
  {10.1093/mnras/stt844}, \href
  {https://ui.adsabs.harvard.edu/abs/2013MNRAS.433.1687M} {433, 1687}

\bibitem[\protect\citeauthoryear{{Mor}, {Netzer}  \& {Elitzur}}{{Mor}
  et~al.}{2009}]{mor09}
{Mor} R.,  {Netzer} H.,   {Elitzur} M.,  2009, \mn@doi [\apj]
  {10.1088/0004-637X/705/1/298}, \href
  {https://ui.adsabs.harvard.edu/abs/2009ApJ...705..298M} {705, 298}

\bibitem[\protect\citeauthoryear{{Mundo} et~al.,}{{Mundo}
  et~al.}{2020}]{mundo20}
{Mundo} S.~A.,  et~al., 2020, \mn@doi [\mnras] {10.1093/mnras/staa1744}, \href
  {https://ui.adsabs.harvard.edu/abs/2020MNRAS.tmp.1878M} {}

\bibitem[\protect\citeauthoryear{{Murphy} \& {Yaqoob}}{{Murphy} \&
  {Yaqoob}}{2009}]{murphy09}
{Murphy} K.~D.,  {Yaqoob} T.,  2009, \mn@doi [\mnras]
  {10.1111/j.1365-2966.2009.15025.x}, \href
  {https://ui.adsabs.harvard.edu/abs/2009MNRAS.397.1549M} {397, 1549}

\bibitem[\protect\citeauthoryear{{Murray}, {Chiang}, {Grossman}  \&
  {Voit}}{{Murray} et~al.}{1995}]{murray95}
{Murray} N.,  {Chiang} J.,  {Grossman} S.~A.,   {Voit} G.~M.,  1995, \mn@doi
  [\apj] {10.1086/176238}, \href
  {https://ui.adsabs.harvard.edu/abs/1995ApJ...451..498M} {451, 498}

\bibitem[\protect\citeauthoryear{{Mushotzky}, {Done}  \& {Pounds}}{{Mushotzky}
  et~al.}{1993}]{mushotzky93}
{Mushotzky} R.~F.,  {Done} C.,   {Pounds} K.~A.,  1993, \mn@doi [\araa]
  {10.1146/annurev.aa.31.090193.003441}, \href
  {https://ui.adsabs.harvard.edu/abs/1993ARA&A..31..717M} {31, 717}

\bibitem[\protect\citeauthoryear{{Nandra} \& {Pounds}}{{Nandra} \&
  {Pounds}}{1994}]{nandra94}
{Nandra} K.,  {Pounds} K.~A.,  1994, \mn@doi [\mnras]
  {10.1093/mnras/268.2.405}, \href
  {https://ui.adsabs.harvard.edu/abs/1994MNRAS.268..405N} {268, 405}

\bibitem[\protect\citeauthoryear{{Nandra}, {George}, {Mushotzky}, {Turner}  \&
  {Yaqoob}}{{Nandra} et~al.}{1997}]{nandra97}
{Nandra} K.,  {George} I.~M.,  {Mushotzky} R.~F.,  {Turner} T.~J.,   {Yaqoob}
  T.,  1997, \mn@doi [\apj] {10.1086/303721}, \href
  {https://ui.adsabs.harvard.edu/abs/1997ApJ...477..602N} {477, 602}

\bibitem[\protect\citeauthoryear{{Netzer} et~al.,}{{Netzer}
  et~al.}{1990}]{netzer90}
{Netzer} H.,  et~al., 1990, \mn@doi [\apj] {10.1086/168594}, \href
  {https://ui.adsabs.harvard.edu/abs/1990ApJ...353..108N} {353, 108}

\bibitem[\protect\citeauthoryear{{Oh} et~al.,}{{Oh} et~al.}{2018}]{oh18}
{Oh} K.,  et~al., 2018, \mn@doi [\apjs] {10.3847/1538-4365/aaa7fd}, \href
  {https://ui.adsabs.harvard.edu/abs/2018ApJS..235....4O} {235, 4}

\bibitem[\protect\citeauthoryear{{Oliva}, {Origlia}, {Maiolino}  \&
  {Moorwood}}{{Oliva} et~al.}{1999}]{oliva99}
{Oliva} E.,  {Origlia} L.,  {Maiolino} R.,   {Moorwood} A.~F.~M.,  1999, \aap,
  \href {https://ui.adsabs.harvard.edu/abs/1999A&A...350....9O} {350, 9}

\bibitem[\protect\citeauthoryear{{Panessa} et~al.,}{{Panessa}
  et~al.}{2011}]{panessa11}
{Panessa} F.,  et~al., 2011, \mn@doi [\mnras]
  {10.1111/j.1365-2966.2011.19268.x}, \href
  {https://ui.adsabs.harvard.edu/abs/2011MNRAS.417.2426P} {417, 2426}

\bibitem[\protect\citeauthoryear{{Parker} et~al.,}{{Parker}
  et~al.}{2019}]{parker19}
{Parker} M.~L.,  et~al., 2019, \mn@doi [\mnras] {10.1093/mnrasl/sly224}, \href
  {https://ui.adsabs.harvard.edu/abs/2019MNRAS.483L..88P} {483, L88}

\bibitem[\protect\citeauthoryear{{Perola}, {Matt}, {Cappi}, {Fiore},
  {Guainazzi}, {Maraschi}, {Petrucci}  \& {Piro}}{{Perola}
  et~al.}{2002}]{perola02}
{Perola} G.~C.,  {Matt} G.,  {Cappi} M.,  {Fiore} F.,  {Guainazzi} M.,
  {Maraschi} L.,  {Petrucci} P.~O.,   {Piro} L.,  2002, \mn@doi [\aap]
  {10.1051/0004-6361:20020658}, \href
  {https://ui.adsabs.harvard.edu/abs/2002A&A...389..802P} {389, 802}

\bibitem[\protect\citeauthoryear{{Peterson}, {Wanders}, {Bertram}, {Hunley},
  {Pogge}  \& {Wagner}}{{Peterson} et~al.}{1998}]{peterson98}
{Peterson} B.~M.,  {Wanders} I.,  {Bertram} R.,  {Hunley} J.~F.,  {Pogge}
  R.~W.,   {Wagner} R.~M.,  1998, \mn@doi [\apj] {10.1086/305813}, \href
  {https://ui.adsabs.harvard.edu/abs/1998ApJ...501...82P} {501, 82}

\bibitem[\protect\citeauthoryear{{Peterson} et~al.,}{{Peterson}
  et~al.}{2002}]{peterson02}
{Peterson} B.~M.,  et~al., 2002, \mn@doi [\apj] {10.1086/344197}, \href
  {https://ui.adsabs.harvard.edu/abs/2002ApJ...581..197P} {581, 197}

\bibitem[\protect\citeauthoryear{{Peterson} et~al.,}{{Peterson}
  et~al.}{2004}]{peterson04}
{Peterson} B.~M.,  et~al., 2004, \mn@doi [\apj] {10.1086/423269}, \href
  {https://ui.adsabs.harvard.edu/abs/2004ApJ...613..682P} {613, 682}

\bibitem[\protect\citeauthoryear{{Petrucci} et~al.,}{{Petrucci}
  et~al.}{2001}]{petrucci01}
{Petrucci} P.~O.,  et~al., 2001, \mn@doi [\apj] {10.1086/321629}, \href
  {https://ui.adsabs.harvard.edu/abs/2001ApJ...556..716P} {556, 716}

\bibitem[\protect\citeauthoryear{{Petrucci}, {Ursini}, {De Rosa}, {Bianchi},
  {Cappi}, {Matt}, {Dadina}  \& {Malzac}}{{Petrucci} et~al.}{2018}]{petrucci18}
{Petrucci} P.~O.,  {Ursini} F.,  {De Rosa} A.,  {Bianchi} S.,  {Cappi} M.,
  {Matt} G.,  {Dadina} M.,   {Malzac} J.,  2018, \mn@doi [\aap]
  {10.1051/0004-6361/201731580}, \href
  {https://ui.adsabs.harvard.edu/abs/2018A&A...611A..59P} {611, A59}

\bibitem[\protect\citeauthoryear{{Petrucci} et~al.,}{{Petrucci}
  et~al.}{2020}]{petrucci20}
{Petrucci} P.~O.,  et~al., 2020, \mn@doi [\aap] {10.1051/0004-6361/201937011},
  \href {https://ui.adsabs.harvard.edu/abs/2020A&A...634A..85P} {634, A85}

\bibitem[\protect\citeauthoryear{{Pounds}, {Nandra}, {Stewart}, {George}  \&
  {Fabian}}{{Pounds} et~al.}{1990}]{pounds90}
{Pounds} K.~A.,  {Nandra} K.,  {Stewart} G.~C.,  {George} I.~M.,   {Fabian}
  A.~C.,  1990, \mn@doi [\nat] {10.1038/344132a0}, \href
  {https://ui.adsabs.harvard.edu/abs/1990Natur.344..132P} {344, 132}

\bibitem[\protect\citeauthoryear{{Poutanen} \& {Svensson}}{{Poutanen} \&
  {Svensson}}{1996}]{poutanen96}
{Poutanen} J.,  {Svensson} R.,  1996, \mn@doi [\apj] {10.1086/177865}, \href
  {https://ui.adsabs.harvard.edu/abs/1996ApJ...470..249P} {470, 249}

\bibitem[\protect\citeauthoryear{{Ramos Almeida} et~al.,}{{Ramos Almeida}
  et~al.}{2011}]{ramos11}
{Ramos Almeida} C.,  et~al., 2011, \mn@doi [\apj] {10.1088/0004-637X/731/2/92},
  \href {https://ui.adsabs.harvard.edu/abs/2011ApJ...731...92R} {731, 92}

\bibitem[\protect\citeauthoryear{{Rees}}{{Rees}}{1984}]{rees84}
{Rees} M.~J.,  1984, \mn@doi [\araa] {10.1146/annurev.aa.22.090184.002351},
  \href {https://ui.adsabs.harvard.edu/abs/1984ARA&A..22..471R} {22, 471}

\bibitem[\protect\citeauthoryear{{Reynolds}}{{Reynolds}}{1997}]{reynolds97b}
{Reynolds} C.~S.,  1997, \mn@doi [\mnras] {10.1093/mnras/286.3.513}, \href
  {https://ui.adsabs.harvard.edu/abs/1997MNRAS.286..513R} {286, 513}

\bibitem[\protect\citeauthoryear{{Reynolds} \& {Begelman}}{{Reynolds} \&
  {Begelman}}{1997}]{reynolds97a}
{Reynolds} C.~S.,  {Begelman} M.~C.,  1997, \mn@doi [\apj] {10.1086/304703},
  \href {https://ui.adsabs.harvard.edu/abs/1997ApJ...488..109R} {488, 109}

\bibitem[\protect\citeauthoryear{{Reynolds} \& {Fabian}}{{Reynolds} \&
  {Fabian}}{1997}]{reynolds97c}
{Reynolds} C.~S.,  {Fabian} A.~C.,  1997, \mn@doi [\mnras]
  {10.1093/mnras/290.1.L1}, \href
  {https://ui.adsabs.harvard.edu/abs/1997MNRAS.290L...1R} {290, L1}

\bibitem[\protect\citeauthoryear{{Reynolds} \& {Fabian}}{{Reynolds} \&
  {Fabian}}{2008}]{reynolds08}
{Reynolds} C.~S.,  {Fabian} A.~C.,  2008, \mn@doi [\apj] {10.1086/527344},
  \href {https://ui.adsabs.harvard.edu/abs/2008ApJ...675.1048R} {675, 1048}

\bibitem[\protect\citeauthoryear{{Reynolds} \& {Nowak}}{{Reynolds} \&
  {Nowak}}{2003}]{reynolds03}
{Reynolds} C.~S.,  {Nowak} M.~A.,  2003, \mn@doi [\physrep]
  {10.1016/S0370-1573(02)00584-7}, \href
  {https://ui.adsabs.harvard.edu/abs/2003PhR...377..389R} {377, 389}

\bibitem[\protect\citeauthoryear{{Ricci}, {Walter}, {Courvoisier}  \&
  {Paltani}}{{Ricci} et~al.}{2011}]{ricci11}
{Ricci} C.,  {Walter} R.,  {Courvoisier} T.~J.~L.,   {Paltani} S.,  2011,
  \mn@doi [\aap] {10.1051/0004-6361/201016409}, \href
  {https://ui.adsabs.harvard.edu/abs/2011A&A...532A.102R} {532, A102}

\bibitem[\protect\citeauthoryear{{Ricci} et~al.,}{{Ricci}
  et~al.}{2017}]{ricci17}
{Ricci} C.,  et~al., 2017, \mn@doi [\apjs] {10.3847/1538-4365/aa96ad}, \href
  {https://ui.adsabs.harvard.edu/abs/2017ApJS..233...17R} {233, 17}

\bibitem[\protect\citeauthoryear{{Ricci} et~al.,}{{Ricci}
  et~al.}{2018}]{ricci18}
{Ricci} C.,  et~al., 2018, \mn@doi [\mnras] {10.1093/mnras/sty1879}, \href
  {https://ui.adsabs.harvard.edu/abs/2018MNRAS.480.1819R} {480, 1819}

\bibitem[\protect\citeauthoryear{{Rivers}, {Markowitz}  \&
  {Rothschild}}{{Rivers} et~al.}{2013}]{rivers13}
{Rivers} E.,  {Markowitz} A.,   {Rothschild} R.,  2013, \mn@doi [\apj]
  {10.1088/0004-637X/772/2/114}, \href
  {https://ui.adsabs.harvard.edu/abs/2013ApJ...772..114R} {772, 114}

\bibitem[\protect\citeauthoryear{{Rivers} et~al.,}{{Rivers}
  et~al.}{2015}]{rivers15}
{Rivers} E.,  et~al., 2015, \mn@doi [\apj] {10.1088/0004-637X/804/2/107}, \href
  {https://ui.adsabs.harvard.edu/abs/2015ApJ...804..107R} {804, 107}

\bibitem[\protect\citeauthoryear{{R{\'o}{\.z}a{\'n}ska}, {Malzac}, {Belmont},
  {Czerny}  \& {Petrucci}}{{R{\'o}{\.z}a{\'n}ska} et~al.}{2015}]{rozanska15}
{R{\'o}{\.z}a{\'n}ska} A.,  {Malzac} J.,  {Belmont} R.,  {Czerny} B.,
  {Petrucci} P.~O.,  2015, \mn@doi [\aap] {10.1051/0004-6361/201526288}, \href
  {https://ui.adsabs.harvard.edu/abs/2015A&A...580A..77R} {580, A77}

\bibitem[\protect\citeauthoryear{{Sanders} \& {Mirabel}}{{Sanders} \&
  {Mirabel}}{1996}]{sanders96}
{Sanders} D.~B.,  {Mirabel} I.~F.,  1996, \mn@doi [\araa]
  {10.1146/annurev.astro.34.1.749}, \href
  {https://ui.adsabs.harvard.edu/abs/1996ARA&A..34..749S} {34, 749}

\bibitem[\protect\citeauthoryear{{Sanders}, {Soifer}, {Elias}, {Madore},
  {Matthews}, {Neugebauer}  \& {Scoville}}{{Sanders} et~al.}{1988}]{sanders88}
{Sanders} D.~B.,  {Soifer} B.~T.,  {Elias} J.~H.,  {Madore} B.~F.,  {Matthews}
  K.,  {Neugebauer} G.,   {Scoville} N.~Z.,  1988, \mn@doi [\apj]
  {10.1086/165983}, \href
  {https://ui.adsabs.harvard.edu/abs/1988ApJ...325...74S} {325, 74}

\bibitem[\protect\citeauthoryear{{Santos-Lle{\'o}} et~al.,}{{Santos-Lle{\'o}}
  et~al.}{1997}]{santos97}
{Santos-Lle{\'o}} M.,  et~al., 1997, \mn@doi [\apjs] {10.1086/313046}, \href
  {https://ui.adsabs.harvard.edu/abs/1997ApJS..112..271S} {112, 271}

\bibitem[\protect\citeauthoryear{{Shappee} et~al.,}{{Shappee}
  et~al.}{2014}]{shappee14}
{Shappee} B.~J.,  et~al., 2014, \mn@doi [\apj] {10.1088/0004-637X/788/1/48},
  \href {https://ui.adsabs.harvard.edu/abs/2014ApJ...788...48S} {788, 48}

\bibitem[\protect\citeauthoryear{{Shu}, {Yaqoob}  \& {Wang}}{{Shu}
  et~al.}{2010}]{shu10}
{Shu} X.~W.,  {Yaqoob} T.,   {Wang} J.~X.,  2010, \mn@doi [\apjs]
  {10.1088/0067-0049/187/2/581}, \href
  {https://ui.adsabs.harvard.edu/abs/2010ApJS..187..581S} {187, 581}

\bibitem[\protect\citeauthoryear{{Sobolewska} \& {Done}}{{Sobolewska} \&
  {Done}}{2007}]{sobolewska07}
{Sobolewska} M.~A.,  {Done} C.,  2007, \mn@doi [\mnras]
  {10.1111/j.1365-2966.2006.11117.x}, \href
  {https://ui.adsabs.harvard.edu/abs/2007MNRAS.374..150S} {374, 150}

\bibitem[\protect\citeauthoryear{{Stern}, {Poutanen}, {Svensson}, {Sikora}  \&
  {Begelman}}{{Stern} et~al.}{1995}]{stern95}
{Stern} B.~E.,  {Poutanen} J.,  {Svensson} R.,  {Sikora} M.,   {Begelman}
  M.~C.,  1995, \mn@doi [\apjl] {10.1086/309617}, \href
  {https://ui.adsabs.harvard.edu/abs/1995ApJ...449L..13S} {449, L13}

\bibitem[\protect\citeauthoryear{{Stern} et~al.,}{{Stern}
  et~al.}{2005}]{stern05}
{Stern} D.,  et~al., 2005, \mn@doi [\apj] {10.1086/432523}, \href
  {https://ui.adsabs.harvard.edu/abs/2005ApJ...631..163S} {631, 163}

\bibitem[\protect\citeauthoryear{{Str{\"u}der} et~al.,}{{Str{\"u}der}
  et~al.}{2001}]{struder01}
{Str{\"u}der} L.,  et~al., 2001, \mn@doi [\aap] {10.1051/0004-6361:20000066},
  \href {https://ui.adsabs.harvard.edu/abs/2001A&A...365L..18S} {365, L18}

\bibitem[\protect\citeauthoryear{{Sulentic}, {Marziani}, {Zwitter}, {Calvani}
  \& {Dultzin-Hacyan}}{{Sulentic} et~al.}{1998}]{sulentic98}
{Sulentic} J.~W.,  {Marziani} P.,  {Zwitter} T.,  {Calvani} M.,
  {Dultzin-Hacyan} D.,  1998, \mn@doi [\apj] {10.1086/305795}, \href
  {https://ui.adsabs.harvard.edu/abs/1998ApJ...501...54S} {501, 54}

\bibitem[\protect\citeauthoryear{{Tortosa} et~al.,}{{Tortosa}
  et~al.}{2017}]{tortosa17}
{Tortosa} A.,  et~al., 2017, \mn@doi [\mnras] {10.1093/mnras/stw3301}, \href
  {https://ui.adsabs.harvard.edu/abs/2017MNRAS.466.4193T} {466, 4193}

\bibitem[\protect\citeauthoryear{{Tortosa}, {Bianchi}, {Marinucci}, {Matt}  \&
  {Petrucci}}{{Tortosa} et~al.}{2018}]{tortosa18}
{Tortosa} A.,  {Bianchi} S.,  {Marinucci} A.,  {Matt} G.,   {Petrucci} P.~O.,
  2018, \mn@doi [\aap] {10.1051/0004-6361/201732382}, \href
  {https://ui.adsabs.harvard.edu/abs/2018A&A...614A..37T} {614, A37}

\bibitem[\protect\citeauthoryear{{Turner} et~al.,}{{Turner}
  et~al.}{2001}]{turner01}
{Turner} M.~J.~L.,  et~al., 2001, \mn@doi [\aap] {10.1051/0004-6361:20000087},
  \href {https://ui.adsabs.harvard.edu/abs/2001A&A...365L..27T} {365, L27}

\bibitem[\protect\citeauthoryear{{U} et~al.,}{{U} et~al.}{2013}]{u13}
{U} V.,  et~al., 2013, \mn@doi [\apj] {10.1088/0004-637X/775/2/115}, \href
  {https://ui.adsabs.harvard.edu/abs/2013ApJ...775..115U} {775, 115}

\bibitem[\protect\citeauthoryear{{Ulrich}, {Maraschi}  \& {Urry}}{{Ulrich}
  et~al.}{1997}]{ulrich97}
{Ulrich} M.-H.,  {Maraschi} L.,   {Urry} C.~M.,  1997, \mn@doi [\araa]
  {10.1146/annurev.astro.35.1.445}, \href
  {https://ui.adsabs.harvard.edu/abs/1997ARA&A..35..445U} {35, 445}

\bibitem[\protect\citeauthoryear{{Urry} \& {Padovani}}{{Urry} \&
  {Padovani}}{1995}]{urry95}
{Urry} C.~M.,  {Padovani} P.,  1995, \mn@doi [\pasp] {10.1086/133630}, \href
  {https://ui.adsabs.harvard.edu/abs/1995PASP..107..803U} {107, 803}

\bibitem[\protect\citeauthoryear{{Vasylenko}, {Zhdanov}  \&
  {Fedorova}}{{Vasylenko} et~al.}{2015}]{vasylenko15}
{Vasylenko} A.~A.,  {Zhdanov} V.~I.,   {Fedorova} E.~V.,  2015, \mn@doi [\apss]
  {10.1007/s10509-015-2585-z}, \href
  {https://ui.adsabs.harvard.edu/abs/2015Ap&SS.360...71V} {360, 37}

\bibitem[\protect\citeauthoryear{{Walsh}, {van den Bosch}, {Barth}  \&
  {Sarzi}}{{Walsh} et~al.}{2012}]{walsh12}
{Walsh} J.~L.,  {van den Bosch} R. C.~E.,  {Barth} A.~J.,   {Sarzi} M.,  2012,
  \mn@doi [\apj] {10.1088/0004-637X/753/1/79}, \href
  {https://ui.adsabs.harvard.edu/abs/2012ApJ...753...79W} {753, 79}

\bibitem[\protect\citeauthoryear{{Wilms}, {Allen}  \& {McCray}}{{Wilms}
  et~al.}{2000}]{wilms2000}
{Wilms} J.,  {Allen} A.,   {McCray} R.,  2000, \mn@doi [\apj] {10.1086/317016},
  \href {https://ui.adsabs.harvard.edu/abs/2000ApJ...542..914W} {542, 914}

\bibitem[\protect\citeauthoryear{{Winter}, {Lewis}, {Koss}, {Veilleux},
  {Keeney}  \& {Mushotzky}}{{Winter} et~al.}{2010}]{winter10}
{Winter} L.~M.,  {Lewis} K.~T.,  {Koss} M.,  {Veilleux} S.,  {Keeney} B.,
  {Mushotzky} R.~F.,  2010, \mn@doi [\apj] {10.1088/0004-637X/710/1/503}, \href
  {https://ui.adsabs.harvard.edu/abs/2010ApJ...710..503W} {710, 503}

\bibitem[\protect\citeauthoryear{{Winter}, {Veilleux}, {McKernan}  \&
  {Kallman}}{{Winter} et~al.}{2012}]{winter12}
{Winter} L.~M.,  {Veilleux} S.,  {McKernan} B.,   {Kallman} T.~R.,  2012,
  \mn@doi [\apj] {10.1088/0004-637X/745/2/107}, \href
  {https://ui.adsabs.harvard.edu/abs/2012ApJ...745..107W} {745, 107}

\bibitem[\protect\citeauthoryear{{Xu}, {Garc{\'\i}a}, {Walton}, {Connors},
  {Madsen}  \& {Harrison}}{{Xu} et~al.}{2021}]{xu21}
{Xu} Y.,  {Garc{\'\i}a} J.~A.,  {Walton} D.~J.,  {Connors} R. M.~T.,  {Madsen}
  K.,   {Harrison} F.~A.,  2021, \mn@doi [\apj] {10.3847/1538-4357/abf430},
  \href {https://ui.adsabs.harvard.edu/abs/2021ApJ...913...13X} {913, 13}

\bibitem[\protect\citeauthoryear{{Yang} et~al.,}{{Yang} et~al.}{2018}]{yang18}
{Yang} Q.,  et~al., 2018, \mn@doi [\apj] {10.3847/1538-4357/aaca3a}, \href
  {https://ui.adsabs.harvard.edu/abs/2018ApJ...862..109Y} {862, 109}

\bibitem[\protect\citeauthoryear{{Zdziarski}}{{Zdziarski}}{1985}]{zdziarski85}
{Zdziarski} A.~A.,  1985, \mn@doi [\apj] {10.1086/162912}, \href
  {https://ui.adsabs.harvard.edu/abs/1985ApJ...289..514Z} {289, 514}

\bibitem[\protect\citeauthoryear{{Zdziarski} \& {Gierli{\'n}ski}}{{Zdziarski}
  \& {Gierli{\'n}ski}}{2004}]{zdziarski04}
{Zdziarski} A.~A.,  {Gierli{\'n}ski} M.,  2004, \mn@doi [Progress of
  Theoretical Physics Supplement] {10.1143/PTPS.155.99}, \href
  {https://ui.adsabs.harvard.edu/abs/2004PThPS.155...99Z} {155, 99}

\bibitem[\protect\citeauthoryear{{Zdziarski}, {Lubi{\'n}ski}  \&
  {Smith}}{{Zdziarski} et~al.}{1999}]{zdziarski99}
{Zdziarski} A.~A.,  {Lubi{\'n}ski} P.,   {Smith} D.~A.,  1999, \mn@doi [\mnras]
  {10.1046/j.1365-8711.1999.02343.x}, \href
  {https://ui.adsabs.harvard.edu/abs/1999MNRAS.303L..11Z} {303, L11}

\bibitem[\protect\citeauthoryear{{Zhao}, {Marchesi}, {Ajello}, {Cole}, {Hu},
  {Silver}  \& {Torres-Alb{\`a}}}{{Zhao} et~al.}{2020a}]{zhao20b}
{Zhao} X.,  {Marchesi} S.,  {Ajello} M.,  {Cole} D.,  {Hu} Z.,  {Silver} R.,
  {Torres-Alb{\`a}} N.,  2020a, arXiv e-prints, \href
  {https://ui.adsabs.harvard.edu/abs/2020arXiv201103851Z} {p. arXiv:2011.03851}

\bibitem[\protect\citeauthoryear{{Zhao}, {Marchesi}, {Ajello}, {Balokovi{\'c}}
  \& {Fischer}}{{Zhao} et~al.}{2020b}]{zhao20}
{Zhao} X.,  {Marchesi} S.,  {Ajello} M.,  {Balokovi{\'c}} M.,   {Fischer} T.,
  2020b, \mn@doi [\apj] {10.3847/1538-4357/ab879d}, \href
  {https://ui.adsabs.harvard.edu/abs/2020ApJ...894...71Z} {894, 71}

\bibitem[\protect\citeauthoryear{{de Rosa}, {Bassani}, {Ubertini}, {Panessa},
  {Malizia}, {Dean}  \& {Walter}}{{de Rosa} et~al.}{2008}]{derosa08}
{de Rosa} A.,  {Bassani} L.,  {Ubertini} P.,  {Panessa} F.,  {Malizia} A.,
  {Dean} A.~J.,   {Walter} R.,  2008, \mn@doi [\aap]
  {10.1051/0004-6361:20078319}, \href
  {https://ui.adsabs.harvard.edu/abs/2008A&A...483..749D} {483, 749}

\bibitem[\protect\citeauthoryear{{de Rosa} et~al.,}{{de Rosa}
  et~al.}{2012}]{deRosa12}
{de Rosa} A.,  et~al., 2012, \mn@doi [\mnras]
  {10.1111/j.1365-2966.2011.20167.x}, \href
  {https://ui.adsabs.harvard.edu/abs/2012MNRAS.420.2087D} {420, 2087}

\makeatother
\end{thebibliography}

\section*{Appendix: Comments on the Spectral Fits}
\textbf{Fairall 9:} We ignore below 0.7 keV for this source due to strong emission near 0.5 keV. Additionally, there is no MOS2 spectrum for this source from XMM-Newton.\\
\textbf{Mrk 359:} We ignore below 0.7 keV for this source due to discrepancies between the PN and MOS spectra from XMM-Newton.\\
\textbf{NGC 1068:} We fit this spectrum above 3 keV to avoid the strong complex of emission lines below this energy. Without this, the fit is completed dominated by the high S/N in the lower energy channels. There is no available PN spectrum for the XMM-Newton epoch we use.\\
\textbf{3C 120:} The XMM-Newton epoch we used had no PN observation. \\
\textbf{Ark 120:} We ignore below 0.7 keV for this source.\\
\textbf{Mrk3:} Due to a strong soft component, with poor quality data, we ignore below 1.0 keV for this source.\\ 
\textbf{IRAS09149-6206:} Because of strong emission lines, we fit above 1.0 keV for this source.\\
\textbf{NGC3227:} We ignore below 0.7 keV for this object.\\
\textbf{NGC4151:} We fit only above 2.0 keV to avoid strong emission lines. \\
\textbf{NGC4593:} For the XMM-Newton observations, only a MOS1 spectrum was available. \\
\textbf{NGC4785:} Because of emission lines and weak constraints on the soft X-rays, we ignore below 3.0 keV.\\
\textbf{Mrk273:} To account for strong soft emission, we add an extra \textsc{zgauss} component with an energy of 0.84 keV and a width of 0.16 keV.\\
\textbf{Mrk1392:} We excluded the MOS spectra from XMM-Newton for this object as fits including this data were unstable.  \\
\textbf{IIZw171:} We ignored below 0.8 keV for this object. \\
\textbf{MR2251-178:} We ignored below 1.0 keV because of evidence for broad emission features. \\
\textbf{Mrk926:} We ignored below 0.8 keV for this object due to strong absorption features.  \\
\textbf{NGC7582:} There is an extremely strong soft excess and line emission below 3.0 keV that we have ignored. \\

\bsp
\label{lastpage}
\end{document}